\documentclass[twocolumn]{aastex631}

\usepackage{natbib}
\usepackage{color}
\usepackage{apjfonts}
\usepackage{xfrac}
\usepackage{ulem}
\usepackage{soul}
\usepackage{amsmath,xspace}
\usepackage{hyperref}
\usepackage{etoolbox}

\newtoggle{longnames}
\toggletrue{longnames}
\def\firstletter#1#2@{#1}
\newcommand{\an}[1]{\iftoggle{longnames}{#1}{\firstletter #1@.}}
\newcommand{\ann}[2]{\iftoggle{longnames}{#1}{#2}}

\hypersetup{colorlinks=true,pdfborder={0 0 0}}
\newif\iflatexml\latexmlfalse
\DeclareGraphicsExtensions{.pdf,.PDF,.png,.PNG,.jpg,.JPG,.jpeg,.JPEG}

\usepackage{graphicx}
\usepackage[space]{grffile}
\usepackage{latexsym}
\usepackage{textcomp}
\usepackage{longtable}
\usepackage{multirow,booktabs}
\usepackage{amsfonts,amsmath,amssymb}
\usepackage{natbib}
\usepackage{url}
\usepackage[T1]{fontenc}
\usepackage{ae,aecompl}

\newif\iflatexml\latexmlfalse
\DeclareGraphicsExtensions{.pdf,.PDF,.png,.PNG,.jpg,.JPG,.jpeg,.JPEG}

\usepackage[utf8]{inputenc}
\usepackage[english]{babel}
\usepackage{graphicx}

\newcommand\MWS{MWS}
\newcommand\insitu{in situ}

\newcommand{\mwswd}{\textsc{mws-wd}}
\newcommand{\nearby}{\textsc{mws-nearby}}
\newcommand{\mwsbhb}{\textsc{mws-bhb}}
\newcommand{\mwsrrlyr}{\textsc{mws-rrlyr}}
\newcommand{\mainred}{\textsc{main-red}}
\newcommand{\mainblue}{\textsc{main-blue}}
\newcommand{\mainbroad}{\textsc{main-broad}}
\newcommand{\faintred}{\textsc{faint-red}}
\newcommand{\faintblue}{\textsc{faint-blue}}

\newcommand{\msun}{\mathrm{M_\odot}}
\newcommand{\Teff}{\mbox{$T_\mathrm{eff}$}}
\newcommand{\feh}{\mathrm{[Fe/H]}}
\DeclareRobustCommand{\kms}{\mathrm{km\,s^{-1}}}

\newcommand{\legacy}{LS}
\newcommand{\redrock}{Redrock}

\newcommand{\svone}{SV1}
\newcommand{\svthree}{SV3}
\newcommand{\SNR}{S/N}

\newcommand{\updatecite}[1]{\textcolor{black}{#1}}

\newcommand{\bugfix}[1]{\textcolor{black}{#1}}
\newcommand{\cwr}[1]{\textcolor{black}{#1}}
\newcommand{\reply}[1]{\textcolor{black}{#1}}
\newcommand{\correct}[1]{\textcolor{black}{#1}}
\newcommand{\proof}[1]{\textcolor{black}{#1}}
\received{2022 September 8}
\revised{2022 December 20}
\accepted{2023 January 16}

\shorttitle{Overview of the DESI Milky Way Survey}
\shortauthors{Cooper et al.}

\begin{document}

\title{Overview of the DESI Milky Way Survey}

\author[0000-0001-8274-158X]{\an{Andrew}~P.~Cooper}
\affil{Institute of Astronomy and Department of Physics, National Tsing Hua University, 101 Kuang-Fu Rd. Sec. 2, Hsinchu 30013, Taiwan}
\affil{Center for Informatics and Computation in Astronomy, NTHU, 101 Kuang-Fu Rd. Sec. 2, Hsinchu 30013, Taiwan}
\affil{Physics Division, National Center for Theoretical Sciences, Taipei 10617, Taiwan}
\email{apcooper@gapp.nthu.edu.tw}
\author[0000-0003-2644-135X]{\an{Sergey}~E.~Koposov}
\affil{Institute for Astronomy, University of Edinburgh, Royal Observatory, Blackford Hill, Edinburgh EH9 3HJ, UK}
\affil{Institute of Astronomy, University of Cambridge, Madingley Rd., Cambridge CB3 0HA, UK}
\author[0000-0002-0084-572X]{\an{Carlos}~Allende~Prieto}
\affil{Instituto de Astrof\'{i}sica de Canarias, C/ Vía L\'{a}ctea s/n, E-38205 La Laguna, Tenerife, Spain}
\affil{Universidad de La Laguna, Dpto.\ Astrof\'{i}sica, E-38206 La Laguna,Tenerife, Spain}
\author[0000-0003-1543-5405]{\an{Christopher}~J.~Manser}
\affil{Imperial College London, South Kensington Campus, London SW7 2AZ, UK}
\author{\an{Namitha}~Kizhuprakkat}
\affil{Institute of Astronomy and Department of Physics, National Tsing Hua University, 101 Kuang-Fu Rd. Sec. 2, Hsinchu 30013, Taiwan}
\affil{Center for Informatics and Computation in Astronomy, NTHU, 101 Kuang-Fu Rd. Sec. 2, Hsinchu 30013, Taiwan}
\author{\an{Adam}~D.~Myers}
\affil{Department of Physics \& Astronomy, University  of Wyoming, 1000 E. University, Dept.~3905, Laramie, WY 82071, USA}
\author{\an{Arjun}~Dey}
\affil{NSF’s NOIRLab, 950 N. Cherry Ave., Tucson, AZ 85719, USA}
\author[0000-0002-2761-3005]{\an{Boris}~T.~G\"ansicke}
\affil{Department of Physics, University of Warwick, Gibbet Hill Rd., Coventry, CV4 7AL, UK}
\author[0000-0002-9110-6163]{\an{Ting}~S.~Li}
\affil{Department of Astronomy \& Astrophysics, University of Toronto, Toronto, ON M5S 3H4, Canada}
\author[0000-0002-6667-7028]{\an{Constance}~Rockosi}
\affil{Department of Astronomy and Astrophysics, University of California, Santa Cruz, 1156 High St., Santa Cruz, CA 95065, USA}
\affil{University of California Observatories, 1156 High St., Sana Cruz, CA 95065, USA}
\author[0000-0002-6257-2341]{\an{Monica}~Valluri}
\affil{Department of Astronomy, University of Michigan, Ann Arbor, MI 48109, USA}
\author{\an{Joan}~Najita}
\affil{NSF’s NOIRLab, 950 N. Cherry Ave., Tucson, AZ 85719, USA}
\author{\an{Alis}~Deason}
\affil{Institute for Computational Cosmology, Department of Physics, Durham University, South Rd., Durham DH1 3LE, UK}
\author[0000-0001-5999-7923]{\an{Anand}~Raichoor}
\affil{Lawrence Berkeley National Laboratory, 1 Cyclotron Rd., Berkeley, CA 94720, USA}
\author{\ann{Mei-Yu}{M.-Y.}~Wang}
\affil{Pittsburgh Supercomputing Center, Carnegie Mellon University, Pittsburgh, PA 15213, USA}
\author{\ann{Yuan-Sen}{Y.-S.}~Ting}
\affil{Research School of Astronomy \& Astrophysics, Australian National University, Cotter Rd., Weston, ACT 2611, Australia}
\affil{School of Computing, Australian National University, Acton, ACT 2601, Australia}
\author[0000-0002-8999-1108]{\an{Bokyoung}~Kim}
\affil{Institute for Astronomy, University of Edinburgh, Royal Observatory, Blackford Hill, Edinburgh EH9 3HJ, UK}
\author{\an{Andreia}~Carrillo}
\affil{Institute for Computational Cosmology, Department of Physics, Durham University, South Rd., Durham DH1 3LE, UK}
\author{\an{Wenting}~Wang}
\affil{Department of Astronomy, School of Physics and Astronomy, Shanghai Jiao Tong University, Shanghai 200240, China}
\author{\an{Leandro}~{Beraldo e Silva}}
\affil{Department of Astronomy, University of Michigan, Ann Arbor, MI 48109, USA}
\author[0000-0002-6800-5778]{Jiwon Jesse Han}
\affil{Center for Astrophysics $|$ Harvard \& Smithsonian, 60 Garden St., Cambridge, MA 02138, USA}
\author{\an{Jiani}~Ding}
\affil{Department of Astronomy and Astrophysics, University of California, Santa Cruz, 1156 High St., Santa Cruz, CA 95065, USA}
\affil{University of California Observatories, 1156 High St., Sana Cruz, CA 95065, USA}
\author{\an{Miguel}~Sánchez-Conde}
\affil{Instituto de F\'{i}sica Te\'{o}rica (IFT) UAM/CSIC, Universidad Aut\'{o}noma de Madrid, Cantoblanco, E-28049, Madrid, Spain}
\author{\an{Jessica}~N.~Aguilar}
\affil{Lawrence Berkeley National Laboratory, 1 Cyclotron Rd., Berkeley, CA 94720, USA}
\author[0000-0001-6098-7247]{\an{Steven}~Ahlen}
\affil{Physics Dept., Boston University, 590 Commonwealth Ave., Boston, MA 02215, USA}
\author[0000-0003-4162-6619]{\an{Stephen}~Bailey}
\affil{Lawrence Berkeley National Laboratory, 1 Cyclotron Rd., Berkeley, CA 94720, USA}
\author{\an{Vasily}~Belokurov}
\affil{Institute of Astronomy, University of Cambridge, Madingley Rd., Cambridge CB3 0HA, UK}
\affil{Center for Computational Astrophysics, Flatiron Institute, 162 5th Ave., New York, NY 10010, USA}
\author{\an{David}~Brooks}
\affil{Department of Physics \& Astronomy, University College London, Gower St., London, WC1E 6BT, UK}
\author{\an{Katia}~Cunha}
\affil{Steward Observatory, University of Arizona, 933 N, Cherry Ave, Tucson, AZ 85721, USA}
\author{\an{Kyle}~Dawson}
\affil{Department of Physics and Astronomy, The University of Utah, 115 South 1400 East, Salt Lake City, UT 84112, USA}
\author{\an{Axel}~de la Macorra}
\affil{Instituto de F\'{\i}sica, Universidad Nacional Aut\'{o}noma de M\'{e}xico, Cd. de M\'{e}xico C.P. 04510,  M\'{e}xico}
\author{\an{Peter}~Doel}
\affil{Department of Physics \& Astronomy, University College London, Gower St., London, WC1E 6BT, UK}
\author{\an{Daniel}~J.~Eisenstein}
\affil{Center for Astrophysics $|$ Harvard \& Smithsonian, 60 Garden St., Cambridge, MA 02138, USA}
\author{\an{Parker}~Fagrelius}
\affil{NSF’s NOIRLab, 950 N. Cherry Ave., Tucson, AZ 85719, USA}
\author{\an{Kevin}~Fanning}
\affil{Department of Physics, The Ohio State University, 191 West Woodruff Ave., Columbus, OH 43210, USA}
\affil{Center for Cosmology and AstroParticle Physics, The Ohio State University, 191 West Woodruff Ave., Columbus, OH 43210, USA}
\author{\an{Andreu}~Font-Ribera}
\affil{Institut de F\'{i}sica d’Altes Energies (IFAE), The Barcelona Institute of Science and Technology, Campus UAB, 08193 Bellaterra Barcelona, Spain}
\author{\an{Jaime}~E.~Forero-Romero}
\affil{Departamento de F\'isica, Universidad de los Andes, Cra. 1 No. 18A-10, Edificio Ip, CP 111711, Bogot\'a, Colombia}
\author{\an{Enrique}~Gaztañaga}
\affil{Institut d'Estudis Espacials de Catalunya (IEEC), 08034 Barcelona, Spain}
\affil{Institute of Space Sciences, ICE-CSIC, Campus UAB, Carrer de Can Magrans s/n, 08913 Bellaterra, Barcelona, Spain}
\author[0000-0003-3142-233X]{\an{Satya}~Gontcho A Gontcho}
\affil{Lawrence Berkeley National Laboratory, 1 Cyclotron Rd., Berkeley, CA 94720, USA}
\affil{Department of Physics and Astronomy, University of Rochester, 500 Joseph C. Wilson Boulevard, Rochester, NY 14627, USA}
\author{\an{Julien}~Guy}
\affil{Lawrence Berkeley National Laboratory, 1 Cyclotron Rd., Berkeley, CA 94720, USA}
\author{\an{Klaus}~Honscheid}
\affil{Center for Cosmology and AstroParticle Physics, The Ohio State University, 191 West Woodruff Ave., Columbus, OH 43210, USA}
\affil{Department of Physics, The Ohio State University, 191 West Woodruff Ave., Columbus, OH 43210, USA}
\author{\an{Robert}~Kehoe}
\affil{Department of Physics, Southern Methodist University, 3215 Daniel Ave., Dallas, TX 75275, USA}
\author[0000-0003-3510-7134]{\an{Theodore}~Kisner}
\affil{Lawrence Berkeley National Laboratory, 1 Cyclotron Rd., Berkeley, CA 94720, USA}
\author[0000-0001-6356-7424]{\an{Anthony}~Kremin}
\affil{Lawrence Berkeley National Laboratory, 1 Cyclotron Rd., Berkeley, CA 94720, USA}
\author[0000-0003-1838-8528]{\an{Martin}~Landriau}
\affil{Lawrence Berkeley National Laboratory, 1 Cyclotron Rd., Berkeley, CA 94720, USA}
\author[0000-0003-1887-1018]{\an{Michael}~E.~Levi}
\affil{Lawrence Berkeley National Laboratory, 1 Cyclotron Rd., Berkeley, CA 94720, USA}
\author[0000-0002-4279-4182]{\an{Paul}~Martini}
\affil{Center for Cosmology and AstroParticle Physics, The Ohio State University, 191 West Woodruff Ave., Columbus, OH 43210, USA}
\affil{Department of Astronomy, The Ohio State University, 4055 McPherson Laboratory, 140 W 18th Ave., Columbus, OH 43210, USA}
\author[0000-0002-1125-7384]{\an{Aaron}~M.~Meisner}
\affil{NSF’s NOIRLab, 950 N. Cherry Ave., Tucson, AZ 85719, USA}
\author{\an{Ramon}~Miquel}
\affil{Instituci\'{o} Catalana de Recerca i Estudis Avan\c{c}ats, Passeig de Llu\'{\i}s Companys, 23, 08010 Barcelona, Spain}
\affil{Institut de F\'{i}sica d’Altes Energies (IFAE), The Barcelona Institute of Science and Technology, Campus UAB, 08193 Bellaterra Barcelona, Spain}
\author[0000-0002-2733-4559]{\an{John}~Moustakas}
\affil{Department of Physics and Astronomy, Siena College, 515 Loudon Rd., Loudonville, NY 12211, USA}
\author[0000-0001-6590-8122]{\ann{Jundan}{J.~D.}~Nie}
\affil{National Astronomical Observatories, Chinese Academy of Sciences, A20 Datun Rd., Chaoyang District, Beijing, 100101, P.R. China}
\author[0000-0003-3188-784X]{\an{Nathalie}~Palanque-Delabrouille}
\affil{Lawrence Berkeley National Laboratory, 1 Cyclotron Rd., Berkeley, CA 94720, USA}
\affil{IRFU, CEA, Universit\'{e} Paris-Saclay, F-91191 Gif-sur-Yvette, France}
\author[0000-0002-0644-5727]{\an{Will}~J.~Percival}
\affil{Waterloo Centre for Astrophysics, University of Waterloo, 200 University Ave W, Waterloo, ON N2L 3G1, Canada}
\affil{Department of Physics and Astronomy, University of Waterloo, 200 University Ave W, Waterloo, ON N2L 3G1, Canada}
\affil{Perimeter Institute for Theoretical Physics, 31 Caroline St. North, Waterloo, ON N2L 2Y5, Canada}
\author{\an{Claire}~Poppett}
\affil{Lawrence Berkeley National Laboratory, 1 Cyclotron Rd., Berkeley, CA 94720, USA}
\affil{Space Sciences Laboratory, University of California, Berkeley, 7 Gauss Way, Berkeley, CA  94720, USA}
\affil{University of California, Berkeley, 110 Sproul Hall \#5800 Berkeley, CA 94720, USA}
\author[0000-0001-7145-8674]{\an{Francisco}~Prada}
\affil{Instituto de Astrof\'{i}sica de Andaluc\'{i}a (CSIC), Glorieta de la Astronom\'{i}a, s/n, E-18008 Granada, Spain}
\author[0000-0002-5683-2389]{\an{Nabeel}~Rehemtulla}
\affil{Department of Astronomy, University of Michigan, Ann Arbor, MI 48109, USA}
\author[0000-0002-3569-7421]{\an{Edward}~Schlafly}
\affil{Lawrence Livermore National Laboratory, P.O. Box 808 L-211, Livermore, CA 94551, USA}
\author{\an{David}~Schlegel}
\affil{Lawrence Berkeley National Laboratory, 1 Cyclotron Rd., Berkeley, CA 94720, USA}
\author{\an{Michael}~Schubnell}
\affil{Department of Physics, University of Michigan, Ann Arbor, MI 48109, USA}
\author{\an{Ray}~M.~Sharples}
\affil{Centre for Advanced Instrumentation, Department of Physics, Durham University, South Rd., Durham DH1 3LE, UK}
\affil{Institute for Computational Cosmology, Department of Physics, Durham University, South Rd., Durham DH1 3LE, UK}
\author[0000-0003-1704-0781]{\an{Gregory}~Tarl\'{e}}
\affil{Department of Physics, University of Michigan, Ann Arbor, MI 48109, USA}
\author{\an{Risa}~H.~Wechsler}
\affil{Kavli Institute for Particle Astrophysics and Cosmology, Stanford University, Menlo Park, CA 94305, USA}
\affil{Physics Department, Stanford University, Stanford, CA 93405, USA}
\affil{SLAC National Accelerator Laboratory, Menlo Park, CA 94305, USA}
\author[0000-0001-7775-7261]{\an{David}~H.~Weinberg} 
\affil{Department of Astronomy, The Ohio State University, 4055 McPherson Laboratory, 140 W 18th Ave., Columbus, OH 43210, USA}
\affil{Center for Cosmology and AstroParticle Physics, The Ohio State University, 191 West Woodruff Ave., Columbus, OH 43210, USA}
\author[0000-0002-4135-0977]{\an{Zhimin}~Zhou}
\affil{National Astronomical Observatories, Chinese Academy of Sciences, A20 Datun Rd., Chaoyang District, Beijing, 100101, P.R. China}
\author[0000-0002-6684-3997]{\an{Hu}~Zou}
\affil{National Astronomical Observatories, Chinese Academy of Sciences, A20 Datun Rd., Chaoyang District, Beijing, 100101, P.R. China}

\begin{abstract}
We describe the Milky Way Survey (MWS) that will be undertaken with the Dark Energy Spectroscopic Instrument (DESI) on the Mayall 4m telescope at the Kitt Peak National Observatory. Over the next 5 yr DESI MWS will observe approximately seven million stars at Galactic latitudes $|b|>20^{\circ}$, with an inclusive target selection scheme focused on the thick disk and stellar halo. MWS will also include several high-completeness samples of rare stellar types, including white dwarfs, low-mass stars within 100\,pc of the Sun, and horizontal branch stars. We summarize the potential of DESI to advance understanding of Galactic structure and stellar evolution. We introduce the final definitions of the main MWS target classes and estimate the number of stars in each class that will be observed. We describe our pipelines for deriving radial velocities, atmospheric parameters, and chemical abundances. We use $\simeq500,000$ spectra of unique stellar targets from the DESI Survey Validation program (SV) to demonstrate that our pipelines can measure radial velocities to $\simeq1\,\kms$ and $\mathrm{[Fe/H]}$ accurate to $\simeq0.2$\,dex for typical stars in our main sample. We find the stellar parameter distributions from $\approx100\,\mathrm{deg}^{2}$ of SV observations with $\gtrsim90$\% completeness on our main sample are in good agreement with expectations from mock catalogs and previous surveys.
\end{abstract}


\section{Introduction}
\label{sec:introduction}

\begin{figure*}
    \centering
    \includegraphics[width=\textwidth]{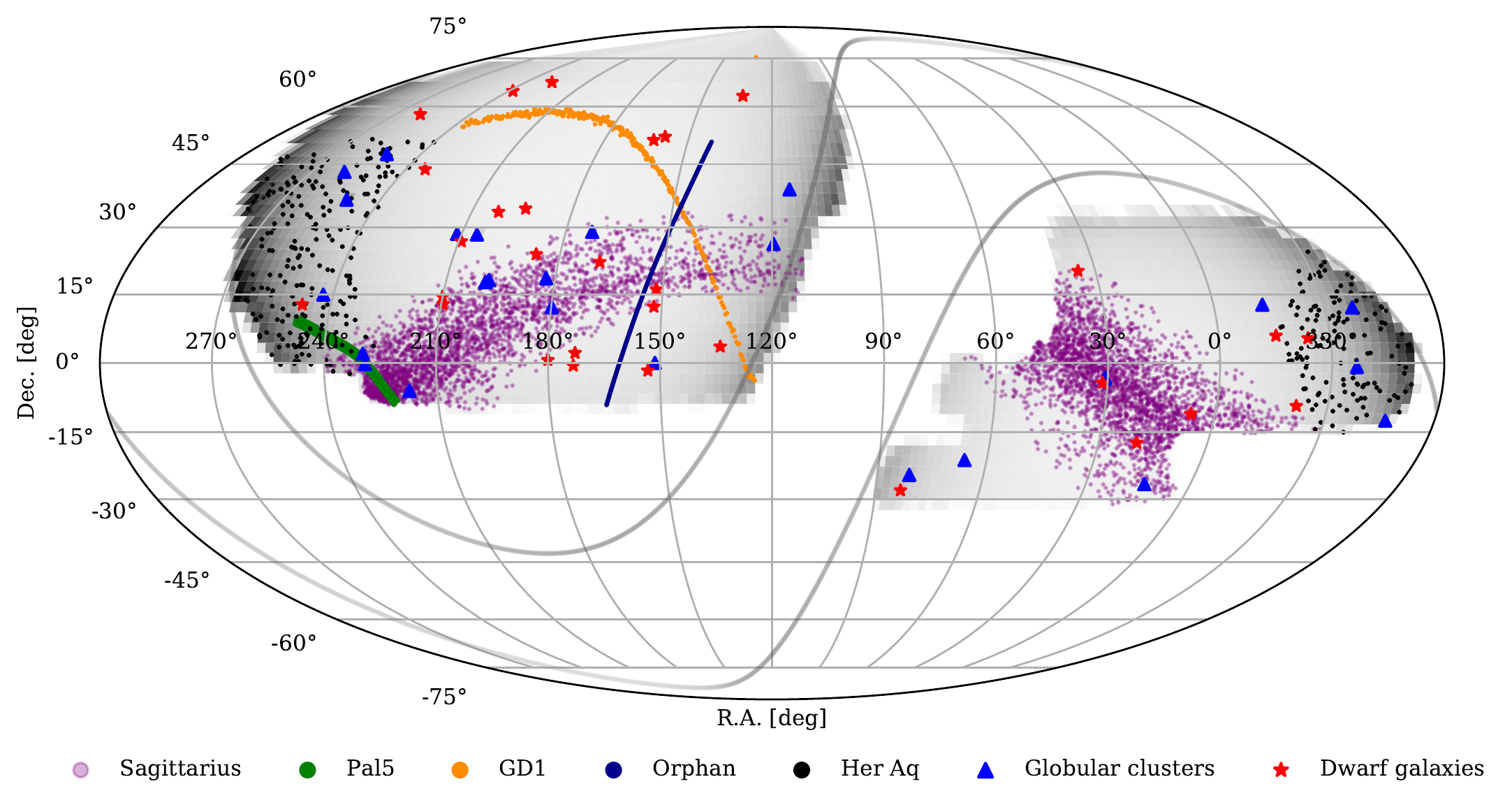}
    \caption{\cwr{The DESI MWS footprint. Gray lines indicate the approximate Galactic latitude limit of the survey, $b\pm20\degr$. The density of MWS targets is shown in gray scale. Colored symbols indicate known Milky Way satellites (stars) and globular clusters (triangles). Points and tracks (colors given in the legend) show the four most prominent streams, as reported in the \texttt{galstreams} compilation \citep{galstreams_software}: Sgr, represented by the \citet{Law_Majewski_2010} model; Palomar 5 \citep{PriceWhelan2019_pal5}; GD 1 \citep{PriceWhelanBonaca2018_gd1} and Orphan \citep{Koposov2019}. We also show the approximate extent of the Hercules--Aquila cloud  as reported in \texttt{galstreams} \citep[based on][]{grillmair_carlin2016}. Many other less prominent streams and stellar overdensities are known in the MWS footprint \citep[see, e.g.][]{mateu2022}.}}
    \label{fig:footprint}
\end{figure*}

The Dark Energy Spectroscopic Instrument (DESI) is currently the premier multiobject spectrograph for wide-field surveys \citep{desiScience,desi-collaboration22a}. DESI deploys 5000 fibers over a $3^\circ.2$ diameter field of view at the prime focus of the Mayall 4m telescope at Kitt Peak National Observatory \citep[][]{desiInstrument, Silber:2023aa, miller22a}. The fibers feed 10 identical three-arm spectrographs, each spanning 3600-9824\AA\ at an FWHM resolution of about 1.8\AA.
Each fiber can be positioned individually by a robotic actuator within a radius of $1'.48$, with a small overlap between the patrol regions of adjacent fibers. 
\correct{The total time to slew the telescope, read out the spectrograph, and reconfigure the focal plane between successive survey exposures can be $\lesssim 2$ minutes  \citep{desi-collaboration22a}.
This allows} large areas of sky to be surveyed very rapidly at a density of $\sim600$ targets per square degree. 

Although DESI is optimized for galaxy redshift surveys, it is also \reply{well suited} to \reply{observing large numbers of Milky Way stars over wide areas at low spectral resolution}. DESI will therefore carry out a Milky Way Survey (MWS) alongside its primary 5-year cosmological program \citep[][]{desiScience}. MWS will operate during bright-sky conditions (when high-redshift galaxy observations are inefficient) and will share the DESI focal plane with a low-redshift Bright Galaxy Survey
\updatecite{\citep[BGS;][]{hahn22a}}. \cwr{Although BGS galaxies will be prioritized for fiber assignment, there is often no BGS target in the patrol region of a DESI fiber. This provides an excellent opportunity to obtain large numbers of stellar spectra, which MWS is designed to exploit.}

The primary MWS program will observe approximately seven million stars to a limiting magnitude of $r=19$ across the full DESI bright-time program footprint. As shown in Fig.~\ref{fig:footprint}, this footprint covers most of the northern Galactic cap region and a significant fraction of the southern cap. It includes many known substructures in the Milky Way stellar halo, such as the Sagittarius, Orphan, and GD 1 streams; the Hercules--Aquila and Virgo overdensities; and many dwarf satellite galaxies and globular clusters. \reply{MWS will provide radial velocities, stellar parameters, and metallicities for an order of magnitude more stars than found in existing samples with similar resolution from the SEGUE and LAMOST surveys, in the same range of magnitudes ($r>16$) and Galactic latitudes (see Section~\ref{sec:context}).}

The MWS selection function is designed to be inclusive, minimally biased, and amenable to forward modeling. The main MWS sample will focus on large-scale spatial and kinematic structures up to 150~kpc from the Sun\cwr{, using three target categories, \mainblue{}, \mainred{} and \mainbroad{}, which in combination cover the full color--magnitude space within $16<r<19$ (see Section~\ref{sec:main_targets}). \mainblue{} randomly samples all point sources in this magnitude range with blue optical colors ($g-r<0.7$). It will be dominated by metal-poor main-sequence turnoff stars in the thick disk and inner halo, and will also contain more distant horizontal branch stars.  We expect these targets to have a spectroscopic completeness of $\approx30\%$ and to comprise 56\% of the main MWS sample. \mainred{} ($\approx30\%$ complete, 12\% of the main sample) applies Gaia proper-motion and parallax criteria to sources with redder colours ($g-r > 0.7$) to boost the probability of observing distant halo giants. Sources with $g-r > 0.7$ that do not meet those astrometric criteria, mostly redder main-sequence stars in the thin disk, are targeted at lower priority in the \mainbroad{} category ($\approx20\%$ complete, 32\% of the main sample). Based on survey forecasts (see Section~\ref{sec:strategy}), we expect to observe $\approx6.6$ million unique targets in the \MWS{} main sample (90\% of all \MWS{} spectra).} These data will constrain the star formation history and chemical and dynamical evolution of the Galactic thick disk and stellar halo, and allow us to identify the low-contrast remnants of ancient dwarf galaxy accretion events. Radial velocities from MWS combined with Gaia astrometry will constrain the three-dimensional distribution of dark matter in the Galaxy.

As part of the primary \MWS{}, the generously selected main sample will be supplemented by several much smaller but highly complete samples of rare stellar types with very low density on the sky. \MWS{} will target a near-complete sample of white dwarfs to the Gaia magnitude limit in order to obtain an independent measurement of the star formation history of the disk and halo populations. It will also collect a highly complete sample of stars in the Gaia catalog within 100\,pc, to investigate the fundamental properties of low-mass stars and measure the stellar initial mass function. Blue horizontal branch stars (BHBs) and RR\,Lyrae variables will be prioritized for their use as tracers of the distant metal-poor halo. In addition to these primary samples, the DESI surveys will allocate a small number of fibers to specialized secondary science programs, some of which will focus on stars. The survey will also operate a backup program for poor observing conditions, which we expect to yield spectra for several million stars brighter than those in the primary \MWS{} sample. The secondary and backup programs will be described in separate publications; here we  provide only brief summaries of their relevance to the goals of \MWS{}.

This paper presents the context (Section \ref{sec:context}) and scientific motivation of MWS (Section \ref{sec:goals}), the target selection scheme (Section \ref{sec:targets}), and the survey strategy (Section \ref{sec:strategy}), including forecasts for the final size of each primary sample. In Section \ref{sec:pipeline} we describe the core components of our spectroscopic analysis pipeline. From 2019 November to 2021 May, DESI carried out a survey validation (SV) campaign consisting of three subprograms (SV1, SV2, and SV3). The SV1 and SV3 datasets each contain spectra for $\sim200,000$ unique stars, covering a superset of the MWS selection function. These observations will be the basis for the first DESI \MWS{} data release. They include observations of calibration fields, such as open and globular clusters, and an $\approx100$ square degree high-completeness minisurvey representative of the northern Galactic cap. We give an overview of these datasets in Section \ref{sec:sv} and use them to demonstrate that our survey design and analysis pipeline meet the requirements set by the ambitious science goals of \MWS{}. We summarize in Section~\ref{sec:conclusion}. 

 Machine-readable tables of the data shown in the figures of this paper are available at \url{https://doi.org/10.5281/zenodo.7013864}. The complete sets of targets selected for the main \MWS{} and the DESI SV programs are available at \url{https://data.desi.lbl.gov/public/ets/target/} as described by \citet{myers22a}.

\section{DESI MWS in context}
\label{sec:context}

\begin{table*}
    \centering
    \caption{\cwr{Summary of completed and ongoing stellar spectroscopic surveys with $\gtrsim 10^{5}$ targets, discussed in Section~\ref{sec:context}. Columns give the number of targets (either in the latest data release or forecast) and the approximate magnitude range, spectral resolution $R$, and wavelength coverage. The rightmost columns give the most recent data release (except where noted, $N_{\mathrm{star}}$ is the number of unique stars that release) and reference for each survey. \reply{For LAMOST, we count the number of unique Gaia source IDs with spectral class \texttt{STAR} and separate the low resolution survey into two rows, respectively for stars in the MWS magnitude range and brighter stars}. The table is divided into low-resolution and medium/high resolution surveys.}}
    \begin{tabular}{lllllll}
      \hline
         Survey & $N_{\mathrm{star}}$/$10^{6}$ & Mag.\ Range & $R$ & $\lambda$ (\AA) & Release & Reference  \\
      \hline
        DESI MWS   & 7.2$^a$ & $16<r< \correct{20}$ & 2500--5000  & 3600--9900     & --   & --\\
        SEGUE I    & 0.2     & $16<r<20$ & 1850--2200 & 3800--9200     & DR17 & \citet{yannyetal09-1} \\
        SEGUE II   & 0.1     & $16<r<20$ & 1850--2200 & 3800--9200     & DR17 & \citet{rockosi22} \\
        LAMOST LRS  & 1.8$^{b}$ & $16<G<18$$^c$ & $\approx 1800$ & 3700--9000 & DR8  & \citet{luo2015} \\
         & 5.0$^{d}$ & $10<G<16$ & $\approx 1800$  & 3700--9000 & DR8  & \citet{luo2015} \\
      \hline
        LAMOST MRS  & 1.1 & $10<G<15$ & $\approx 7500$ & 4900--6800 & DR8  & \citet{luo2015} \\
         RAVE      & 0.5       & $9 < I < 12$     & 7500      & 8410--8795      & DR6  & \citet{steinmetz2020} \\
         Gaia RVS  & 33$^e$  & $G < 14$         & 11,500     & 8450--8720      & DR3  & \citet{Katz2022} \\
         Gaia-ESO  & 0.1       & $17 < r < 18$   & 20,000     &  3700--9500     & DR5  & \citet{Gilmore2012} \\
         GALAH     & 0.6       & $9 < V < 14$ & 20,000--50,000 & 4718--7890$^d$  & DR3  & \citet{buder2021} \\
         APOGEE    & 0.7   & $10<G<17$    & 22,500     & 15,140--16,960  & DR17 & \citet{APOGEE} \\
         H3        & 0.3$^f$   & $15 < r < 18$    & 32,000     & 5150--5300      & --   &  \citet{H3} \\
         \hline
         \hline
         \multicolumn{7}{l}{{\it Notes:} $^a$Forecast. $^b$1.2M at $|b| > 20$, of which 0.6M have stellar parameters. $^c$69,233 with $G>18$. $^d$2.9M at $|b|>20$, of which}\\
         \multicolumn{7}{l}{ 2.3M have stellar parameters; $^e$\reply{5.6M have stellar parameters \citep{Recio-Blanco2022}}. $^f$Four bands of width 200\AA.}\\
    \end{tabular}
    \label{tab:rvsurveys}
\end{table*}

DESI is the first on-sky instrument among a new generation of multiobject survey spectrographs with a 4m aperture, high fiber density over a wide field of view, and rapid fiber positioning. These advances allow a new approach to large-scale stellar spectroscopy, which seeks to minimize selection biases and to provide uniformly high spectroscopic completeness over a large fraction of the sky. As we describe below, even subject to operational constraints imposed by the DESI cosmological programs, MWS expects to assemble an effectively flux-limited sample of $\simeq7$ million stellar spectra to a magnitude limit $r\lesssim19$ over 5 yr. \cwr{In comparison to existing surveys, DESI MWS will provide a much higher density of faint stars to search for substructures and to probe the kinematics and chemistry of the outer thin disk, thick disk, tidal streams, and diffuse stellar halo.}

\cwr{Table~\ref{tab:rvsurveys} lists the parameters of other recent spectroscopic surveys of $>10^{5}$ stars with spectral resolution comparable to or greater than that of DESI. The closest existing counterparts to DESI MWS in scientific scope and spectral resolution ($R \equiv \lambda/\mathrm{FWHM} \simeq 2000$) are the Sloan Digital Sky Survey (SDSS) optical surveys \citep{yorketal00-1, eisensteinetal11-1, blantonetal17-1} and the LAMOST survey \citep{cuietal12-1}.}

\cwr{The primary SDSS low-resolution stellar surveys were SEGUE-1 \citep[][]{yannyetal09-1} and SEGUE-2 \citep[][]{rockosi22}, which together observed $\simeq300,000$ stars. Stars were also observed as calibration targets and serendipitously by other SDSS programs, including $\sim380,000$ stars by the BOSS/eBOSS cosmological surveys \citep{dawsonetal13-1}. The SEGUE surveys each covered $\lesssim 1500$ $\mathrm{deg}^2$ in total, with individual fields distributed over the SDSS imaging footprint. The SEGUE-2 target selection focused on the distant stellar halo, with similar scientific goals to \MWS{}. The most significant differences of MWS from SEGUE its are contiguous sky coverage, much larger number of spectra and broader, simpler selection function.}

LAMOST \citep{cuietal12-1, luo2015} is an ongoing survey covering 17,000 $\mathrm{deg}^2$ in the northern hemisphere ($0^{\circ}< \delta <60^{\circ}$) \cwr{including} the Galactic anticenter. To date, LAMOST has obtained more than 11 million $R\approx1800$ stellar spectra \correct{in its LRS}, primarily for stars brighter than $r\sim18$. The most recent public data release \correct{(DR8 v2) contains $10.3$ million LRS stellar spectra. Stellar parameters have been derived for $6.7$ million of these, corresponding to $\approx4.8$ million unique sources in the Gaia catalog. LAMOST also includes a medium-resolution survey ($\approx1.1$M stars in LAMOST DR8). In comparison to LAMOST, MWS will provide many more spectra for fainter stars in the outer thick disk and stellar halo. Of the 4.8 million unique stars with stellar parameters in LAMOST LRS (DR8), only $\approx580,000$ are at Galactic latitudes $|b|>20^{\circ}$ and fainter than Gaia $G=16$.}

\cwr{High-resolution Milky Way surveys, such as Gaia-ESO \citep{gaia-eso},  GALAH \citep{GALAH}, \reply{RAVE \citep{steinmetz2020}}, APOGEE \citep{zasowski2013_apogee, APOGEE}, and the ongoing H3 survey \citep{H3}, have collected spectra for substantial samples of bright stars ($\lesssim1$ million) with $R \simeq 20,000$-$50,000$. APOGEE has observed $\approx400,000$ stars at $|b|>20^\circ$, of which $\sim95\%$ are brighter than $G=17$. Although APOGEE observations have concentrated on the Galactic bulge and disk, a number of fields have been observed in the halo. Some of these fields targeted have known halo substructures, including the Sagittarius stream \citep[e.g.][]{Hasselquist_2019}, and include target selections that overlap in distance with MWS. The brightest MWS targets observed with the APOGEE halo dataset will be useful for calibrating MWS parameter and abundance measurements.}

\cwr{The H3 survey is using MMT Hectochelle \citep{Szentgyorgyi11} to observe $300,000$ high resolution optical spectra ($R=32,000$) in the northern hemisphere sky at $|b|>20^{\circ}$ and $\delta>-20^{\circ}$, sparsely sampling $15,000$ $\mathrm{deg}^2$. H3 targets a magnitude range similar to that of \MWS{}, and has similar science goals focused on the stellar halo. Like the MWS selection function, the H3 selection function is close to magnitude-limited, with a weak parallax selection, and priority given to sparsely distributed halo giant candidates, BHBs, and RR Lyrae variables. In many respects H3 therefore provides a complementary (slightly shallower) high-resolution counterpart to MWS.}

Radial velocity surveys with intermediate resolution, such as RAVE \citep[][]{RAVE}, have also focused on brighter targets than those in \MWS{}. By far the largest of these is the Gaia Radial Velocity Spectrometer survey (RVS) \citep[][]{GaiaRVS}. \reply{Both RAVE and Gaia RVS use a narrow spectral window around the Ca \textsc{ii} triplet ($8400 \lesssim \lambda \lesssim 8800$\AA). Despite this limited wavelength coverage, multiple individual abundances have been recovered for the majority of the 0.5 million $R\approx7,500$ RAVE spectra, including [Fe/H], to an accuracy of $\approx0.2$~dex \citep{Boeche_RAVE_abundances, steinmetz2020_b_abundances}. \citet{Recio-Blanco2022}, using an extension of the same core pipeline, report abundances for $\sim5$ million Gaia RVS spectra with $G<14$}. MWS will provide both radial velocities and chemical abundance information, \reply{based on measurements over a wider range of wavelengths,} for a significant fraction of much fainter stars, \reply{including many with} Gaia astrometric measurements but no useful Gaia RVS data.

It has \cwr{previously} been shown that multiple elemental abundances can be extracted from low-resolution spectra \citep[e.g.][]{fernandez2015,ting17,xiang19}. 
These approaches can be further developed with DESI\cwr{, which, compared to SDSS and LAMOST, has better} sensitivity at $\lambda \lesssim 4000$\AA\ and slightly higher resolution. Precise spectrophotometric distances can also be estimated based on spectra at the resolution of DESI, to better than $10\%$ at signal-to-noise ratios (\SNR{}) $> 50$ \citep{hogg19,xiang2021} 
and 20\% at \SNR{} $> 20$. Furthermore, DESI's broad optical wavelength range will provide useful information about stellar ages \cwr{for red giants} inferred through \cwr{mass estimates from} C and N abundances \cwr{\citep{2015MNRAS.453.1855M, martig15, 2016ApJ...823..114N, 2018MNRAS.481.4093S,2019ApJ...872..137S}}.

\section{Science Goals of DESI MWS}
\label{sec:goals}

DESI MWS will assemble an extremely large sample of radial velocities and chemical abundances, predominantly for distant stars at high Galactic latitude. In this section we review how these data, combined with spectrophotometric distance estimates and Gaia astrometry, will advance understanding of the dark and luminous structure of the Milky Way, the history of stellar mass growth through tidal stripping, \insitu\ star formation, and stellar (and planetary) astrophysics.

\subsection{Probing the Dark Matter and Accretion History of the Milky Way}
\label{sec:goals_halo}

Dark matter constitutes approximately 86\% of the gravitating mass in the present-day universe. Galaxies form in dark matter potential wells. 
While the large-scale distribution of galaxies traces the dark matter distribution on cosmological scales, the internal kinematics of galaxies provides a uniquely powerful laboratory in which to study the small-scale structure of dark matter. 

Galaxies build their stellar halos through mergers with satellite galaxies, which may be accompanied by their own globular cluster systems. As dwarf galaxies are tidally disrupted, their stars and globular clusters are pulled out into thin tidal streams. Mergers with lower mass ratios may also generate or enhance a galactic thick disk component. Because dynamical times in the  \cwr{outer} halo are long \cwr{assuming approximate global dynamical equilibrium, the kinematics of halo stars and globular clusters} probe the mass distribution and dynamics of the dark matter halo \citep[for recent reviews, see][and references therein]{2016ARA&A..54..529B,2020SCPMA..6309801W}. \cwr{The merger history of the Milky Way is preserved in the clustering of stellar debris from past accretion events, and correlations between halo stars in phase space and in abundance space encode the assembly history of the Galaxy \citep[see, e.g.][]{Helmi_2020_ARAA}.}

\subsubsection{The Shape and Mass of the Dark Matter Halo}
\label{sec:goals_dmhalo}

Characterizing the dark matter halo of the Milky Way is crucial for understanding our Galaxy in its cosmological context. This requires accurate constraints on its total mass ($M_{\mathrm{vir}}$) within the virial radius ($r_{\mathrm{vir}}$), the 
\cwr{form} of the radial mass density profile, and its three-dimensional shape (spherical, axisymmetric, or triaxial). 

$\Lambda$CDM cosmological simulations
predict that the smooth components of stellar halos have power-law or broken-power-law stellar density profiles \citep[e.g.][]{cooper2010, deason2013, Amorisco2017a,font2020} and predominantly radially anisotropic velocity dispersion profiles \citep{bullock_johnston_2005, abadi2006, loebman_etal_2018}. Current observations are in broad agreement with these predictions; namely, the Milky Way stellar halo has a ``broken'' density profile \citep{deason2011, sesar2011}, and the orbits of halo stars, at least in the inner $\sim 30-50$~kpc, are highly eccentric \cwr{\citep{deason2018, bird2019, cunningham2019, iorio2019, iorio2021, lancaster2019, Hattori_etal_2021}}. 

Dark matter halos are believed to depart significantly from spherical symmetry. Cosmological simulations assuming collisionless dark matter produce halos that are triaxial and have almost constant shape at all radii \citep{jing_suto_00}. Simulations that include baryons produce halos that are oblate-axisymmetric within the inner one-third of the virial radius, but become triaxial or prolate at larger radii \citep{kazantzidis_etal_04shapes, deb_etal_08,  zemp_etal_11,Prada2019}. Cosmological simulations with warm dark matter ~\citep[sterile neutrinos,][]{bose_frenk_16_WDM} and self-interacting dark matter ~\citep[][]{peter_13_SIDM,Vargya_etal_2021} also predict triaxial dark matter halos, albeit with small but quantifiable differences in the variation of shape with radius.

Estimates of the shape of the Milky Way’s dark matter halo from halo field stars, based on samples within $\approx 30$~kpc, imply a nearly spherical inner halo \citep{Wegg2019,Hattori_etal_2021}. 
Recent attempts to probe the mass and shape of the dark matter halo have been greatly advanced by all-sky Gaia proper-motion measurements for individual stars to $G\sim20$ and averaged proper-motions for more distant satellites and globular clusters
\citep[e.g.][]{2019MNRAS.484.5453C,2020ApJ...894...10L,Deason_etal_2021}. 
Studies combining Gaia proper motions and radial velocities from the H3 survey suggest that the dark matter halo of the Milky Way may be triaxial or tilted relative to the Milky Way disk \citep{Han2022}, although self-consistent modeling is required to confirm this.
More generally, the total virial mass, inner density slope, concentration parameter, and local dark matter density of the halo remain uncertain \citep{2019MNRAS.484.5453C} and depend quite strongly on the assumed properties of the baryonic components \citep{deSalas_etal_2019}. 

\cwr{Despite evidence that the Milky Way experienced a significant merger event (the ``Gaia Sausage/Enceladus'' (GSE); see Section~\ref{sec:goals_assembly}) approximately $8-11$~Gyr ago and is currently interacting with the Large Magellanic Cloud (LMC; see below), most estimates of the global mass distribution of dark matter in the Galaxy assume that it is in dynamical equilibrium. In many contexts, this assumption is justified;  tests with mock data from cosmological simulations of Milky Way analogs including those with ongoing interactions \citep[e.g.][]{Kafle_etal_2018, Hattori_etal_2021,Rehemtulla_etal_2022} show that both the mass and average flattening of the halo can be estimated with 15-25\% accuracy using datasets that cover large areas of the sky and extend to large distances from the Galactic center.} 

DESI MWS survey will greatly improve our understanding of the mass and shape of the dark matter halo by significantly increasing the samples of halo tracers ($\sim1$ million turnoff and subgiant branch stars between 10 and 30~kpc and $\sim10,000$ giant stars in the halo beyond 30~kpc; see below) with precise radial velocities (velocity errors $\sigma_{v_{los}} \sim 1-20\,\kms$) and spectrophotometric distances.  \MWS{} will also obtain spectra for half of all Gaia-detected RR\,Lyrae variables with brightness $14<G<19$ in the survey footprint and increase the sample of known BHBs with radial velocities by a factor of $\approx 3$ (see sections \ref{sec:rrlyrae_targets} and \ref{sec:bhb_targets}).
Radial velocities, in conjunction with better proper-motion measurements from future Gaia data releases and proper motions from the Roman Observatory’s High Latitude Survey \citep{WFIRST_astrometry_2019}, will enable us to more fully characterize the dark matter halo of the Milky Way, \cwr{significantly reducing the current factor of $\sim 2$ uncertainties on the cumulative mass of the Milky Way within $\sim100$~kpc.}

\paragraph{Mapping the interaction with the LMC. \label{sec:goals_lmc}}

\cwr{It has become clear in the past 15 yr that the Milky Way can no longer be modeled to high accuracy as an isolated galaxy. There is considerable evidence that the ongoing gravitational interaction with the LMC may have distorted the halo and displaced the stellar component of the Galaxy away from the center of the dark matter potential. Hubble Space Telescope proper-motion measurements of the LMC have been used to show that the LMC is on its first infall and has only recently passed its pericenter \citep{Besla_etal_2007}. Furthermore, the LMC is much more massive than previously believed, $\sim 2\times 10^{11}\,\msun$  \citep{Besla_etal_2010}, massive enough to cause  a significant reflex motion of $\sim 40-60\,\kms$ of the Milky Way disk \citep{Gomez_etal_2015}.}

\cwr{The dynamical response of the Galactic stellar and dark matter halo to the LMC has two primary components. First, a classical dynamical friction wake is produced behind the LMC in the Galactic southern hemisphere as it orbits the Milky Way, accompanied by a density enhancement in the Galactic northern hemisphere referred to as the ``collective response.’’ Second, the fact that both the LMC and the Milky Way orbit their common barycenter results in a ``reflex velocity’’ of the stellar halo relative to the Milky Way disk. This was predicted to appear as a  dipole in the velocity distribution of halo stars \citep{Garavito-Camargo_etal_2019} and has been detected \citep{Petersen_Penarrubia_2020}. More recently, \citet{Conroy_etal_2021} have  used star counts from Gaia to measure the stellar overdensity due to the dynamical friction wake and collective response. Both of these overdensities have been observed in nearly the same locations predicted by $N$-body simulations. Although the wake is in the Galactic southern hemisphere and outside the DESI footprint, the ``collective response'' in the Galactic north lies partially within the footprint of the main survey and will also be covered by the DESI Backup Program. A velocity accuracy for DESI MWS of $1-20\kms$ will be sufficient to yield constraints on the predicted dark matter wake and large-scale motions in the stellar halo associated with the ongoing gravitational interaction between the Milky Way and the LMC.}

\subsubsection{Small-scale Substructure in the Dark Matter Halo} 
\label{sec:goals_streams}

The distribution of dark matter on small scales, including the subhalo mass function and the central density profiles of dwarf galaxies, is an extremely important probe of the nature of dark matter \citep[e.g.][]{zavalafrenk2019}. \MWS{} can constrain these properties in the Milky Way and Local Group through observations of tidal streams and dwarf galaxies.

\paragraph{Tidal Streams. \label{sec:goals_tidalstreams}}

Kinematically cold structures like stellar tidal streams are ideal for probing small-scale substructures in the Milky Way's dark matter halo. \cwr{Interactions between cold streams and dark matter subhalos can produce stream gaps, overdensities and characteristic off-stream structures (``S-shaped'' kinks and ``spurs'') \citep{Siegal-Gaskins_Valluri_2008, Carlberg2013, Erkal_2015a}, as well as perturbations to the line-of-sight velocities and proper motions of stars near the gaps \citep{Erkal_2015b}. The velocities of stars near gaps in cold, thin globular cluster streams (such as the gaps, kink and spur observed in GD 1, or the ATLAS and Aliqa Uma streams; e.g.\ \citealt{WhelanBonacaGD12018, Li2021}) will constrain the clumpiness of the dark matter distribution encountered by the stream \citep[e.g.][]{Carlberg2012,Erkal2016_gaps,Bovy2017,Bonaca_spur_2018}.}
In addition, \cwr{it has been recently shown from simulations that secular tidal evolution} of accreted globular cluster streams can be used to probe the density profiles of the parent dwarf galaxies from which the clusters were accreted by the Milky Way \citep{Malhan_etal_2021}. \cwr{Such secular tidal evolution produces broader ``cocoon'' components around thin streams \citep{Carlberg_2018} and increases the velocity dispersion over large swaths of the stream \citep{Malhan_etal_2021}. DESI spectra will provide the stellar chemical abundances that are crucial for determining stream membership and will enable discrimination between localized perturbations from subhalos and secular tidal evolution. Furthermore,} the chemical abundance and radial velocity gradients along dwarf galaxy streams will  constrain the rate of stellar mass loss, providing further clues to their dark matter distributions \citep{Errani2015}.

Many known tidal streams are covered by the \MWS{} footprint and target selection criteria (e.g., Sagittarius, GD 1, Palomar~5 and the Orphan stream), including 43 of the 73 streams and other halo substructures in the \texttt{galstreams} catalog\footnote{\url{https://github.com/cmateu/galstreams}, revision \texttt{35b71b1}.} \citep{galstreams_software, mateu_galstreams_2018} and tens of more recently discovered Gaia streams \citep[e.g.][]{Ibata2019_Abyss, mateu2022}.
For each of these, we expect MWS to identify \cwr{on the order of} $\sim10$-$100$ member stars per stream. For example, based on the selection criteria and fiber assignment forecasts described below, we expect to obtain spectra for $\sim600$ candidate members of the GD 1 stream (satisfying a broad cut in position and proper motion), of which $\sim10-30$\% may be true members. This number may be even greater for the Sagittarius stream: we expect to obtain spectra for $\sim6,600$ of the $\sim33,600$ candidate stream members identified by \citet{Antoja_etal_2020} in our footprint with magnitudes $15<G<20$, and a further $\sim1,800$ Sagittarius RR\,Lyrae from the catalog of \citet{Ramos_Mateu_2020} in the same magnitude range. \cwr{For comparison, \citet{Vasiliev_etal_2021_tango} find that 4,465 stars in their catalog of 55,192 probable Sagittarius members have existing radial velocities. We expect to observe $\approx 2,800$ of the 5,600 \citeauthor{Vasiliev_etal_2021_tango} candidate members in the MWS footprint, of which 306 have existing radial velocities.} Of course, \MWS{} may also observe members of Sagittarius, GD 1 and other streams that are not identified in existing catalogs. \MWS{} radial velocities, chemical abundances, and spectrophotometric distances for candidate stream members, in conjunction with Gaia proper motions, will enable us to better assess membership and chemical homogeneity (or the opposite) within streams, and to search for yet unknown phase-space substructures.

Since tidal streams approximately trace the orbits of their progenitors, they have  been widely used for measuring the shape of the Milky Way’s dark matter halo \citep{Johnston_etal_1999, Johnston_etal_2005, Helmi_2004, Fellhauer2006, Koposov2010, Law_Majewski_2010, Sanders2013, Bovy2014, Bovy2016, Gibbons2014, Bowden2015, Kupper2015,  Malhan2019, Vasiliev_etal_2021_tango}. A \cwr{high-precision} model of the Sagittarius stream \citep{Vasiliev_etal_2021_tango} has \cwr{been derived from} Gaia proper motions \citep{Antoja_etal_2020}, \cwr{following} the realization that the LMC gives rise to a significant reflex motion of the Milky Way's center of mass \citep{Gomez_etal_2015}. This nonequilibrium model implies a radially varying and time-dependent shape for the Milky Way halo. However, even this model relies on a limited sample of stars with radial velocities. \MWS{} will provide 6-dimensional kinematic information from multiple stellar streams, which can be jointly used to constrain the Milky Way's dark halo density profile to even greater precision \citep{Bonaca2018ColdStreams}. Furthermore, some of these streams can be used to probe the mass of the LMC at large radii. \citet{Erkal2018_tucana} first argued that the LMC could induce a substantial proper motion perpendicular to the track of the stream on the sky, and this offset could be used to measure the mass of the LMC. Such a misalignment of proper motions with the stream track has been found in the Orphan--Chenab stream \citep{Koposov2019} and used to constrain the LMC and Milky Way potential simultaneously \citep{Erkal2019}. A similar analysis has been carried out for the Sagittarius stream \citep{Vasiliev_etal_2021_tango}. Recently, \citet{Shipp2021} extended this approach to five stellar streams with proper motions measured by Gaia~EDR3 and radial velocities measured by the Southern Stellar Stream Spectroscopic Survey \citep[$S^5$;][]{Li2019, Li2022}. Using this 6-dimensional kinematic information for an ensemble of streams, they found a mass of the LMC ranging within $\sim 1.4-1.9 \times 10^{11}\,\mathrm{M_\odot}$. 

\paragraph{Satellite dwarf galaxies.
\label{sec:goals_satellites}}

The satellite galaxies of the Milky Way provide an additional testing ground for the nature of dark matter on small scales \citep{sg07,nadler21}. The particle physics governing dark matter may have observable effects on the luminosity function of satellite galaxies, the density profiles of their dark matter halos (cusps versus cores), and the production of energetic Standard Model particles through annihilation or decay. Satellites also provide a window into the formation of the oldest and least massive galaxies, inaccessible to direct observation \citep{kirby08, blandhawthorn15}. Many open questions remain concerning the structure, stellar populations, and star formation histories of these galaxies, answers to which would shed light on their past interactions with the Milky Way and complement surveys of the diffuse stellar halo.

Although significant resources have been invested to discover and characterize dwarf galaxies, very few spectroscopic observations have been taken in their outskirts, mainly due to the limited field of view of existing spectroscopic facilities. Recent work suggests at least some of these galaxies may be surrounded by significant low surface brightness structure, which simulations suggest may occur even for galaxies that are not presently losing significant mass through tidal stripping \citep[e.g.][]{WangMeiYu:2017}. For example, \citet{Chiti2020} used deep SkyMapper UV photometry to find candidate members of the ultrafaint dwarf galaxy Tucana II out to 7 half-light radii. Spectroscopic follow-up is now required to confirm their association, and more generally to address the prevalence and properties of similar features in other satellite galaxies.

The DESI MWS footprint includes 31 known dwarf spheroidal and ultrafaint dwarf galaxies (according to \citealt{McConnachie2012_dwarfdatabase}, see Figure~\ref{fig:footprint}). Sixteen of these have distance modulus smaller than $20$ and hence, potentially, red giant branch (RGB) stars within the \MWS{} main survey selection. The primary MWS target selection categories will therefore provide a sparse but extremely wide-field sampling of potential members in the outskirts of these galaxies, most of which have half-light radii well within the $3^{\circ}.1$ field-of-view of DESI. In addition, a DESI secondary program will allocate fibers at higher priority to potential members (selected in a broad window of proper-motion and color--magnitude space) around dwarf galaxies in the DESI footprint, in both the bright- and dark-time DESI surveys. The confirmation of distant members will enable the study of large-scale metallicity gradients and tidal effects.

\subsubsection{The Assembly History of the Milky Way Halo}
\label{sec:goals_assembly}

\cwr{The accretion and star formation histories of the Milky Way are encoded in its stellar populations. The Milky Way was formed through the  accumulation of stars, gas, and dark matter through mergers, which continues to the present. The stellar halo and thick disk still retain kinematic signatures of this process, which MWS seeks to recover through measurements of line-of-sight velocity, metallicity, and $\alpha$-abundance, in combination with Gaia proper motions. DESI also has the potential to measure individual} elemental abundances \cwr{including C, Mg, Ca, and Fe abundances. These measurements will enable} the identification of fossil remnants of the assembly process in a multidimensional phase space, including the relating of tidally-stripped halo stars \cwr{to} their parent objects (i.e., dwarf galaxies and globular clusters). 

\cwr{Our view of the stellar halo has changed dramatically since the first data releases of the Gaia mission. It is now understood that the inner halo (within 20--30 kpc) is dominated by one massive dwarf progenitor, namely the GSE \citep{Belokurov_2018, Helmi_2018_GSE}. This major event in the Milky Way's assembly history was predicted from the observed halo star counts by \cite{deason2013}, who argued that the ``break" in the stellar halo density profile (at 20--30 kpc) signifies the apocenter of a massive accretion event. Our view of the (inner) halo is thus intricately linked with the properties of the GSE event. The large increase in stellar halo tracers provided by DESI in the inner region of the halo dominated by the GSE will be essential to quantifying the details of this accretion event and the properties of the progenitor dwarf galaxy. For example, the two-dimensional phase space of Galactocentric distance and radial velocity can be used to identify shells (overdensities around common orbital apocenters) linked to the GSE. These in turn can be used to constrain the orbit and accretion time of the progenitor, while the metallicity distribution function (MDF) and chemical abundances can be used to infer its stellar mass and star formation history.}

\cwr{The DESI footprint covers many well-known Milky Way globular clusters. Finding and characterizing tidal features around globular clusters is crucial to understanding their orbits and mass-loss histories. Estimates of the fraction of mass contributed to the Milky Way stellar halo by disrupted globular clusters range from 50\% \citep{Martell:2010,Martell:2011} to negligible \citep[e.g.][]{deason2015}}. Significant mass loss ($\geq 90\%$) from \cwr{globular clusters} has been invoked to explain the presence of multiple stellar populations within them \citep[see, e.g.,][and references therein]{Bastian:2018}; however, such extreme mass loss requires globular clusters to be born with very much larger effective radii than currently observed. 

Several imaging studies, especially those carried out in combination with Gaia proper motions, have led to the discovery of tidal tails associated with a few globular clusters \citep[e.g.,][]{Shipp2020, Sollima:2020}. At the same time, \cwr{the lack} of convincing signs of tails around \cwr{other} Milky Way GCs \citep[e.g.,][]{Leon:2000, Kuzma:2016} \cwr{has} led to speculation that \cwr{they may be embedded in their own dark matter halos} \citep[e.g.][]{penarrubia17, Carlberg:2021} \cwr{which} prevents stars from escaping at the expected rate \citep{Moore:1996}. However, it is likely that the extremely low surface brightness of these features limits the discovery potential of imaging alone. DESI's spectroscopic observations can complement these results by identifying stripped stars through kinematic and chemical tagging. This will allow a better characterization of the tidal features themselves and the frequency with which they occur around globular clusters.

More generally, \cwr{DESI MWS radial velocities and stellar abundances, in conjunction with Gaia proper motions, will enable} us to determine what fraction of the halo is made up of merger or tidally stripped remnants, and whether there exists a smooth, isotropic component that is chemically and dynamically old. \cwr{As demonstrated by \citet{belokurov_kravtsov_2022} using APOGEE abundances, such data can also constrain the properties, and hence the origin, of an `in situ' stellar halo component. Distinguishing between the wide range of processes that produce in situ stellar halos in cosmological simulations \citep[including early chaotic gas accretion and disk collapse, dynamical heating, and star formation in cooling instabilities or accreted gas; see, e.g.][]{cooper2015} would be a significant advance in the understanding of early galaxy formation as a whole, and would address a major source of uncertainty in the galactic archaeology of the Milky Way and other galaxies.}

\subsubsection{The Formation History of the Milky Way Thick Disk}
\label{sec:goals_thick_disk}

The disk of the Milky Way, visible from any dark site, is the feature that defines its morphology. The disk is made up of at least two components of different thickness, and there is evidence that the outer disk may show warps and flares \citep{Djorgovski_Sosin_1989, Drimmel_Spergel_2001, Lopez_Corredoira_2002, Mackereth_2017}. Past studies of stars in the Milky Way disk have focused on the inner disk (R $\lesssim$ 13~kpc) and generally been restricted to the thin (and young) disk component \citep[e.g.][]{martig14, aumer16, mackereth19, ting19a}. DESI MWS will focus primarily on the thick disk.

The thick disk of the Milky Way is believed to be an ancient structure that has been dynamically heated over time. Several scenarios have been proposed for its origin: secular evolution due to scattering with giant molecular clouds \citep[e.g.][]{aumer16, ting19a} or dark matter substructures \citep{church19}, in situ formation from cooling of a thick and turbulent interstellar medium \citep[so called ``upside-down growth'';][]{bird13, grand16}, heating of a protodisk by a massive merger \citep[e.g.][]{1996ApJ...460..121W, 1999MNRAS.304..254V}, and early scattering by massive clumps \citep[][]{2009ApJ...707L...1B, Clarke2019, Beraldo_e_Silva2020}. A combination of dynamical and chemical data on thick disk stars is required to distinguish between these various scenarios. In particular, the long dynamical times, especially in the outer disk, result in a persistent memory of past perturbations, either secular or external, which have created lasting Galactic warps and flares \citep{monany06, minchev15, laporte21}. 

DESI MWS will yield chemodynamical measurements of $\sim 4$ million thick disk stars. The DESI spectral resolution is sufficient to determine \correct{the abundance of} C, Mg, Ca and Fe \reply{to an internal statistical precision of $\approx 0.1 - 0.2$~dex} for the majority of stars, and \correct{of additional elements} (e.g. Al, Si, Cr) for a subset with higher \SNR{} \citep[see][]{ting17,xiang19,xiang20,sandford20}. Chemical and age estimates \citep[derivable, for giant stars, using C and N features;][]{2015MNRAS.453.1855M, martig15,  2016ApJ...823..114N, 2018MNRAS.481.4093S, ting19a, vincenzo21} as a function of the scale height and orbital properties will help constrain the origin of the thick disk and the relative ages of the main disk components \citep[e.g.][]{belokurov2020, Beraldo_e_Silva2021, Ciuca2021, Montalban2021, 2022Natur.603..599X}. \MWS{} will also study open clusters (OCs) in the thick and thin disks. Although most OCs are at below $|b|<20$, \cwr{several fall} within the \MWS{} footprint, including M67 and M44. \cwr{These well-studied clusters can be used to calibrate abundance measurements and assess systematic differences with other surveys.} DESI spectra can\cwr{, in principle,} be used to explore chemical abundance inhomogeneities in \cwr{these} clusters, probing the diffusion of heavy elements in stellar atmospheres \citep[e.g.][]{Souto2019} and perhaps constraining cluster formation timescales. The broad sky coverage of \MWS{} will enable searches for former cluster members as they disperse through the Galaxy, and potentially contribute \cwr{to} the discovery and characterization of old OCs at high latitude \citep[e.g.][]{schmeja2014,cantatgaudin2020}. 

\subsubsection{Primordial Stars in the Milky Way}
\label{sec:goals_empstars}

Primordial Milky Way stars are expected to form with extremely low metal abundance (perhaps zero). Stars that form after the pristine interstellar medium has been metal-enriched by the first supernova are predicted to reflect exotic abundance ratios and can inform models of yields from the first generation of supernovae. For example, dramatically high carbon-to-iron ratios have been found and are interpreted to be the result of fallback of the innermost layers of supernova progenitors \citep{2014ApJ...792L..32I,2019MNRAS.488L.109N}.

Studies of the MDF show that only 1 in 10,000 halo stars has a metallicity $\mathrm{[Fe/H]}<-4$ \citep{2005ARA&A..43..531B, 2017MNRAS.471.2587S}. This, together with the decrease in metallicity in the outer halo, makes large and distant samples a necessity in order to search efficiently for the first-generation low-mass stars that have survived. Surprisingly, a significant fraction of the stars with the lowest iron abundances {\cwr have their orbits in the Galactic plane \citep{sestito20, sestito22}.}

Most of the extremely metal-poor stars known have been identified from deep surveys of the Galactic halo, such as the Hamburg/ESO Survey \citep{schork_hes_mdf} and SDSS \citep{yannyetal09-1,rockosi22}. \reply{Although metallicity precision is likely to be poor at  $\mathrm{[Fe/H]}<-4$ for spectra at the resolution of DESI, such stars can, at least, be identified efficiently for follow-up observations \citep[e.g.][]{2015_cap_lowz_1,2015_frebel_lowz_2}}. Given the large sample size of \MWS{} and its focus on faint stars at high latitude, we expect to uncover over 100 new stars with $\mathrm{[Fe/H]}<-4$, which would triple the number of currently known stars in this regime. We also hope to identify new stars with iron abundances $\mathrm{[Fe/H]}< -7$, comparable to the current lowest known measurements \citep{2019MNRAS.488L.109N}. The statistics of the parameters and chemical abundances of these stars will shed light on the first generation of stars, and their supernovae \citep{2014ApJ...792L..32I,2019MNRAS.488L.109N}.

\subsection{The DESI White Dwarf Survey}
\label{sec:goals_wd}

\begin{figure}
	\includegraphics[width=\columnwidth]{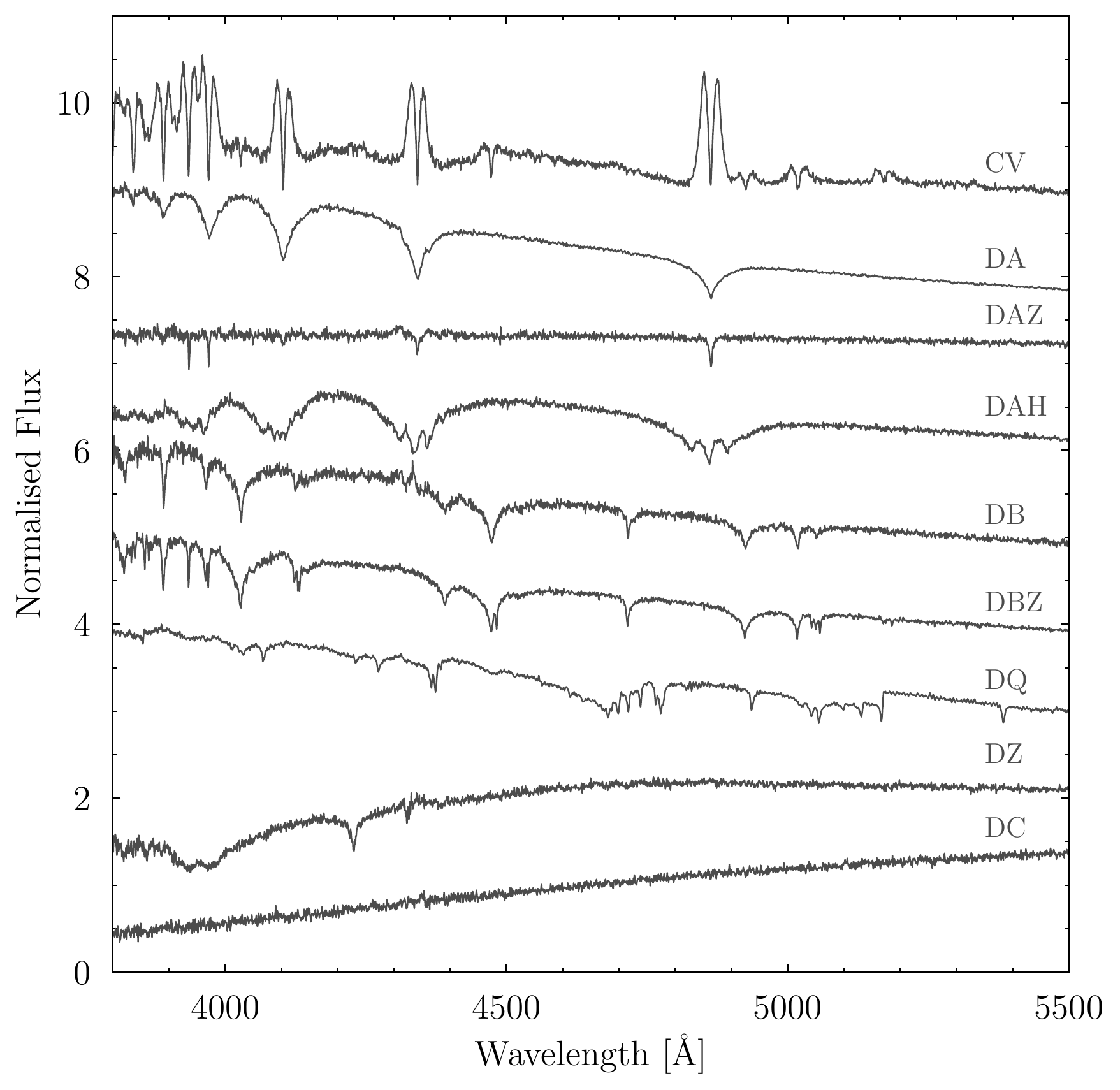}
    \caption{Various types of white dwarf systems observed by DESI. Most white dwarfs have atmospheres dominated by hydrogen or helium, and their spectra contain only Balmer (DA) or He \textsc{I} (DB) lines. As the white dwarfs cool, they eventually cease to produce these lines, resulting in featureless spectra (DC).  However, $\simeq20\%$ of white dwarfs exhibit spectroscopic peculiarities. Accretion of planetary material results in photospheric contamination by metals (DAZ, DBZ, and DZ). The presence of carbon in the atmospheres of white dwarfs can indicate the dredge up of material from the white dwarf core, or, in higher-mass white dwarfs, is an indication of a merger, possibly descending from R\,Corona Borealis stars (DQ). Up to 10\% show magnetic fields (up to $10^{9}\,\mathrm{MG}$) via Zeeman splitting across all atmospheric compositions (e.g.\ DAH), and serve as laboratories for atomic physics under extreme conditions. Finally, white dwarf binaries span a wide range of evolutionary channels. One common type of white dwarf binary is a cataclysmic variable (CV), where a main-sequence companion accretes onto a white dwarf via an accretion disk that can be identified from double-peaked line profiles.}
    \label{fig:wdspectra}
\end{figure}

White dwarfs are the final stage of evolution for stars with masses $\lesssim 8 - 10\,\msun$ \citep{ibenetal97-1, dobbieetal06-1}, a destiny that the majority of A/F-type stars in the Milky Way have already reached. As such, white dwarfs play a central role across a variety of areas in astrophysics. These dense stellar remnants are chemically stratified with atmospheric compositions dominated by hydrogen and/or helium (e.g., \citealt{eisensteinetal06-1, giammicheleetal12-1}), although $\simeq$\,20\% of white dwarfs display spectroscopic peculiarities (see Figure~\ref{fig:wdspectra} for examples). Spectroscopy spanning the full optical range is therefore critically important for the study of white dwarfs.

Homogeneous samples of white dwarfs with accurate physical parameters are essential for constraining and calibrating stellar evolution theory \citep{williamsetal09-1, cummingsetal18-1}, internal rotation profiles and loss of angular momentum \citep{hermesetal17-1}, and fundamental nuclear reaction rates \citep{kunzetal02-1}, with important implications for stellar population synthesis and galaxy evolution theory \citep{marastonetal98-1, kaliraietal14-1}. Because of their well-constrained cooling ages, white dwarfs provide an insight into the age of the Galactic disk \citep{wingetetal87-1, oswaltetal96-1}, OCs \citep{garcia-berroetal10-1}, and globular clusters \citep{hansenetal07-1}, and can even trace variations in the Galactic star-formation rate \citep{tremblayetal14-1}. White dwarfs also indirectly allow the investigation of other areas of astrophysics, such as main-sequence stars in binaries with white dwarfs \citep{zorotovicetal10-1, rebassa-mansergasetal16-1, toonenetal17-1}, planetary systems \citep{zuckermanetal07-1, farihietal09-1, gaensickeetal12-1, koesteretal14-1, hollandsetal18-1,vanderburg20,guidry21}, and extreme physics \citep{guan06-1, kowalski06-1}.

\citet{gentile-fusilloetal19-1} assembled the first unbiased all-sky magnitude-limited ($G\simeq20$) sample of $\simeq260,000$ white dwarf candidates using Gaia~DR2. While the Gaia data are sufficient to identify white dwarfs with high confidence, follow-up spectroscopy is required (Figure~\ref{fig:wdspectra}) to determine their physical properties and derive fundamental properties that are necessary to address the science areas outlined above. 

The DESI white dwarf survey will target $\simeq$70,000 white dwarfs largely from the \citet{gentile-fusilloetal19-1} sample, and will roughly double the number of white dwarfs with high-quality spectroscopy in the northern hemisphere \citep{kleinmanetal13-1, gentile-fusilloetal15-1}. This will be the first large and homogeneous spectroscopic sample that is not subject to complex selection effects, and is therefore ideally suited for detailed statistical analyses of white dwarfs in the context of galactic, stellar, and planetary structures and evolution. The large sample size will also result in the identification of rare white dwarf species, tracing the extremes of parameter space of short-lived phases in their evolution. The combination of accurate Gaia parallaxes and photometry with spectroscopic mass determinations will result in an extremely stringent test of the mass--radius relation of white dwarfs \citep{tremblayetal17-1}. The DESI spectroscopy will also provide radial velocity measurements, which will complement the Gaia proper-motions and produce the first large sample of white dwarfs with full 3D kinematics. With this sample, the thin disk, thick disk, and halo populations can be distinguished \citep{paulietal03-1, paulietal06-1, anguianoetal17-1}, with an expected $\simeq1\%$ of halo white dwarfs. Furthermore, the kinematics will provide constraints on the age--velocity dispersion relation, and insight into the mass distribution of white dwarfs that formed via binary mergers \citep{wegg+phinney12-1}.

\subsection{The DESI Nearby Star Survey}
\label{sec:goals_nearby}

Prior to Gaia, our knowledge of the solar neighborhood was extremely limited: the RECONS survey \citep{henryetal94-1,henry:18} provided only a nearly complete view of the stellar population within 10\,pc, comprising $\simeq300$ stellar systems, of which about a third are binaries or higher multiples. Thanks to Gaia, the inventory of nearby stars has considerably increased, and is now $\simeq90$\% complete for stars down to the hydrogen burning limit, $M_G\simeq15$, within 100\,pc \citep{gaia100pc-21}. 

\MWS{} will include a highly complete spectroscopic census of stars within 100\,pc of the Sun. As described below, these stars will be selected in the DESI footprint based on Gaia parallaxes, in the magnitude range $16 \le G \le 20$~mag (i.e. below the faint limit of the Gaia RVS spectrograph). This sample will be heavily dominated by M-dwarfs, which represent the bulk of the Galactic stellar mass. Much of the current framework for the star formation history of the Milky Way disk and the low-mass end of the stellar initial mass function is based on smaller and brighter analogs of this sample \citep{freeman:02, nordstrom:04, henry:18}.  The \MWS{} 100~pc sample will provide the most complete census of the kinematics, chemical evolution, and initial mass function in the extended solar neighborhood to date.  DESI will cover two major indicators of stellar activity, H$\alpha$ emission and the Ca\,\textsc{II} H and K doublet, and the 100\,pc MWS sample will provide detailed constraints on the fraction of active M-dwarfs; on empirical relations between magnetic activity, rotation rate (which can be determined from, e.g., Zwicky Transient Facility light curves), and age \citep{skumanich_72, pizzolato_03, barnes_07, mh_08}; and on the role of multiplicity in magnetic activity (notably for older stars; \citep{newton_16, stauffer_18}. Correlations between kinematics (i.e.\ tangential velocity, velocity dispersion, or vertical action) and stellar activity \citep{gizis_00, schmidt_07, kiman_19, angus_20} can also be explored with this sample.  

Our knowledge of the coldest (late T and Y type) brown dwarfs ($350 \lesssim T_{\rm eff} \lesssim 500$~K) is similarly restricted to this volume \citep{kirkpatrick2019,kota2022}. In addition to the Gaia 100\,pc sample, a DESI secondary program will target fast-moving high-proper-motion identified using combinations of multi-epoch data from SDSS, the NOIRLab Source Catalog \citep{nidever2021AJ}, and the Wide-field Infrared Survey Explorer (WISE) data \citep[e.g.,][]{meisner2021}. These data will sample the faint end of the brown dwarf luminosity function, and provide better constraints on the low-mass end of the initial mass function. 

\subsection{Survey requirements}

The scientific goals above impose common requirements on the sample of stars to be observed, and on the performance of DESI  and its pipelines, particularly on the accuracy of key measurements (stellar parameters, radial velocities, and abundances) on the final dataset. The rest of this paper presents the design of the survey and our first evaluation of our pipeline against those requirements.

In the next section, we describe our target selection procedure and forecast the size and content of the full sample. We determine the completeness of each target class by applying the DESI fiber assignment algorithm, accounting for the
prioritization of targets from BGS. For the bulk of our targets the completeness is $\approx 30$~per cent averaged over the footprint. Using a simple mock catalog based on an empirical model of the Galactic structure, we demonstrate that our target selection recovers tracers with sufficient density and distance coverage to address the core dark matter and stellar halo structure science goals. In future work, we will apply our target selection to mock catalogs derived from ab initio cosmological simulations, to better understand the effects of halo-to-halo variation and substructures. 

We then introduce our analysis pipelines and use data from early DESI observations to demonstrate that we can recover radial velocities and $\mathrm{[Fe/H]}$ to the necessary precision of $1\,\mathrm{km\,s^{-1}}$ and $\approx0.2$~dex respectively. We identify systematic offsets with respect to other surveys and literature $\mathrm{[Fe/H]}$ measurements for globular clusters and dwarf galaxies. We also demonstrate that metal pollution signatures of planetary systems can be identified in DESI spectra of white dwarfs. 

We identify priorities for development of the current pipelines, which we expect to include $\mathrm{[\alpha/Fe]}$ measurements, more robust fitting of very cool stars, and potential improvements to radial velocity accuracy below $1\,\mathrm{km\,s^{-1}}$.  These improvements will be addressed in future publications, along with more detailed studies of the white dwarf and nearby samples, spectrophotometric distance estimates, and the recovery of individual element abundances.

\section{Selecting Milky Way Targets for DESI}
\label{sec:targets}

This section describes the selection criteria for \MWS{} targets. This description updates the summary given in \citet{mwsTSNote} and corresponds to the final design of \MWS{} at the start of the main survey in 2021 May. We refer the reader to \updatecite{\cite{myers22a}} for further details of the DESI target selection process, the \texttt{desitarget} software, and associated data products.

\subsection{Overview of MWS target categories}

\paragraph{\cwr{Primary targets.}} The MWS \textit{main sample} is designed to address the dark matter halo, stellar halo, and thick disk science cases described in sections \ref{sec:goals_dmhalo}, \ref{sec:goals_assembly} and \ref{sec:goals_thick_disk}. \cwr{It comprises} the bulk of the stellar targets that will be observed by DESI \cwr{\MWS{}}, split into three categories\footnote{These names refer to the \texttt{mws\_mask} (and in some cases \texttt{scnd\_mask} and \texttt{desi\_mask}) bitmask fields described in \updatecite{\citet{myers22a}}. \cwr{They are used in} the DESI data products to identify categories of targets selected according to specific criteria.}, \mainblue{}, \mainred{} and \mainbroad{}. The union of these \cwr{categories} is essentially \cwr{a magnitude-limited selection} and therefore well suited to characterizing stellar halo substructure\cwr{s} associated with disrupted Milky Way satellites and globular clusters (section \ref{sec:goals_streams}) and to the serendipitous discovery of rare types of stars (such as extremely metal poor stars; section~\ref{sec:goals_empstars}).

\cwr{The main sample selection function is deliberately generous. Any starlike source in the survey footprint with magnitude $16<r<19$ has a finite probability of being observed in one of the three \MWS{} main target categories (excluding only sources that are masked or have poor data quality in the input imaging). As we describe in detail below, \mainblue{} (defined by the simple color criterion $g-r<0.7$) will provide a highly complete sample of metal-poor main-sequence, turnoff, and bluer subgiant stars, with distances in the range 5 -- 30~kpc and no kinematic selection bias. This will be a dense, high-fidelity dataset with which to study the thick disk and inner halo over the likely extent of the GSE debris, as well as Sagittarius and many other known streams.}

\cwr{We assign redder stars ($g-r>0.7$) in the same $16<r<19$ magnitude range to the \mainred{} category if Gaia astrometry indicates they are more likely to be distant halo giants, and to the \mainbroad{} category otherwise. We give \mainred{} targets the same fiber assignment priority as \mainblue{} targets. We give \mainbroad{} targets lower priority in order to select more \mainred{} giant candidates to probe the kinematics and substructure of the outer stellar halo, particularly to provide dynamical constraints on the dark matter halo mass over a large fraction of the virial radius. Of course, not all stars in \mainred{} will be distant giants. The astrometric selection criteria are "mild"; as described in more detail below, they accept relatively high contamination from nearby red main-sequence stars as a trade-off to minimize kinematic bias. Based on the current DESI bright-time observing plan, focal plane state, and fiber assignment strategy, we expect \mainred{} spectra to account for $\approx12\%$ of the main sample and to have a completeness of $\approx32\%$ (similar to that of \mainblue{}). The fraction of true distant halo stars in \mainred{} will be sensitive to the form of the stellar halo density profile at large distances; below, we present a fiducial model of the halo density which predicts $\sim5,000$--$10,000$ \mainred{} giants beyond 50~kpc and $\sim1,000$ beyond 100~kpc. Although \mainbroad{} targets will have lower spectroscopic completeness than \mainred{} $(\sim19\%)$, those spectra will still account for $\approx32\%$ of the main \MWS{} sample. They will serve both as a constraint on forward models of the \MWS{} dataset and as an important scientific sample in their own right, providing radial velocities for thin disk stars without Gaia RVS spectra, as well as more detailed abundance measurements.}

\cwr{In addition to the main sample,} MWS includes several \textit{high-priority samples} of targets with high scientific value and low surface density ($\lesssim 10$ per square degree), comprising white dwarfs (\mwswd{}, section \ref{sec:goals_wd}), faint nearby stars (\nearby{}, section \ref{sec:goals_nearby}), RR\,Lyrae variables (\mwsrrlyr{}), and horizontal branch stars (\mwsbhb{}). These target categories are given higher fiber assignment priority than the main sample to ensure high completeness, as discussed in Section~\ref{sec:strategy}. Two further categories, \faintblue{} and \faintred{}, \cwr{target fainter stars in the same color ranges as the \mainblue{} and \mainred{} samples,} at lower priority and hence much lower completeness.

F-type stars in a similar magnitude range to the main \MWS{} sample will be targeted as \textit{flux standards} across all bright- and dark-time DESI programs\footnote{White dwarfs were also targeted as standards in the DESI SV programs, and consequently received a larger number of repeat and dark-time observations than they would have as \MWS{} targets alone. White dwarfs are not currently being targeted as standards in the main DESI survey program.}. The algorithm for selecting flux standards is more complex than \cwr{that for} the \mainblue{} sample, of which they are \cwr{essentially} a subset \cwr{(flux standards can be slightly brighter than the flux limit of the main sample)}. Their targeting and prioritization is independent of \MWS{}; survey, hence they may also be observed as part of the \MWS{} main sample or as one of the high-priority \MWS{} samples.  Main \MWS{} targets will be observed only once by design\cwr{; they will only be reobserved if no unobserved MWS targets  or targets from any other program are available to a fiber}. However, stars \cwr{may be observed as} flux standards multiple times in both bright and dark conditions.

We refer to the \MWS{} main, high priority and faint samples \cwr{collectively} as the \textit{primary MWS target categories}. All these targets are selected within the bright-time program footprint, shown in Fig.~\ref{fig:footprint}. The following subsections give the specific target selection criteria for each category. Most criteria are based on optical photometry from Data Release 9 of the DESI Legacy Imaging Survey \updatecite{\citep[\legacy{} DR9;][]{zou_bass_2017,ls_overview,schlegel22a}}, which also includes infra-red photometry derived from WISE data \citep{meisner17}, and is matched to photometry from Gaia DR2 and astrometric measurements from Gaia EDR3 \citep{EDR3_2021}.

Table~\ref{tab:dr9_target_summary} lists the number of targets per category in the bright-time footprint (categories are not mutually exclusive). Approximately 30.5 million unique \legacy{} sources meet the primary sample selection criteria. Section~\ref{sec:strategy} describes the relative prioritization among the categories and the resulting fiber assignment efficiencies over four passes of the footprint, which are also listed in Table~\ref{tab:dr9_target_summary}. The complete survey will contain $\simeq7.2$ million \cwr{spectra of unique stars} from the primary MWS categories, \cwr{of which} $\simeq6.5$ million \cwr{are} from the three main sample categories.

\paragraph{\cwr{Secondary targets.}} As noted in Section~\ref{sec:introduction}, the DESI survey also defines a number of \textit{secondary science programs}, each of which corresponds to one or more \textit{secondary target classes}\footnote{These are distinguished in the DESI data products by a nonzero entry in the \texttt{SCND\_TARGET} column corresponding to one or more fields in the \texttt{scnd\_mask} bitmask; see \updatecite{\citet{myers22a}}.}. The secondary programs will use spare fibers or dedicated observations of specific areas (not necessarily within the main survey footprint) to complement the goals of the primary dark- and bright-time DESI surveys. Several secondary programs are complementary to \MWS{}, including observations of fainter BHBs and white dwarf binary candidates using dark time, higher prioritization for proper-motion and color outliers, surveys of M31 and M33 \updatecite{\citep{DeyM31}}, and high-completeness observations of selected Milky Way dwarf satellites and globular clusters. These secondary programs will be described separately. The targets for these secondary programs may overlap with primary \MWS{} targets. \updatecite{\citet{myers22a}} provide a more detailed explanation of the primary and secondary target categories, how they can be identified in DESI target catalogs, and the implications of targets being selected by multiple programs.

\paragraph{\cwr{Backup targets.}} Finally, in poor weather conditions and bright skies (including twilight), the DESI \textit{backup program} will target stars brighter than those in the \MWS{} main sample. The backup program will cover a larger footprint, extending to lower Galactic latitude ($|b|\gtrsim7^{\circ}$). Backup targets will be prioritized in three categories by magnitude and color, to favor stars for which high \cwr{\SNR{}} can be achieved in poor conditions, with two additional categories to favor brighter halo giants\footnote{In the DESI data products, backup program targets can be identified using the MWS target column ({\tt MWS\_TARGET}) and associated bitmask ({\tt mws\_mask}), but the backup program is in most respects treated separately from the other DESI surveys \updatecite{\citep{sv22a,myers22a}}.}. This sample will be suited to detailed chemodynamical modeling of the Galactic disk and nearby stellar halo. The total number of spectra obtained by the backup program will depend on observing conditions. Extrapolating from the first months of the main survey, we expect the backup program will obtain at least five million additional stellar spectra. A full description of the backup program will be given \cwr{in a separate publication}.

\subsection{Main sample selection}
\label{sec:main_targets}

\begin{table}
 \caption{Summary of main \MWS{} target classes. From left to right, columns give the name of the class in the \texttt{mws\_mask} bitmask, the total number of DR9 targets in the class available to a DESI fiber and the total number and fraction of fibers assigned to those targets in a fiducial 4-pass survey (with the footprint and focal plane state defined in March 2022; see text). Targets may be counted in more than one class; the final row gives the total number of unique targets.}
\label{tab:dr9_target_summary}
\begin{tabular}{lrrr}
Class & $N_{\mathrm{footprint}}$ & $N_{\mathrm{fiber}}$ & $\%_\mathrm{4\,pass}$ \\
\hline
\mainred{} & 2,589,211 & 805,794 & 31 \\
\mainblue{} & 11,450,969 & 3,693,518 & 32 \\
\mainbroad{} & 10,783,830 & 2,077,222 & 19 \\
\mwswd{} & 66,811 & 66,365 & 99 \\
\nearby{} & 33,788 & 20,892 & 62 \\
\mwsrrlyr{} & 17,303 & 8981 & 52 \\
\mwsbhb{} & 32,353 & 17,706 & 55 \\
\faintred{} & 960,134 & 71,771 & 7 \\
\faintblue{} & 4,606,314 & 486,568 & 11 \\
\hline
Total (unique) & 30,470,033 & 7,197,722 & 24 \\
\end{tabular}
\end{table}

The \MWS{} main sample is selected from the \legacy{} DR9 source catalog combined with astrometric measurements from Gaia~EDR3. It is divided into three subcategories: \mainblue{}, \mainred{} and \mainbroad{}. All three categories share the following definition of a \MWS{} main sample stellar target based on fields in the \legacy{} DR9 catalog\footnote{\url{https://www.legacysurvey.org/dr9/description}}:

\begin{itemize}
\item \texttt{$16< r <19$}
\item $r_{\mathrm{obs}}<20$
\item \texttt{type = PSF}
\item \texttt{gaia\_astrometric\_excess\_noise} $< 3$
\item \texttt{gaia\_duplicated\_source} = \texttt{False}	
\item \texttt{brick\_primary} = \texttt{True}	
\item \texttt{nobs\_\{g,r\}} $>0$ 
\item \texttt{\{g,r\}\_flux} $>0$ 
\item \texttt{fracmasked\_\{g,r\}} $<0.5$
\end{itemize}

The observed $r$-band magnitude is obtained from the \legacy{} flux in nanomaggies as $r_{\mathrm{obs}} = 22.5-2.5\log_{10}$ \texttt{r\_flux}. \cwr{The extinction-corrected magnitude is computed} as $r = r_{\mathrm{obs}} + 2.5 \log_{10}$ \texttt{mws\_transmission\_r}. We do not put any requirements on the availability or quality of data in the \legacy{} $z$ band. 

The \mainblue{} subsample is further defined by the color criterion 
\begin{itemize}
    \item $g-r < 0.7$ 
\end{itemize}
with $g$ defined in the same way as $r$ above. \cwr{\mainblue{} is therefore an approximately magnitude-limited selection, comprised mainly of main-sequence turnoff stars and bluer subgiants. Stars redward of the $g-r=0.7$ cut include redder main-sequence stars, subgiants, and giants. To improve the sampling of the more distant galaxy, higher priority is given to \mainred{} targets, which have astrometric cuts designed to reduce nearby main-sequence contaminants. However, Gaia astrometry becomes less precise near the faint limit of MWS, so the lower priority \mainbroad{} target category loosens these astrometric cuts. More specifically,} the \mainred{} subsample is defined by the following additional criteria:
\begin{itemize}
\item $g-r \ge 0.7$
\item \texttt{astrometric\_params\_solved} $\ge 31$
\item Gaia parallax $\pi < 3\sigma_{\pi} + 0.3\; \mathrm{mas}$
\item Gaia proper-motion $|\mu| < 5 \sqrt{f_{r}/f_{19}}   \;\mathrm{mas/yr}$ $\left(f_{x} = 10^{(22.5-x)/2.5}\right)$
\end{itemize}

The \mainbroad{} sample comprises stars with color $g-r \ge 0.7$ that satisfy the same magnitude and data quality requirements as the other two categories, but do not satisfy one or more of the astrometric criteria that define \mainred{} (stars that do not have well-measured astrometric parameters in the Gaia catalog are therefore included in \mainbroad{}). 

\cwr{The parallax cut applied to \mainred{} with respect to \mainbroad{}} aims to remove nearby red contaminants \cwr{in order to recover a larger sample of distant halo giants}. However, because {\it Gaia} parallax accuracy is not good enough at the faint limit of the survey, $r=19$, we also apply a proper-motion cut to reject stars with high tangential velocities that are more likely to be nearby contaminants. The proper-motion cut is made to be a function of the $r$-band magnitude, corresponding to $5\, \mathrm{mas\,yr^{-1}}$ at $r=19$ and $\sim 20\,\mathrm{mas\,yr^{-1}}$ at $r=16$. This corresponds to different limits on tangential velocity, depending on the absolute magnitude of the target. For example, for a 10 Gyr old stellar population with $\mathrm{[Fe/H]}<-0.5$, the faintest stars on the giant branch with $g-r>0.7$ will have $M_r\lesssim 1$, thus the proper-motion cut will select all stars with tangential velocities below $950\,\kms$. For stars with fainter $M_r$, for example the subgiant branch stars of a 10 Gyr old stellar population with $\mathrm{[Fe/H]=0}$ that have $M_r\sim 3.5$, the tangential velocity cut will correspond to $\sim 300\,\kms$.

The lower probability of observing stars with high tangential velocities in the \mainred{} sample at distances $\lesssim20$~kpc does not compromise the science goals of MWS. As shown below, the MWS sample at these distances is overwhelmingly dominated by \mainblue{} targets, which have no astrometric selection bias. It reduces the chance of discovering extremely high-velocity red stars around the base of the giant branch at these intermediate distances, but not to zero, because such objects are still targeted in the \mainbroad{} selection. 

Finally, the \faintblue{} and \faintred{} samples extend the \mainblue{} and \mainred{} classes, respectively, to the fainter magnitude range $19 < r < 20$. The selection criteria are otherwise identical to those for the corresponding main classes, except for a slightly fainter limit on the observed $r$-band magnitude, $r_{\mathrm{obs}}<20.1$. These faint categories are intended as targets of last resort and are given the lowest fiber assignment priority of any unobserved target\footnote{\cwr{We treat \faintblue{} and \faintred{} targets as part of our primary sample, but for technical reasons, they are implemented in the DESI target selection software as secondary program targets} (and hence are identifiable in the data products using \texttt{scnd\_mask} and \texttt{SCND\_TARGET} rather than \texttt{mws\_mask} and \texttt{MWS\_TARGET}). See \updatecite{\citet{myers22a}}}. Although \faintblue{} and \faintred{} will be much lower completeness and lower \cwr{\SNR{}} samples, they will provide opportunities for serendipitous discovery and may boost the detection significance of faint substructures. 

\subsection{Main sample targets in the Galaxia model}

\begin{figure}
    \centering
    \includegraphics[width=\linewidth]{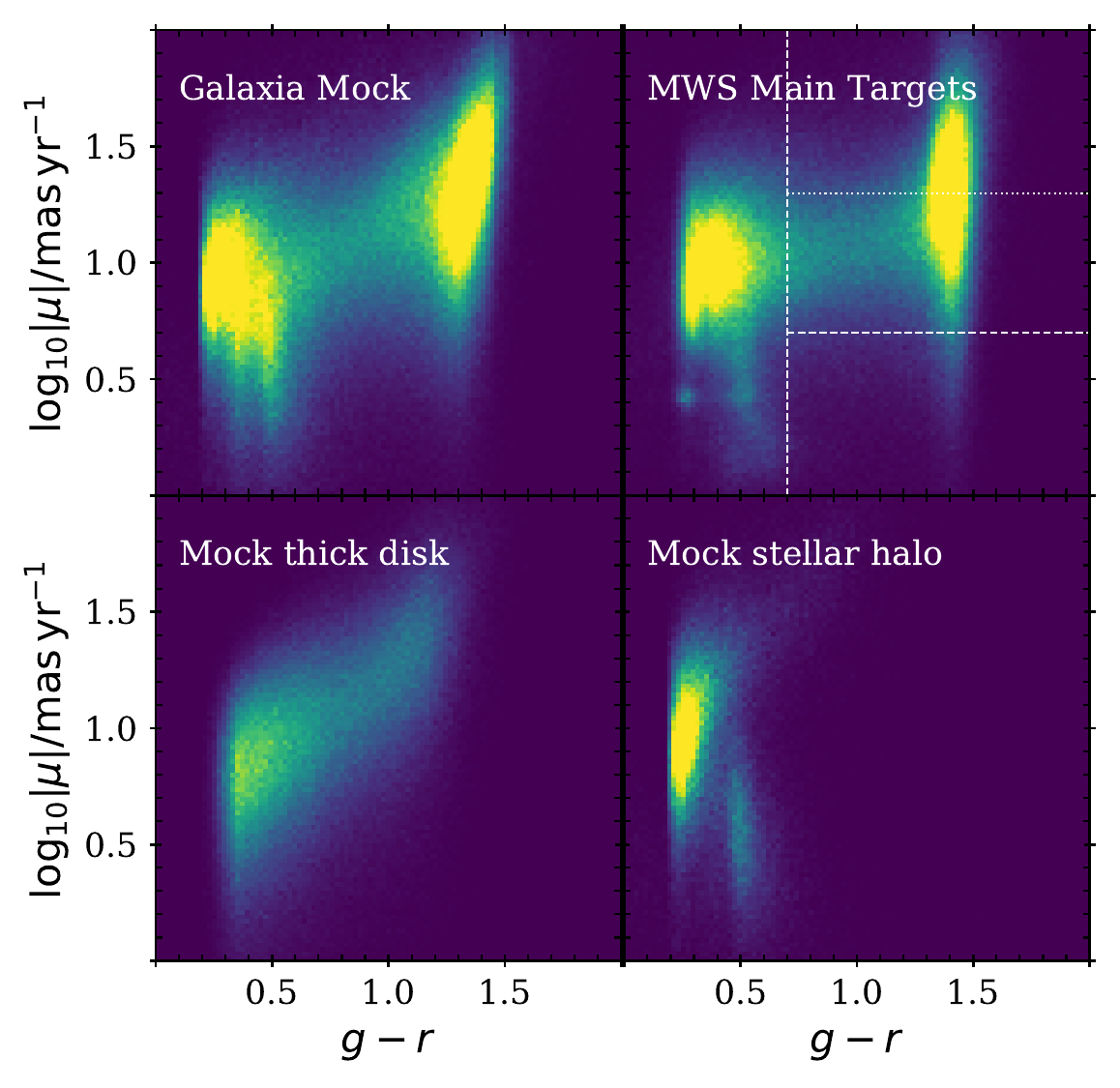}
    \caption{Top panels: Density of MWS  main sample targets at high Galactic latitude ($b > 70\,\deg$) in the space of $g - r$ color and proper-motion, in our broken-power-law halo Galaxia mock catalog (left) and the real MWS target catalog (right). The vertical line marks the color separation between the \mainblue{} and \mainred{}/\mainbroad{} samples. The horizontal lines show the separation in proper-motion between \mainred{} and \mainbroad{} at $r=16$ (upper dotted line) and $r=19$ (lower dashed line). The main-sequence turnoff and giant branch of the globular cluster M15 are visible as density peaks at $\log_{10}\,|\mu|/\mathrm{mas\,yr^{-1}}\sim0.4$. Bottom panels: The contribution of thick disk (left) and halo stars (right) to the density distribution in the Galaxia mock catalog.}
    \label{fig:pm_color_4panel}
\end{figure}

Fig.~\ref{fig:pm_color_4panel} shows the color and proper-motion distribution of \mainblue{}, \mainred{} and \mainbroad{} targets\footnote{Note that stars with low proper-motion can also be assigned to \mainbroad{} if they have low parallax, or if their proper-motion or parallax is not measured by Gaia.} at high Galactic latitude ($b > 70\degr$). We compare these distributions to a mock catalog derived from the Galaxia Milky Way model \citep{sharma2011}, which describes the phase-space distribution of stellar populations in the Milky Way's thin disk, thick disk, and halo. The parameters of the default Galaxia model were calibrated to surveys of relatively bright and nearby stars. More recent studies have improved our knowledge of these parameters, particularly for the stellar halo. The original Galaxia model uses a power-law density profile with a logarithmic slope of $-2.44$, which is inconsistent with recent evidence that the stellar halo density breaks to a much steeper slope at a distance of $\sim25$~kpc \citep[e.g.][]{bell2008,deason2013,deason2014}. \cwr{To make more accurate predictions for \MWS{}, we modify the Galaxia model such that the logarithmic slope of the stellar halo density profile\footnote{\cwr{For these simple estimates of target counts, we only modify the Galaxia density profile. This is a simplification. If the break is associated with the characteristic apocenter of debris from a single progenitor that dominates the inner halo, the velocity (and hence proper-motion) distribution in the model should also be adjusted.}} changes from $-2.5$ to $-4$ at a Galactocentric distance of 25~kpc}. To turn the output of Galaxia into a mock MWS survey, we convert magnitudes from Galaxia to the LS photometric system via the PS1 color terms given on the LS website, and assign proper-motion and parallax errors using \texttt{PyGaia}\footnote{\url{https://github.com/agabrown/PyGaia}}. We then apply exactly the same footprint and target selection functions (except for the LS data quality flags) to obtain mock \mainblue{}, \mainred{} and \mainbroad{} samples.

The upper panels of Fig.~\ref{fig:pm_color_4panel} \cwr{show} two broad peaks in both the MWS target catalog (top right) and our Galaxia mock catalog (top left). These correspond to the metal-poor halo (bluer and low proper-motion, mostly in the \mainblue{} sample) and the thin disk (redder and high proper-motion, mostly in the \mainbroad{} sample). In the distribution of real \mainblue{} targets, two smaller, concentrated peaks visible at $\log_{10}|\mu / \mathrm{mas\,yr^{-1}}|\sim0.4$ correspond to the main-sequence and giant branch of the globular cluster M15. In the Galaxia mock, the lower-density sequence that bridges the disk and halo regions of the diagram corresponds to the thick disk, as \cwr{shown} in the lower left panel, where we isolate the thick disk stars in Galaxia. The lower right panel shows only the halo stars in Galaxia, demonstrating that the three distinct sequences visible in \mainblue{} correspond to the halo main-sequence, the metal-poor thick disk main sequence and the halo giant branch (from blue to red). 

\begin{figure*}
    \centering
    \includegraphics[width=\textwidth]{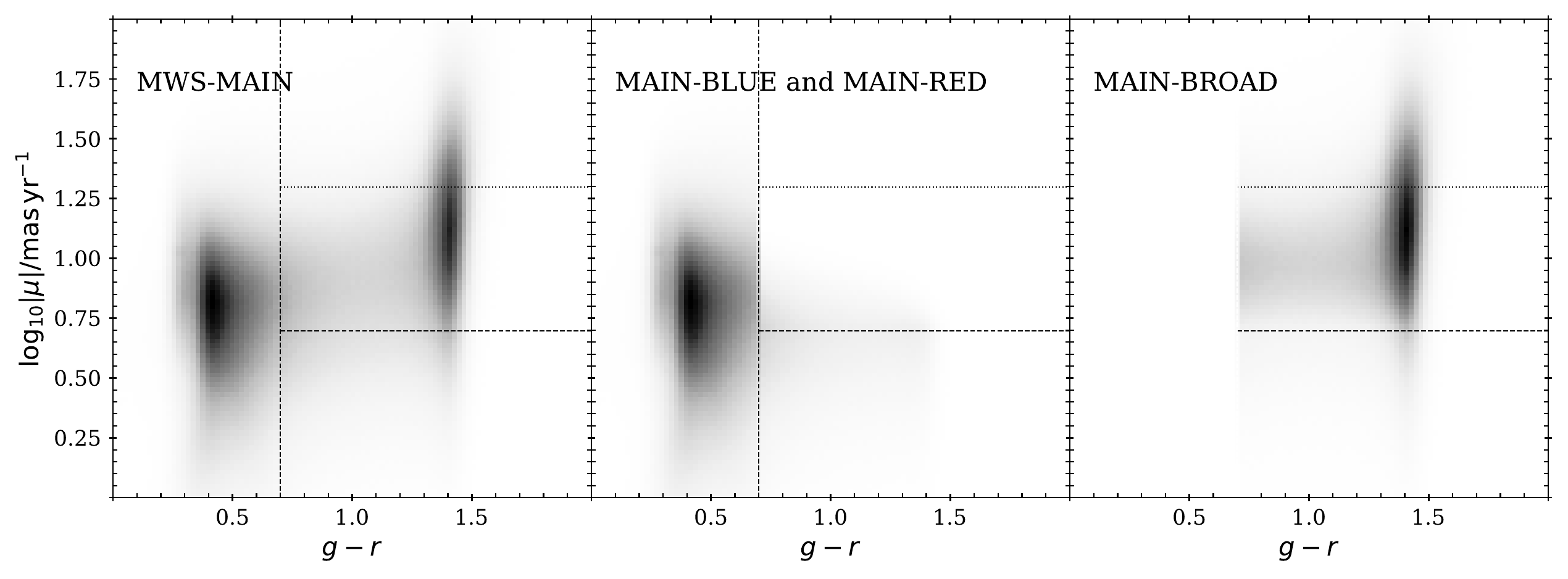}
    \caption{The color and proper-motion distribution of all MWS main sample targets, as shown in the top right panel of Fig.~\ref{fig:pm_color_4panel}, here for the full MWS footprint (left). The middle and right panels show the separation of targets into \mainblue{}/\mainred{} (separated by $g-r=0.7$, vertical dashed line) and \mainbroad{} (separated according to the parallax and magnitude-dependent proper-motion criteria described in the text; the proper-motion cuts at the bright and faint limits of the sample are shown by horizontal lines).}
    \label{fig:pm_color_targets}
\end{figure*}

Fig.~\ref{fig:pm_color_targets} illustrates the magnitude-dependent proper-motion cut that divides the $g-r>0.7$ sample into \mainred{} and \mainbroad{} (this figure includes all targets in the survey footprint, not just those at high latitude). Thick disk stars are included in \mainred{} at $r=16$ but almost completely relegated to \mainbroad{} at the faint limit of $r=19$. 

\begin{figure*}
    \centering
    \includegraphics[width=\textwidth]{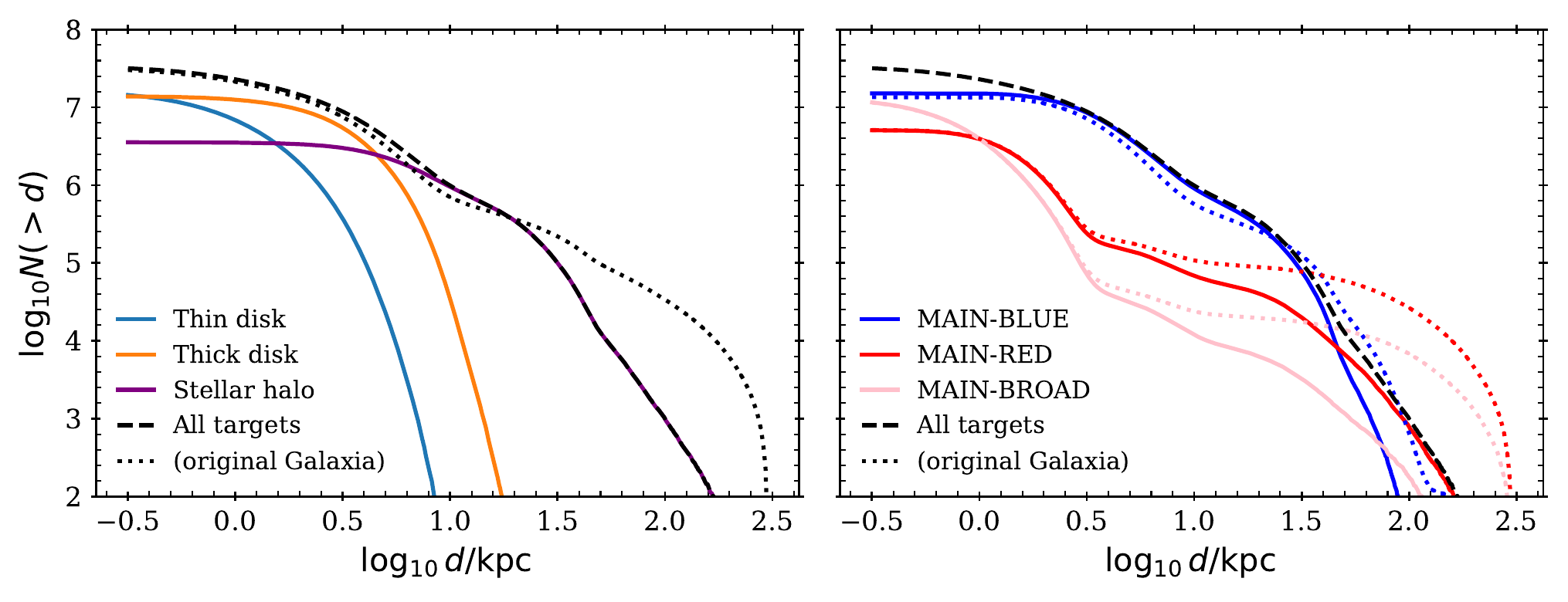}
    \caption{Number of \MWS{} \cwr{main sample} target stars beyond a given heliocentric distance in our broken power-law variant of the Galaxia model (see text), for stars selected in the MWS footprint \cwr{according to the \mainblue{}, \mainred{} and \mainbroad{}} criteria. In the left panel, mock MWS targets are separated by Galaxia structural component (thin disk, thick disk and halo), and in the right panel they are separated by MWS target category. Profiles for the original single-power-law halo model \citep{sharma2011} are shown by dotted lines of the same color.}
    \label{fig:galaxia-distance-distribution}
\end{figure*}

Fig.~\ref{fig:galaxia-distance-distribution} shows the distribution of heliocentric distances in our mock catalog. The mock \mainblue{} targets are dominated by thick disk stars between 1 and 10~kpc. \mainblue{} contains a further $\sim~500,000$ stars between 10 and 20~kpc, which corresponds to the transition between the thick disk and halo in Galaxia. In this distance range the mock \MWS{} sample is dominated by the metal-poor turnoff and subgiant branch. Beyond 20~kpc the mock sample is dominated by the stellar halo. For comparison, the dotted lines in Fig.~\ref{fig:galaxia-distance-distribution} show the number of distant giants predicted by the original Galaxia model \citep{sharma2011}, which has a much shallower density profile in the outer stellar halo. Our fiducial model predicts \MWS{} will target \cwr{$\sim1,000$} giants beyond 100~kpc, whereas the original Galaxia model predicts $\sim10,000$. The right-hand panel of Fig.~\ref{fig:galaxia-distance-distribution} shows that most of the difference is in the \mainred{} and \mainbroad{} samples (which include all but the most metal poor giants at large distances). 

\begin{figure}
    \centering
    \includegraphics[width=\linewidth]{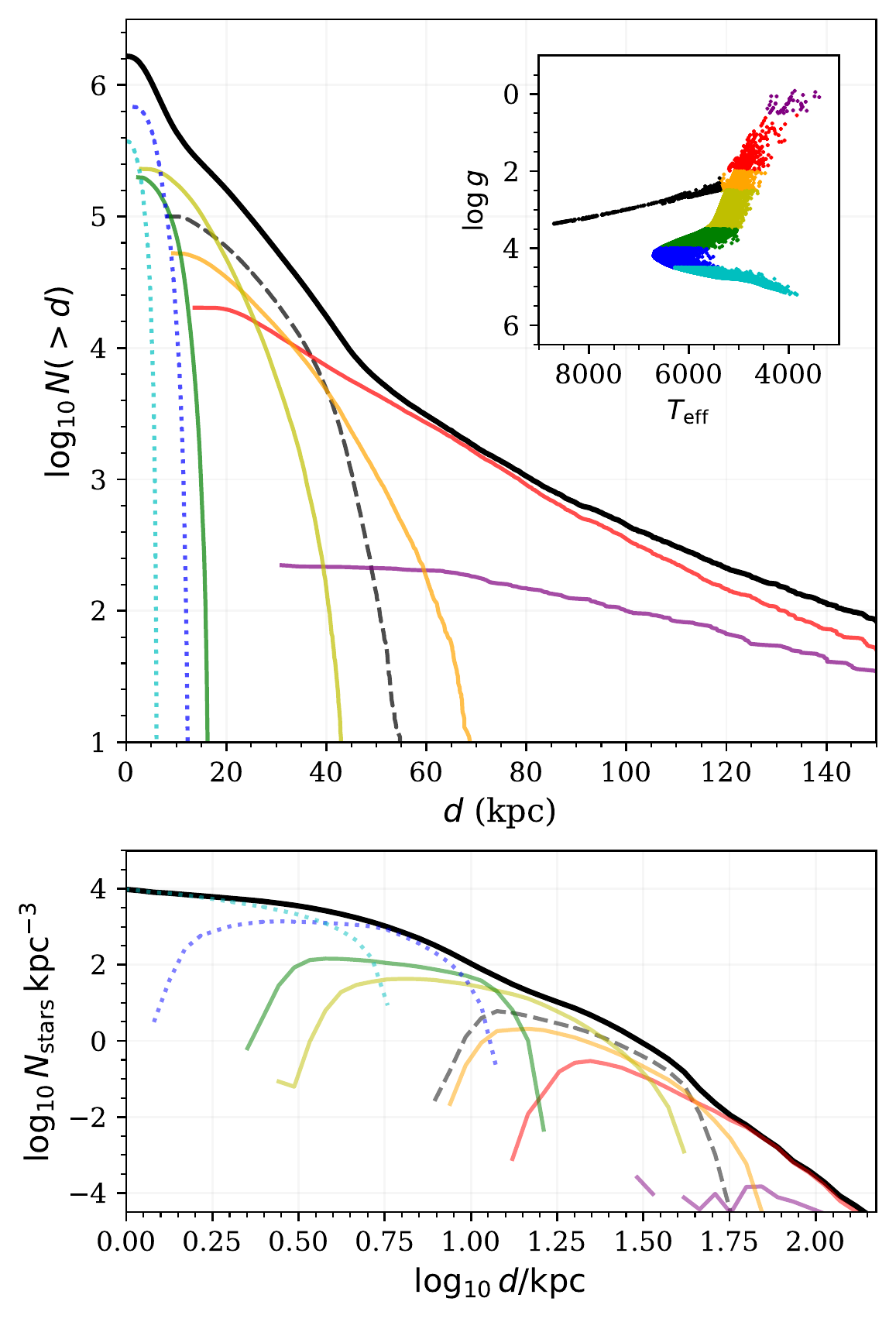}

    \caption{The heliocentric distances of \MWS{} \cwr{main sample} \bugfix{stellar halo} targets in our Galaxia model variant with a broken-power-law stellar halo density profile. The top panel shows the number of \MWS{} targets beyond a given heliocentric distance. The bottom panel shows the volume density of targets at a given distance. Colors correspond to the eight regions of the $T_{\mathrm{eff}} - \log g$ diagram (inset). Dotted lines correspond to main-sequence and turnoff stars, solid lines to subgiants and giants, and dashed lines to horizontal branch stars. The solid black lines correspond to the whole sample.}
    \label{fig:distances_cmd_galaxia_halo}
\end{figure}

Finally, Fig.~\ref{fig:distances_cmd_galaxia_halo} illustrates how \bugfix{halo} stars of different types enter the \MWS{} selection function at different distances. For example, the figure shows that, in this \MWS{} mock survey, \bugfix{halo} turnoff and subgiant branch stars dominate the sample up to \bugfix{$\sim10$~kpc}. The \bugfix{mock} survey targets $\sim100,000$ horizontal branch stars between \bugfix{10} and \bugfix{50~kpc} at a volume density of 1 to 10 per $\mathrm{kpc}^{-3}$, and comparable numbers of RGB and red clump stars in the same region. These are target densities; they can be converted to densities of observed spectra with reference to the fiber assignment simulations described in Sec.~\ref{sec:strategy}. Those simulations show that, in a 5 yr survey sharing the DESI focal plane with BGS, we expect to observe $\sim30\%$ of the \mainblue{} and \mainred{} targets (e.g.\ $\sim300$ of the $\sim1000$ giants beyond \bugfix{80}~kpc).

\subsection{High-priority Sample Selections} 

\subsubsection{\mwswd{}}
\label{sec:wd_targets}

We select white dwarfs using a set of $BP-RP$ color and $G$-band absolute magnitude criteria based on \citet{gentile-fusilloetal19-1} that identify the white dwarf cooling sequence in the Gaia photometric catalog alone. This selection is applied to the \legacy{} catalog using the properties of cross-matched Gaia~EDR3 sources; the \legacy{} photometry is not used in this selection. Photometric measurements are taken from Gaia DR2 and astrometric measurements from EDR3.

\begin{itemize}
\item $G_{\mathrm{abs}} > 5$
\item $BP-RP < 1.7$
\item $G_{\mathrm{abs}} > 5.93 + 5.047(BP-RP)$
\item $G_{\mathrm{abs}} > 6(BP-RP)^3 - 21.77(BP-RP)^2 + 27.91(BP-RP) + 0.897$
\end{itemize}
These criteria are applied to a sample defined by extinction-corrected Gaia $G$ flux (\texttt{gaia\_phot\_g\_mean\_mag} in the \legacy{} catalog) at high latitude, with mild parallax and proper-motion cuts to select against nearby \cwr{luminous blue stars (early-type main-sequence stars, horizontal branch stars, and subdwarfs) as well as QSOs}:
\begin{itemize}
\item $G < 20.0$
\item $|b| > 20$ degrees
\item $\pi/\sigma_{\pi} > 1\;\mathrm{mas}$
\item \texttt{astrometric\_params\_solved} $\ge 31$
\item $|\mu| > 2\;\mathrm{mas/yr}$ 
\end{itemize}
We impose the following photometric quality cut, because failing this criterion results in poor astrometry:
\begin{itemize}
\item \texttt{phot\_bp\_rp\_excess\_factor} $< 1.7 + 0.06(BP-RP)^2$	
\end{itemize}
However, we retain objects \cwr{that have reliable parallaxes and significant proper-motions that meet} either of the following criteria:
\begin{itemize}
\item \texttt{astrometric\_sigma5d\_max} $< 1.5$
\item (\texttt{astrometric\_excess\_noise} $<1$) \&  ($\pi/\sigma_{\pi} >4$) \& ($|\mu| > 10\;\mathrm{mas/yr}$)
\end{itemize}

\subsubsection{\mwsrrlyr{}}
\label{sec:rrlyrae_targets}

We target Gaia~DR2 sources with magnitudes $14<G<19$ that are classified as RR\,Lyrae by the general variability classifier and the Special Object Studies classifier 
\citep{Holl2018, Clementini2019}, comprising all objects from the table \texttt{vari\_rrylrae} and any sources from the table \texttt{vari\_classifier\_result} for which \texttt{best\_class\_name} includes "RR". For technical reasons, \mwsrrlyr{} targets are associated with the secondary target bitmask in the DESI data products, rather than the \MWS{} bitmask \updatecite{\citep[see][]{myers22a}}.

\subsubsection{\nearby{}}
\label{sec:nearby_targets}

To address the goals described in Section~\ref{sec:goals_nearby}, we select dwarf stars within $100$~pc of the Sun based only on apparent Gaia magnitude and parallax, allowing for moderate parallax uncertainties:
\begin{itemize}
\item $16 < G < 20$
\item $\pi + \sigma_{\pi} > 10\;\mathrm{mas}$
\item \texttt{astrometric\_params\_solved} $\ge 31$
\end{itemize}
No extinction correction is necessary to select stars at these distances. This sample corresponds to fewer than 10 targets per square degree. As described below, to ensure high completeness, \nearby{} targets are prioritized over all other MWS targets except \mwswd{} and \mwsrrlyr{}.

\subsubsection{\mwsbhb{}}
\label{sec:bhb_targets}

We select BHBs starting from the basic definition of main sample stars given at the start of this section (common to \mainblue{} and \mainred{}), with the following additional criteria:
\begin{itemize}
\item \correct{$G > 10$}
\item $\pi \leq 0.1 + 3\sigma_{\pi}\;\mathrm{mas}$
\item $-0.35 \leq g-r \leq -0.02$
\item $ -0.05 \leq X_{\mathrm{BHB}} \leq 0.05$
\item $ r - 2.3(g-r) - W1_{\mathrm{faint}} < -1.5$
\end{itemize}
These criteria exclude nearby stars and select around the BHB locus in a combined \legacy{} and WISE color space, defined by
\begin{align}
X_{\mathrm{BHB}} & = & (g-z) - [1.07163(g-r)^5 - 1.42272(g-r)^4 \nonumber \\
                 &   & +\, 0.69476(g-r)^3 - 0.12911(g-r)^2 \nonumber \\
                 &   & +\, 0.66993(g-r) - 0.11368] 
\end{align}
and
\begin{equation}
W1_{\mathrm{faint}} = 22.5 - 2.5\log_{10}(W1 - 3\sigma_{W1}),
\end{equation}
where $W1$ and $\sigma_{W1}$ are the WISE $3.4\mathrm{\mu m}$ flux and its error, respectively.

\subsection{Flux standards}
\label{sec:std_targets}

Stellar flux standards for all DESI dark and bright programs are selected as point sources having blue colors consistent with F-type stars \cwr{around the metal-poor main-sequence turnoff}:
\begin{itemize}
    \item $16<G<18$
    \item $0<g-r<0.35$
    \item $r-z<0.2$ 
\end{itemize}
\cwr{Stars in the magnitude range $18<G<19$ that meet the same color criteria are also considered for use as standards, at lower priority}. \updatecite{\citet{myers22a}} describe \cwr{the} full set of criteria for selecting flux standards \cwr{for DESI}, including \cwr{additional} requirements on photometric quality in the LS. Since approximately 100 flux standards are assigned at higher priority than \cwr{all other targets} on every configuration of the DESI focal plane, this \cwr{subset of the \mainblue{} sample will be observed at a higher sampling rate and have a higher fraction of stars with repeat observations. These stars ($6000 \lesssim T_\mathrm{eff} \lesssim 7000$K dwarfs) will have distances up to $\lesssim 20$~kpc, with most in the range $5 \lesssim d \lesssim 10$~kpc (see Fig.~\ref{fig:distances_cmd_galaxia_halo})}. Flux standards can be identified in the DESI data products using the \texttt{DESI\_TARGET} bitmask  \updatecite{\citep[see][]{myers22a}}.

\section{Observing Strategy }
\label{sec:strategy}

\subsection{The DESI bright-time program}

\cwr{The DESI bright-time program comprises both \MWS{} and BGS. The observing strategy and fiber assignment strategy for the bright-time program are primarily designed to meet the goals of BGS. BGS galaxies are prioritized over all \MWS{} targets when assigning fibers, with the exception of \mwswd{} targets ($\sim1$ per square degree). The higher priority of BGS galaxies is therefore the main limitation on the fraction of \MWS{} targets that will be allocated a DESI fiber. We describe the relative prioritization of the different \MWS{} target categories in the next subsection.}

The bright-time program defines 5675 tiles to cover a footprint of \cwr{more than 14,000} $\mathrm{deg}^2$ \updatecite{\citep[a minimum coverage of $9000$ $\mathrm{deg}^2$ is a design requirement of BGS; see][]{hahn22a}}. Each tile corresponds to a single configuration of DESI fibers. The tiles are organized into four passes of $\sim1400$ tiles each. Tiles within a pass do not overlap. Tiles on different passes are offset such that all \cwr{points on the sky} will be covered by three tiles on average. A detailed description of the tiling strategy is given in \updatecite{\citet{raichoor22b}}. 

On successive passes covering a particular area of the sky, fibers are preferentially assigned to unobserved targets. In general, a fiber will only be assigned to reobserve an \MWS{} target if no unobserved target is available\footnote{\mwsbhb{} targets are an exception. They will be observed for a second time at a priority only slightly lower than that at their first observation, to obtain higher \cwr{\SNR{}}.}. All bright-time tiles will be observed to an effective exposure time of $T_{\mathrm{eff}}=180\,\mathrm{s}$, defined as the time required to achieve \cwr{the same \SNR{} that would be obtained for a typical BGS emission line galaxy in an exposure of $180$s} under nominal dark conditions \updatecite{\citep{hahn22a}}. Actual exposure times are scaled on the fly, based on real-time measurements of sky brightness, seeing, and transparency; the airmass; and the Milky Way dust attenuation of the tile \updatecite{\citep{kirkby22a,schlafly22a}}. The decision to observe dark, bright, or backup program tiles is also made according a real-time measure of survey speed, the rate at which \cwr{\SNR{}} increases for a fiducial target under the prevailing observing conditions \updatecite{\citep{schlafly22a}}. The total open-shutter time on each tile may be split over several exposures, possibly on different nights, until $T_{\mathrm{eff}}$ (computed from the spectra \cwr{themselves}) is reached \updatecite{\citep{guy22a}}. In section \ref{sec:sv} we show examples of co-added MWS spectra with $T_{\mathrm{eff}}\simeq180$s.

\subsection{Target density and fiber assignment}

\begin{figure}
    \centering
    \includegraphics[width=\linewidth]{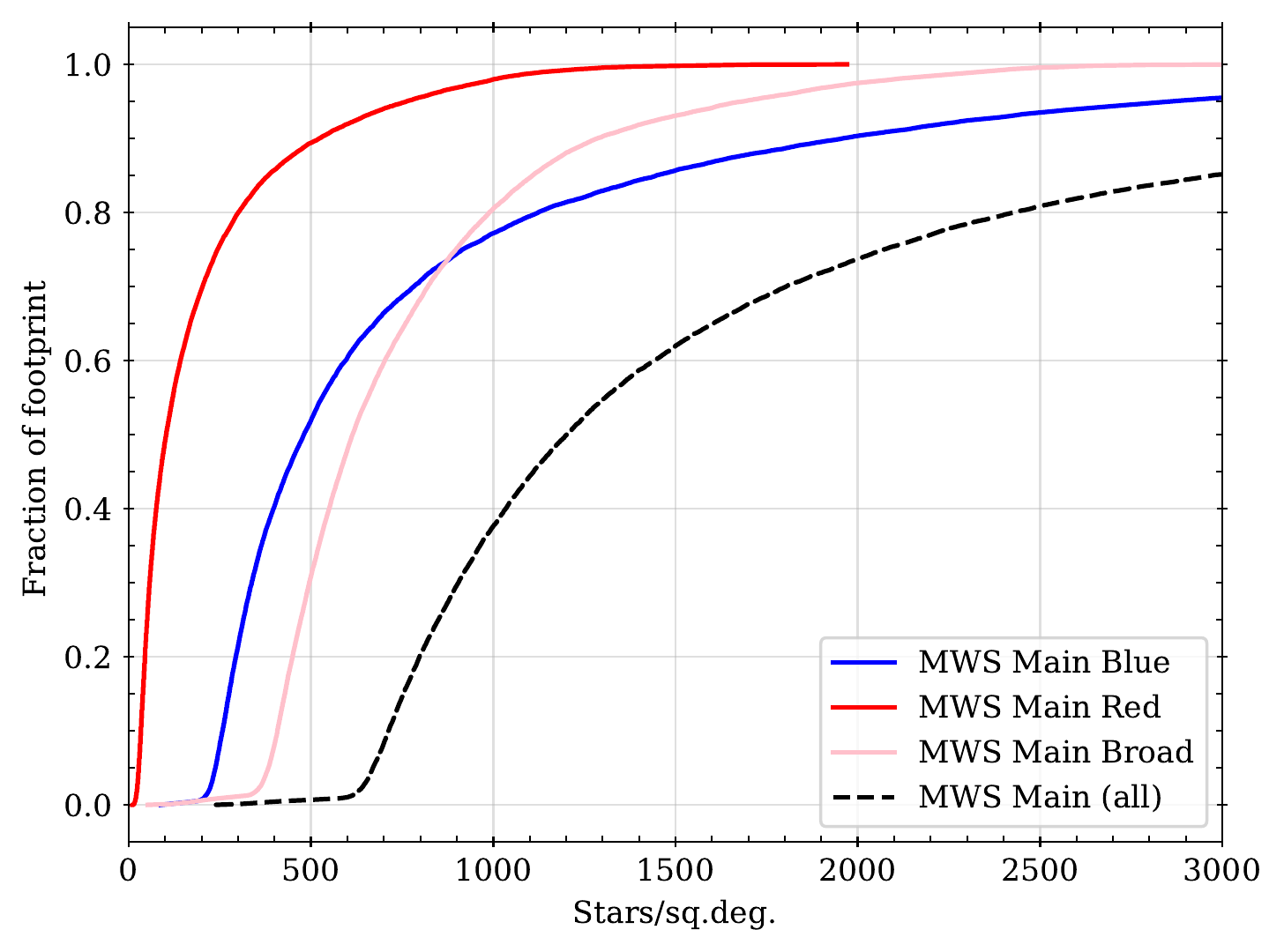}
    \caption{Fraction of survey footprint having less than a given surface density of targets in the three main MWS target categories individually (solid lines) and in total (dashed line). }
    \label{fig:cumulative-area-fraction-at-density}
\end{figure}

The density of \MWS{} targets varies significantly across the bright-time footprint, which ranges from the edge of the Galactic plane ($|b|\simeq20^{\circ}$) to the Galactic poles. This is illustrated in Fig.~\ref{fig:cumulative-area-fraction-at-density}, which shows the range of surface densities of \mainred{}, \mainblue{}, and \mainbroad{} targets across the survey footprint. The median density of \mainred{} (\mainblue{}, \mainbroad{}) targets is $\gtrsim 150$ (500, 600) per square degree. 

For comparison, the average density of DESI fibers on a single tile (i.e. one pass of the survey) is at most $\simeq667$ per square degree (if all positioners are operational). \cwr{However, not all fibers will be available for allocation to MWS targets. A fiber will be preferentially assigned to a BGS target in its patrol region. Moreover, on} each tile, DESI allocates 50 fiber positioners to flux standards and 40-65 fibers to blank sky locations, per spectrograph petal, which leaves at most 4500 fibers available for science targets. This will be reduced \cwr{further} by nonoperational fiber positioners, the number of which will vary over the course of the survey and is $\sim$15\% as of 2021 December \citep{Silber:2023aa}. 

Operational positioners are allocated to targets in a predetermined priority order, based on target category. Each category is assigned an integer base priority that establishes the ordering of classes in fiber assignment. Fiber collisions between targets of the same category (and between classes with the same base priority, such as \mainblue{} and \mainred{}) are resolved by a continuous subpriority value, randomly assigned to each target at the start of the survey \updatecite{\citep[see][]{myers22a}}. Further details of the DESI fiber assignment \cwr{algorithms} are given in \updatecite{\citet{raichoor22b}}.

The highest-priority target category in the bright-time program overall is \mwswd{}, followed by all BGS target categories, and then by all other \MWS{} categories. The order of all \MWS{} categories from highest to lowest priority is therefore \mwswd{}, \mwsrrlyr{}, \nearby{}, \mwsbhb{}, \mainblue{} and \mainred{} (at equal priority), \mainbroad{} and finally \faintblue{} and \faintred{} (at equal priority). This scheme prioritizes the most scientifically valuable but sparsely distributed targets above the bulk of the main \MWS{} sample.

Table~\ref{tab:dr9_target_summary} gives estimates of the number of targets in each \MWS{} category that will be allocated to a fiber over the course of the survey. \cwr{These estimates are made by running the DESI fiber assignment algorithm on all bright-time program tiles, assuming the state of the focal plane current in 2022 March. The estimates further assume that all tiles will be observed to their planned effective exposure time, $T_\mathrm{eff}$}. Over the entire footprint, we expect to assign a fiber to approximately 30\% of \mainblue{} and \mainred{} targets, falling to 19\% for the lower-priority \mainbroad{} sample. The higher priority \mwswd{}, \nearby{}, \mwsrrlyr{} and \mwsbhb{} samples have correspondingly higher completeness (99\%, 62\%, 52\% and 55\% respectively). Completeness increases as target density falls toward the Galactic poles, such that $\sim40\%$ of the footprint has $\sim50\%$ completeness for \mainblue{} and \mainred{}. 

BGS targets have a mean density of $\sim1,400$ per square degree with tile-to-tile fluctuations corresponding to the large-scale structure in the low-redshift galaxy distribution. Since DESI can deploy at most $\sim600$ fibers per square degree (after accounting for standards and sky fibers), more \MWS{} targets will be observed in the later passes of the survey, when the density of remaining unobserved BGS targets is lower. Approximately 35\% of the \MWS{} sample will be observed on the fourth pass, alongside the final 15\% of BGS targets. Approximately 11\% of the \MWS{} targets will be observed more than once; for the majority of these no other target is available to a fiber. \reply{The tile-to-tile fluctuations in BGS density give rise to corresponding fluctuations in the MWS completeness of $\approx \pm5\%$ at high latitude. The clustering pattern of the extragalactic large-scale structure will therefore be imprinted on the apparent clustering of observed MWS targets.}

\section{The DESI MWS Data Pipeline}
\label{sec:pipeline}

A massive software development effort has been carried out to ensure that the main data products from DESI, such as extracted flux-calibrated spectra and galaxy redshifts, are of the highest possible quality and processed on a daily basis from the beginning of the survey. The main DESI pipeline uses the \redrock{} spectral fitting code to determine redshifts  \updatecite{\citep{bailey22a}}. \redrock{} classifies spectra into broad categories (galaxy, star, white dwarf, quasar, etc.) but is not optimized for stellar radial velocity measurement at the accuracy required for \MWS{} science. \MWS{} has therefore developed additional software to extract information from stellar spectra, which we refer to collectively as the \MWS{} pipeline.

There are three main components to the \MWS{} pipeline, which we refer to as RVS, SP, and WD. The RVS component derives stellar radial velocities and atmospheric parameters. The SP component deals with the determination of stellar atmospheric parameters and complements the RVS branch by providing a way for inferring abundances for individual elements through the definition of spectral windows. The WD component is focused on determining parameters for white dwarf stars. Each of these components is based on improvements to existing tools for the analysis of stellar spectra.

\subsection{RVS}
\label{sec:rv}

The RVS pipeline determines the radial velocities of all stellar objects and finds the best stellar atmospheric parameters. The pipeline algorithms are based on ideas first presented in \citet{Koposov2011}, where stellar spectra, without flux calibration, are fit by stellar templates multiplied by a polynomial continuum. The code has been expanded to perform interpolation across templates for the Gaia-ESO project \citep{Gilmore2012}. A \texttt{python} implementation of these algorithms, \texttt{RVSpecfit}, is publicly available \citep{rvscpecfit2019}\footnote{\url{https://github.com/segasai/rvspecfit}}, including a version specific to DESI (which we describe here) and variants for other surveys, such as S$^{5}$ \citet[][]{Li2019} and the WHT Extended Aperture Velocity Explorer survey \citep[WEAVE;][]{Dalton2012}.

We briefly summarize the algorithm. A stellar spectrum ${\bf S}$ evaluated at pixels with wavelengths $\lambda_k$ is modeled as 
\begin{equation}
M(\lambda_k|{\bf \alpha},v,\phi)= \sum_i \alpha_i P_i(\lambda_k) T\left(\lambda_k\sqrt{\frac{1-v/c}{1+v/c}}|\phi \right) \label{eqn:rvs_model}
\end{equation}
where $T(\lambda|\phi)$ is the interpolated stellar template from the library convolved to the resolution of the observed spectrum. $P_i(\lambda)$ denotes basis functions, such as polynomials $\lambda^i$, Chebyshev polynomials, or radial basis functions (RBFs), which take care of modifying the continuum shape. The vector ${\bf \alpha}$ is a vector of linear coefficients, $v$ is the radial velocity, and $\phi$ is the vector of stellar atmospheric parameters, abundances, and potentially stellar rotations $V {\sin i}$.
As shown in \citet{Koposov2011} the vector of linear coefficients, $\alpha$, can be determined through simple linear algebra operations. This allows us to compute the log-likelihood for each spectrum given only the stellar atmospheric parameters and radial velocities, $\log P({\bf S}|\phi, v)$. In the case of DESI, the total log-likelihood function is the sum of the log-likelihoods for each of the three arms of the spectrograph.

The RVS pipeline optimizes the log-likelihood combined across the three arms simultaneously through Nelder--Mead optimization \citep{NeldMead65}. Since the likelihood surface is extremely complex in the radial velocity dimension, we need to provide a reasonable initial guess. \cwr{This is obtained from  cross-correlation of the continuum normalized spectrum with a 0.5\% subset of continuum-normalized templates from the PHOENIX grid. We continuum-normalize both the observed spectra and the model templates by a spline-based continuum with a regular grid of knots in wavelength separated by $1000\,\kms$. The cross-correlation uses the uncertainties and is computed as $ ({\bf S}/{\bf E}^2 \circledast {\bf T})^2 / (1/{\bf E}^2 \circledast {\bf T}^2) $, where {\bf T} is the template vector, {\bf S} is the spectrum, {\bf E} is the uncertainty vector, and $\circledast$ is a convolution operator that is computed using a fast Fourier transform. The cross-correlation functions from the three arms of the instrument are added together.} The best cross-correlation velocity and the stellar template parameters are then used as a starting point for the likelihood optimization.

Both the initial cross-correlation step and the nonlinear optimization steps are conducted in the radial velocity interval of $\pm 1500\,\kms$.

The interpolated stellar templates are constructed from the PHOENIX grid of models in a two-step process. First, global RBF interpolation is used to interpolate the original grid from \citet{Husser2013} onto a regular grid without holes, using steps of 0.2 in $\mathrm{[Fe/H]}$, 0.25 in $\mathrm{[\alpha/Fe]}$ and the original grid points in $\log g$ and $T_\mathrm{eff}$. This grid is then convolved to the average resolution of DESI in the $B$, $R$ and $Z$ arms. The resulting templates are then interpolated using multilinear interpolation. 

\begin{table*}
    \centering
    \begin{tabular}{llllr}
    Type of Star & Models & Bands Fit & Parameters & Reference \\
    \hline
    Cool and very cool ($2,300 < T_{\rm eff} < 5,100$ K)          & Phoenix & $R, Z$ & $T_{\rm eff}, \log g$, [Fe/H], [$\alpha$/Fe] & \citet{Husser2013} \\
    Cool and warm ($3,500 < T_{
    \rm eff} < 30,000 $K)  & Kurucz & $B, R, Z$ & $T_{\rm eff}, \log g$, [Fe/H], [$\alpha$/Fe], $\xi$ & \citet{allendeprietoetal18} \\
    White dwarf    & Koester & $B, R, Z$ & $T_{\rm eff}, \log g$ &  \citet{koester10-1}
    \end{tabular}
    \caption{Summary of the families  of grids of model spectra employed in the SP branch of the DESI-MWS pipeline. }
    \label{spmodels}
\end{table*}

We note that, because we are fitting the product of the template model and a polynomial or RBF continuum modifier, at this stage we are not using large-scale flux calibration information in the spectra. Currently we use 10 basis functions in each arm for the continuum modification. This approach may be refined in the future. 

The RVS pipeline quantifies the uncertainty in the parameters of each fit through an estimate of the Hessian matrix at the minimum. However, the uncertainty in radial velocity is estimated somewhat differently, by evaluating the log-likelihood from the best-fit template $P({\bf S}|v,\phi_0)$ on a finely spaced grid of radial velocities and computing the standard deviation of the radial velocity (which is equivalent to assuming a uniform prior on the radial velocity and computing the standard deviation over the posterior).

The RVS outputs are the best-fit stellar atmospheric parameters [Fe/H], $\log g$, \cwr{$\mathrm{[\alpha/Fe]}$}; the best-fit radial velocity; the stellar rotation $V \sin i$; and their uncertainties. We also return the best-fit chi-square values per arm $\chi^2_{min,B}$, $\chi^2_{min,R}$ and $\chi^2_{min,Z}$ and the combined $\chi^{2}$. \cwr{To facilitate the identification of sources that are not well fit by stellar templates, such as galaxies, quasars, or unknown types of objects, we also compute the $\chi^{2}$ values for a continuum-only model, i.e.\ a template $T(\lambda)=1$ in Eq.~\ref{eqn:rvs_model}, providing $\chi^2_{cont,B}$, $\chi^2_{cont,R}$ and $\chi^2_{cont,Z}$. The rationale for computing these chi-squares is that they give a reference value for the goodness of fit, with respect to which we can assess the stellar fits.}

The outputs of the pipeline are also used to populate the 64 bit warning bitmask {\tt RVS\_WARN}, as follows. If the difference in $\chi^2$ of a stellar model with respect to the continuum model is small, $(\chi^2_{min,B}+\chi^2_{min,R}+\chi^2_{min,Z})-(\chi^2_{cont,B}+\chi^2_{cont,R}+\chi^2_{cont,Z})<50$, the first bit of {\tt RVS\_WARN} is set to 1. If the radial velocity is within $5\,\kms$ of the edges of the interval of radial velocities considered, $[-1500,1500] \mathrm{\,\kms}$, the second bit of {\tt RVS\_WARN} is set to 1. If the radial velocity error is larger than $100\,\mathrm{\,\kms}$ the third bit of {\tt RVS\_WARN} is set to 1. A good spectrum without any of those issues has $\mathtt{RVS\_WARN}=0$.

The RVS pipeline, by default, does not require either targeting information or \redrock{} results. However when processing the main survey data, if \redrock{} outputs are available we use the classification from \redrock{} to avoid fitting spectra of quasars and galaxies. Thus we only process through RVSpecfit the spectra that either 
are classified as type \texttt{STAR} by \redrock{} or have been targeted as either a Milky Way target, a stellar secondary target, or a spectrophotometric standard.

\subsection{SP}
\label{sec:sp}

The SP pipeline is based on  FERRE\footnote{\url{https://github.com/callendeprieto/ferre}} \citep{2006ApJ...636..804A}, a FORTRAN code that fits numerical models to data by optimization and interpolation of a precomputed model grid. FERRE has been used in the analysis of SDSS optical spectra \citep{2009AJ....137.4377Y,rockosi22}, and SDSS/APOGEE near-IR high-resolution spectra \citep{APOGEE}, and in other surveys such as Gaia-ESO \citep{Gilmore2012}. 

A new \texttt{python} software package, \texttt{piferre}\footnote{\url{https://github.com/callendeprieto/piferre}}, has been written specifically to handle DESI MWS data. This package reads the reduced spectra, corrects for the radial velocity (measured by the MWS RVS pipeline or the primary DESI pipeline), resamples the spectra, and prepares submission scripts for batch processing, taking advantage of FERRE's OpenMP parallelism. The \texttt{piferre} package manages the independent processing of DESI tiles, exposures, and petals; collects the results. and creates the final data products.

Only the spectra that have been successfully fitted by the RVS pipeline are processed by the SP pipeline. The input data are the observed spectra and associated inverse covariance arrays.

We set up FERRE to work with grids of model spectra having between two and five parameters: mostly, effective temperature $T_{\rm eff}$, surface gravity $\log g$ and metallicity $\mathrm{[Fe/H]}$, and, in some instances, microturbulence $\xi$ and [$\alpha$/Fe] as well. We adopt several sources for the template models, including the same PHOENIX models \citet{Husser2013} of the RVS pipeline (Section \ref{sec:rv}) for very cool stars ($2300 < T_{\rm eff} < 5100$ K),  Kurucz ATLAS9 models \citep{kurucz79-1, kurucz93-1} for warmer stars ($ 3500 < T_{\rm eff} < 30,000$ K) \citep{meszarosetal12-1,allendeprietoetal18} and models for DA and DB white dwarfs \citep{koester10-1}. Table \ref{spmodels} describes the parameter range, steps, and origin of the models we use to analyze DESI commissioning and SV data. In some cases there are multiple grids for each family: one for PHOENIX models, five for the Kurucz models, and four for the Koester models.

FERRE derives all the atmospheric parameters simultaneously and provides error covariance matrices for every spectrum by inverting the Hessian matrix. As in the case of the RVS code, the optimization is based on the Nelder-Mead algorithm \citep{NeldMead65}. {\cwr As expected for a purely spectroscopic analysis,  there are significant correlations between the  uncertainties in $T_{\mathrm{eff}}$ and $\log g$, and especially between $T_{\mathrm{eff}}$ and [Fe/H].} 

We analyze the continuum-normalized spectra, after dividing the data by a running mean with a width of 500 spectral bins (equivalent to 250 \AA). This is similar to the normalization approach described by \citet{2017A&A...604A...9A}, 
which ignores the continuum information and focuses on the absorption lines. The entire spectral range is fitted for all models, except for the lowest-temperature stars in the Phoenix models, for which the blue ($B$) spectrograph arm is ignored because including it leads to significant systematic errors. All the spectra are 
evaluated against 
all the model libraries in Table \ref{spmodels}. The solution with the smallest value of the reduced chi-square is kept and the others are discarded. The results 
comprise the parameter estimates, the error covariance matrix, and the best-fitting model, which can be directly compared with the continuum-normalized observation. We also store a version of the best model without continuum normalization, which may be used for flux calibration.
A {\it success} flag is set for fits with $\chi^{2} < 1.5$ and median \cwr{\SNR{}}~$>5$.

The SP pipeline performs a number of iterations holding the atmospheric parameters fixed, except for [Fe/H] or [$\alpha$/Fe], to infer the abundances of several elements by evaluating the $\chi^2$ in regions of the spectrum dominated by transitions associated with those elements. This procedure resembles the methodology followed in APOGEE \citep{2016AJ....151..144G,2020AJ....160..120J}, and requires a careful definition of the windows to be used for each element. These are created by assigning weights  $W_{\lambda}$ ($X$) proportional to the derivative of the flux ($F$) with respect to the abundance of the element of interest ($X$), subtracting possible contributions from the rest of the elements,
\begin{equation}
W_{\lambda} (X) = \frac{\partial F_{\lambda}}{\partial X} - \sum_{Y \neq X} \frac{\partial F_{\lambda}}{\partial Y},
\label{weights}
\end{equation}
where the derivatives are computed using a solar model atmosphere. The weights are set to zero when Eq.~\ref{weights} gives a value smaller than a predetermined threshold.

\subsection{WD}
\label{sec:wd_pipeline}

In addition to the white dwarf fitting performed by the SP branch of the pipeline described above, we fit DA white dwarfs with a bespoke code using models provided by \cite{koester10-1}, which span $6\,000\,\textrm{K} \leq T_{\textrm{eff}} \leq 100\,000\,\textrm{K}$ and $4.00 \leq \log g \leq 9.75$. The fitting procedure employed is similar to that outlined in detail in \cite{liebertetal05-1}. We first continuum-normalize the flux of the observed and convolved model spectra around the Balmer lines, H$\beta$ to H8, which are all contained within the blue arm of the DESI spectrograph (see Table 5.5 of \citealt{desiInstrument}). This fitting procedure allows us to use white dwarfs as an independent test of the absolute flux calibration of DESI obtained from F- and A-type stars across the blue, red, and NIR spectrograph arms, because the best-fit white dwarf models are based solely on fits to the blue arm spectra.

After the continuum normalization, we determine the best-fitting models from our grid. As the equivalent width (EW) of the Balmer lines go through a maximum at approximately $T_{\textrm{eff}}\simeq 12\,000\,\textrm{K}$ (dependent on $\log g$, \citealt{daouetal90-1, bergeronetal92-1}), we determine two solutions, one hotter and one cooler than this $T_\textrm{eff}$. We use both initial best-fit grid solutions as our first estimates for a $\chi^2$ minimization between the observed and linearly interpolated model grid spectra. The degeneracy between the two solutions can be broken either with the continuum flux of the flux-calibrated DESI spectrum, or through use of Gaia photometry and parallaxes. The current WD pipeline uses the DESI continuum flux, but as this can be affected by flux calibration (e.g. \citealt{tremblayetal11-1}), we plan to incorporate the Gaia photometry and parallaxes in a future code development. We also plan to extend this code to DB and DBA white dwarf model fitting, using data from both the blue and red arms of the DESI spectrograph.

\begin{figure*}
    \centering
    \includegraphics[width=\textwidth]{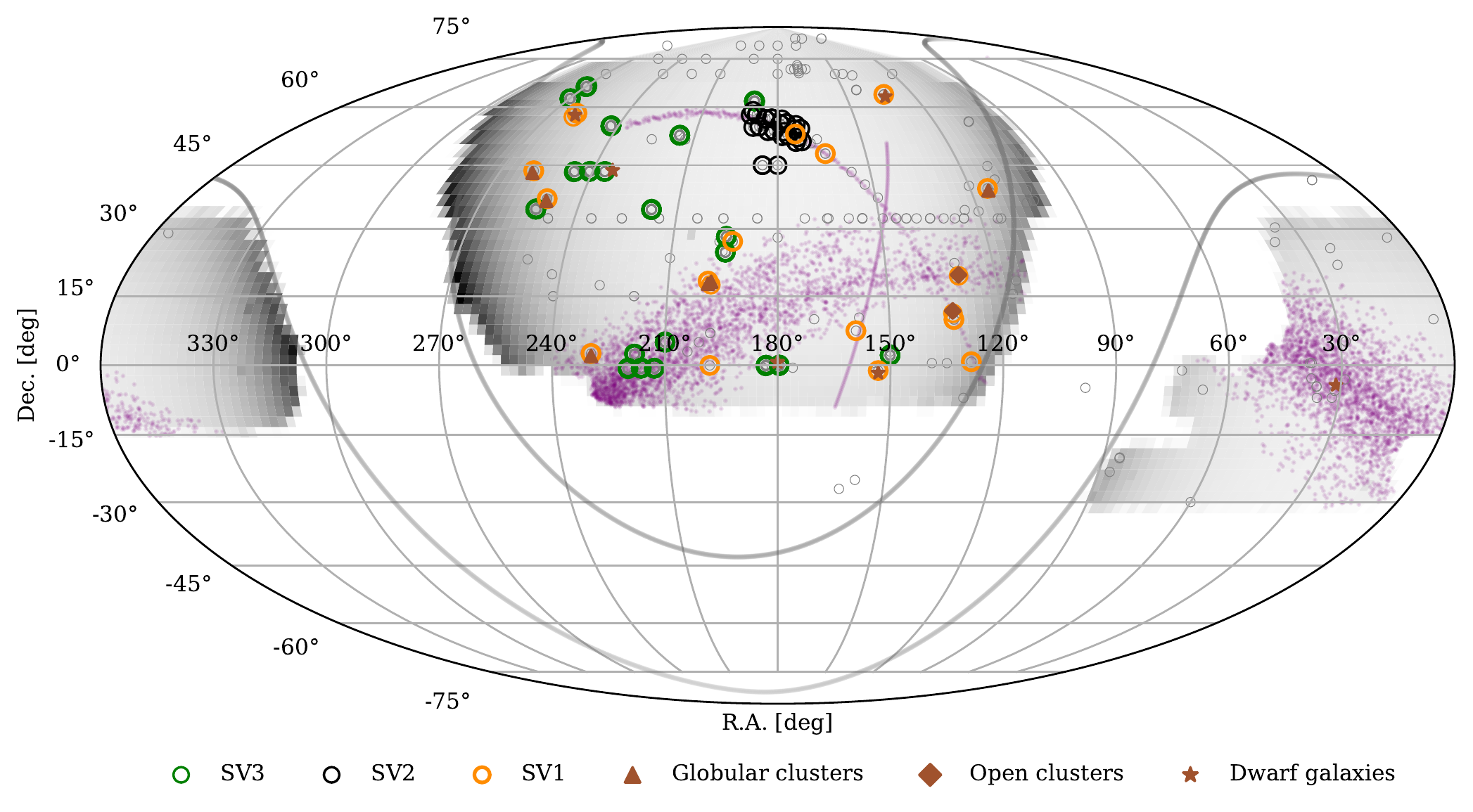}
    \caption{SV fields discussed in the text. Orange and green circles indicate fields with large numbers of stellar targets in the SV1 and SV3 programs, respectively, which were selected specifically for the validation of the MWS main survey and prioritized stellar targets. Black circles correspond to the SV2 program, which will be presented separately. Gray circles show SV fields not associated with the stellar SV program. \cwr{Purple points and tracks show the extent of the Sagittarius, GD 1 and Orphan streams as reported by the \texttt{galstreams} compilation; references are given in Fig.~\ref{fig:footprint}}. Gray lines indicate the approximate Galactic latitude limit of the survey, $|b|\pm20\degr$.}
    \label{fig:svfootprint}
\end{figure*}

\section{SV}
\label{sec:sv}

The DESI SV campaign \updatecite{\citep[SV;][]{sv22a}} included observations dedicated to MWS. These were split into two major subprograms, \svone{} and \svthree{}. Fig.~\ref{fig:svfootprint} shows all the DESI SV fields and highlights those associated with MWS. The MWS target selection for SV was previously described in \citet{mwsTSNote}; other DESI SV programs were introduced in \citet{raichoor2020_elg_rn}, \citet{zhou2020_lrg_rn}, \citet{yeche2020_qso_rn}, and \citet{ruizmacias2020_bgs_rn}. The results of these other programs and the final target selections for the DESI cosmological surveys are given in companion papers by \citet{sv22a}, \citet{raichoor22a}, \citet{zhou22a}, \citet{chaussidon22a} and \citet{hahn22a}.

The following \cwr{subsections summarize} the MWS SV observations. We then use these data to test our analysis pipelines (\cwr{introduced} in Section~\ref{sec:pipeline}) and to evaluate our survey selection function. \cwr{These tests reflect the current state of our pipelines, which we will continue to develop ahead of the first MWS data release.} The quantity and quality of the DESI SV spectra also allow us to address a number of interesting scientific questions, including the detection of extended features around star clusters and satellite galaxies, the structure of the GD 1 stream, the discovery of rare sources, and the characterization of radial velocity variability in our sample. We will present these scientific results from the MWS SV program in separate publications. 

\subsection{SV target selection}

In all the \svthree{} fields and all but a few of the \svone{} fields, targets were selected using a simpler version of the scheme described in Section~\ref{sec:targets} \citep[see also][]{mwsTSNote}. This selection was magnitude-limited between $r=16$ and $r=19$ and did not include the astrometric criteria we use to separate \mainred{} and \mainbroad{} targets in the main survey. The \mwswd{} and \nearby{} samples were selected as in Sections~\ref{sec:wd_targets} and \ref{sec:nearby_targets}, except that the astrometric criteria were applied to the Gaia DR2 catalog rather than to EDR3.

For the SV, stars in the range $19 < r < 20$ were also included at lower priority. A subset of MWS \svone{} tiles included additional sets of high-priority targets specific to the region observed (for example, likely members of satellite galaxies and star clusters selected using Gaia astrometry, or stars observed by previous  surveys).

\subsection{SV1 MWS Data Set}

In \svone{}, the 14 regions listed in Table~\ref{tab:mws_sv_areas} were observed to significantly greater depth than the main survey\footnote{These fields are different from those observed in the SV1 `BGS+MWS' program, which prioritized a larger set of BGS targets and filled spare fibers with MWS targets. For more details of the SV1 program see \updatecite{\citet{sv22a}}.}. In most cases, these observations consisted of a single fiber configuration (tile) for which multiple exposures were accumulated under different observing conditions. The completeness of MWS targets in these regions is therefore somewhat higher than that for the main survey (\cwr{because, in \svone{}, all fibers were allocated to stars}) but lower than that for the \svthree{} fields (because the same targets were observed on all exposures regardless of their accumulated \cwr{\SNR{}}). We use these high-\cwr{\SNR{}} and multi-epoch spectra to validate our analysis pipelines.

Most MWS \svone{} regions (orange circles in Fig.~\ref{fig:svfootprint}) correspond to known features in the Milky Way halo, including the GD 1 stream, several globular clusters, and the satellite dwarf galaxies Draco, Sextans, and UMa II. Other tiles were placed on blank regions. Since the angular sizes of most of the star clusters and dwarf galaxies are much smaller than the DESI field of view, much of the area on those tiles is also effectively blank. Additional tiles were placed on the UMa II satellite galaxy and the globular clusters M53 and NGC 5053 with customized target selections favoring likely members of those systems. \svone{} included many other observations dedicated to testing the DESI cosmological surveys (gray circles in Fig.~\ref{fig:svfootprint}), which yielded a small number of additional stars. 

\reply{The RVS and SP pipelines are run on the co-added spectra of 45,738 unique SV1 targets, of which 38,051 were targeted by the MWS SV1 program, mostly in the MWS SV1 fields but also with spare fibers in the BGS SV1 program. Many of these MWS targets are observed more than once. Among the MWS targets, 1209 ($\sim3\%$) cannot be fit reliably by RVS.}

\reply{The remaining 7687 SV1 spectra comprise brighter flux standards and serendipitous (potentially) stellar contaminants to the galaxy and QSO selections of other DESI SV programs\footnote{Particularly along the Sagittarius stream, which was targeted under dark conditions in SV1 to quantify contamination of the QSO target selection.}. We attempt to fit all targets classified as stars by the \redrock{} pipeline even if they would not be selected in the main MWS survey, although many of these are faint extragalactic sources and therefore have a higher failure rate in our fitting.}

\begin{table*}
    \centering
    \begin{tabular}{lrrllrl}
         Region Name & R.A. [$^\circ$] & Decl. [$^\circ$] & Objective &  $T_{\mathrm{eff}}$ (s) & $N_\mathrm{exp}$ & Target Selection\\
         \hline
         NGC 2419             & 114.54  & 39.38   & Globular cluster  & 1424 & 14 & Standard\\
         GD 1 Low Latitude A   & 128.50  & 0.80     & Tidal stream      & 1034 & 23 & Standard\\
         UMa II               & 132.87	& 63.73	  & Dwarf galaxy      & 1071 & 10 & Standard\\
         UMa II (extra)       & 132.87	& 63.73	  & Dwarf galaxy      & 85   & 2  & Customized\\
         Sextans              & 153.26  & -1.11   & Dwarf galaxy      & 250  & 3  & Standard\\
         Orphan               & 159.08	& 7.50    & Tidal stream      & 1118 & 5  & Standard\\
         GD 1 High Latitude A  & 163.74	& 47.86   & Tidal stream      & 1536 & 18 & Standard\\
         GD 1 High Latitude B  & 173.55  & 52.86   & Tidal stream      & 2038 & 14 & Standard\\
         NGP                  & 192.86	& 27.13   & Blank sky         & 914 & 21  & Standard\\
         BOSS 7456            & 198.04	& 0.00    & Blank sky/calibration & 1647  & 12 & Customized\\
         M53+N5053            & 199.10  & 18.30   & Globular cluster  & 1829 & 16 & Standard\\
         M53+N5053 (deep)     &	198.30  & 17.50   & Globular cluster  & 6178 & 8  & Customized\\
         M5                   & 229.64	& 2.58    & Globular cluster  & 2698 & 17 & Standard\\
         M13                  & 250.42  & 36.96	  & Globular cluster  & 2787 & 16 & Standard\\
         M92                  & 259.28  & 43.64   & Globular cluster  & 3446 & 14 & Standard\\
         Draco                &	260.07	& 58.42   & Dwarf galaxy      & 2258 & 16 & Standard\\
         \hline
    \end{tabular}
    \caption{Summary of dedicated MWS observations in SV1. From left to right, columns give the name and central coordinates of the field, the objective of the observation (in most cases the main feature of interest), the total effective exposure time calculated according to the bright-time metric (see text), the number of exposures taken in SV1, and the target selections used for those configurations (\textit{standard} refers to the MWS SV1 target selection criteria described in the text, \textit{customized} to cases in which one or more additional high-priority target categories were added to the standard selection\cwr{)}.}
    \label{tab:mws_sv_areas}
\end{table*}

\subsection{SV3 MWS Data Set}

In \svthree{} \cwr{(also referred to as the One-percent Survey in other DESI publications)}, 20 regions with the area of a single DESI field were observed using multiple tiles dithered in a rosette pattern around a common center, each with total coverage comparable to a single DESI tile. Details of these regions and the corresponding DESI observations are given in \updatecite{\citet{sv22a}}. The total area covered is $100\,\mathrm{deg}^{2}$, mostly distributed over Galactic latitudes $25 < b < 65$ degrees, with two tiles covering the north Galactic pole. 

Unlike the case of the \svone{} observations, multiple exposures of each \svthree{} tile were accumulated up to an effective exposure time only 20\% greater than that to be used for each main survey tile, with fibers on subsequent tiles in the dither pattern being reallocated to unobserved targets. Fibers were assigned to BGS and MWS targets on the same focal plane in the same priority order that will be used in the main survey, but with the broader target selection criteria used in \svone{} (see below). This strategy mimics the full DESI survey, but with $\sim6$ bright-time passes on each region rather than 4 (the other tiles in each region were observed with the dark-time and backup program target selections). The result is a highly complete sampling of each region, typically yielding spectra for $\gtrsim90\%$ of MWS targets (see Section~\ref{sec:sv3areas}). Fig.~\ref{fig:svfootprint} shows that, in contrast to the \svone{} fields, most of the \svthree{} fields do not overlap significant Milky Way substructures (most are cosmological deep fields). However, four are close to the midline of the Sagittarius stream and a further four are in its outskirts. Three more regions lie close to the GD 1 stream. 

\reply{The MWS SV3 dataset contains 213,414 spectra taken under bright conditions, almost all of which ($99.6\%$) are for MWS main survey targets. Of these, 982 ($\approx0.5\%$) are not fit successfully by the RVS pipeline. Visual inspection suggests the majority of these failures are for QSOs. Other classes of targets for which the current version of the RVS pipeline fails or produces unreliable results include white dwarfs (which we fit separately) and cool M-dwarfs, which we discuss further in Sec.~\ref{sec:fit_quality}.}


\subsection{Validating the pipelines}
\subsubsection{Radial velocities} 

\begin{figure}
    \centering
    \includegraphics{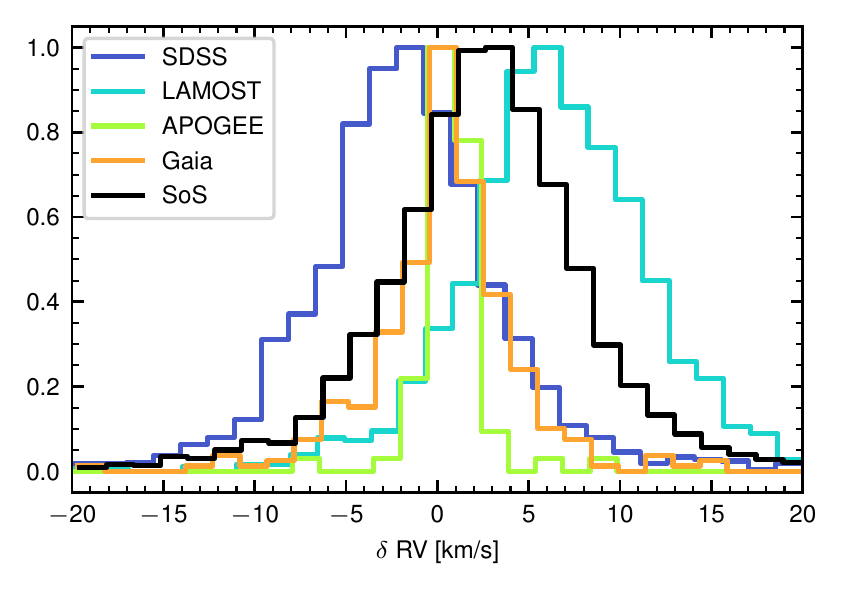}
    \caption{The comparison of radial velocities measured in \cwr{the DESI} SV program in bright conditions to those from other surveys. Colored curves show comparisons to SDSS DR14, APOGEE DR17, LAMOST DR7, and Gaia DR3, while the black curve shows the comparison to the homogenized SoS dataset \citep{Tsantaki2021}. }
    \label{fig:rv_offsets}
\end{figure}

We validate the radial velocities from the RVS pipeline by comparing them to radial velocities from other large surveys. Figure ~\ref{fig:rv_offsets} shows the distribution of radial velocity residuals for the stellar sources observed during the SV program in bright observing conditions. The different curves show comparisons to different individual surveys, while the black curve shows a comparison to the Survey of Surveys (SoS) dataset \citep{Tsantaki2021}, a homogenized set of radial velocities from various surveys. We remark that the comparison to high-resolution surveys such as APOGEE and Gaia shows offsets within $1-2$ km\,s$^{-1}$, \cwr{while} the SoS dataset (dominated by \cwr{low-resolution} LAMOST and SDSS spectra) shows a systematic offset of $\sim$ 2.5\,km\,s$^{-1}$. We are investigating the cause of this offset, which is likely associated with the \cwr{DESI} wavelength calibration and adjustments using skylines.

\subsubsection{Radial velocity accuracy}

\begin{figure}
    \centering
    \includegraphics{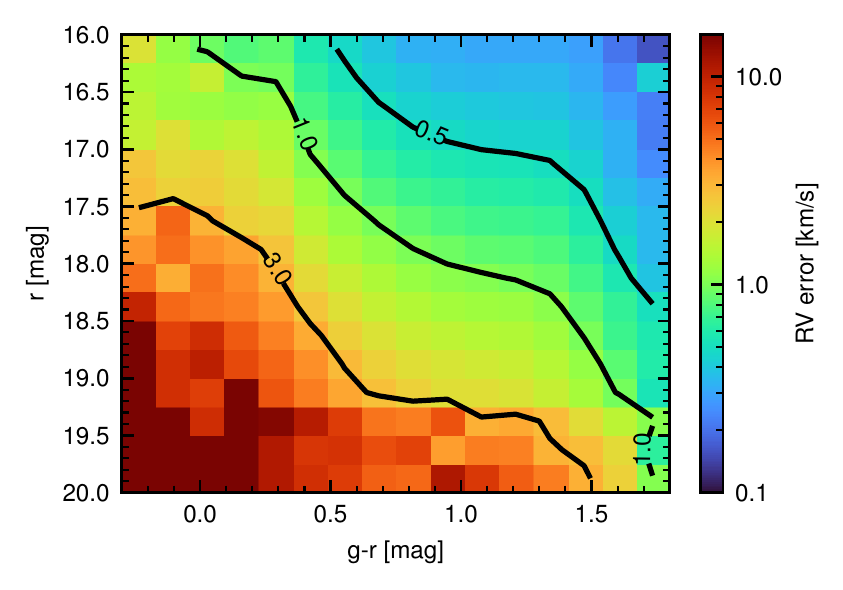}
    \caption{The typical radial velocity uncertainty in bright observing conditions for total exposure times below 1000s. Each bin of the color-magnitude diagram shows the radial velocity uncertainty reported by the RVS pipeline. The lines indicate the radial velocity errors of 0.5, 1, and 3 $\kms$.}
    \label{fig:rv_errors}
\end{figure}

Figure~\ref{fig:rv_errors} shows the median radial velocity error reported by the RVS pipeline for SV spectra, as a function of stellar color and magnitude. The SV data provide a representative sampling of the color and magnitude range, and hence velocity accuracy, expected for the main survey.

\begin{figure}
    \centering
    \includegraphics{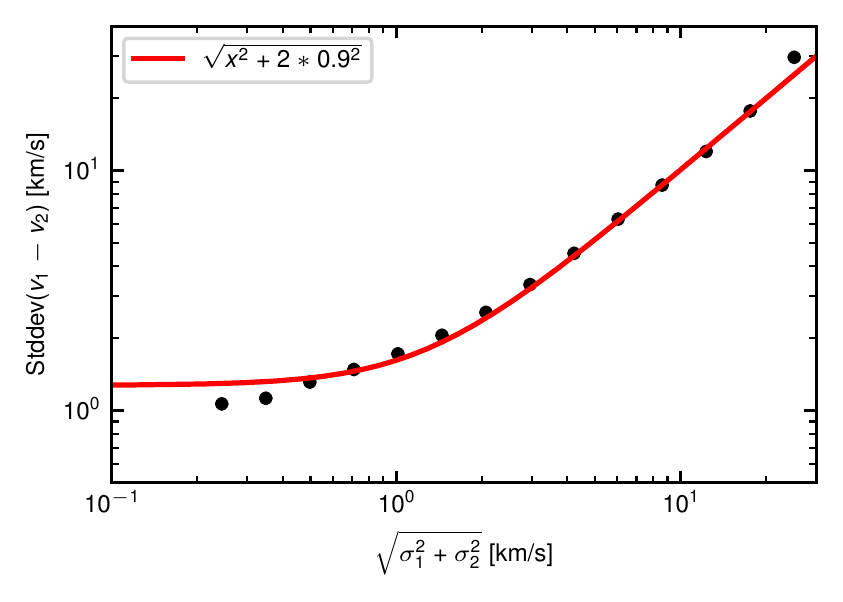}
    \caption{The accuracy of radial velocity errors from individual DESI exposures. Black points show the robust standard deviation estimate (from the 16\textsuperscript{th} and 84\textsuperscript{th} percentiles) for pairwise radial velocity measurements as a function of their combined formal uncertainties $\sqrt{\sigma_1^2+\sigma_2^2}$. If the radial velocity errors reported by the pipeline are accurate, the black points should lie on a one-to-one line. The red line shows the model curve after a $0.9\,\kms$ systematic error has been added in quadrature to the reported errors.}
    \label{fig:rv_prec_vs_error}
\end{figure}

\begin{figure}
    \centering
    \includegraphics{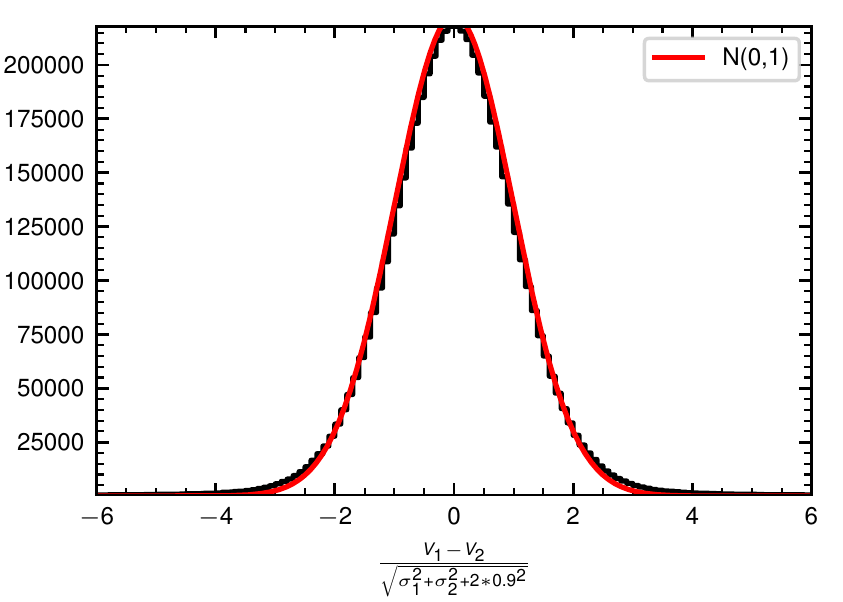}
    \caption{The distribution of pairwise radial velocity differences normalized by the radial velocity errors with a $0.9\mathrm{\,km\,s^{-1}}$ systematic error added. The red curve shows a Gaussian distribution with zero mean and unit variance.}
    \label{fig:rv_prec_gau}
\end{figure}

To validate the radial velocity errors, we analyze the pairwise velocity differences from several tiles that were observed multiple times in SV.
Figure~\ref{fig:rv_prec_vs_error} shows the distribution of pairwise velocity differences as a function of the combined velocity uncertainty. We also show the robust estimate of the standard deviation of pair-wise differences obtained using the 16th and 84th percentiles in bins of velocity uncertainty. The red curve is the expected behavior if the radial velocity errors have an additional systematic component of $\sim0.9\,
\kms$. This extra systematic error component is likely due to the wavelength calibration. Adding this systematic error \cwr{in quadrature} to the radial velocity errors reported by our pipeline, we see that the distribution of pairwise radial velocity errors normalized by the total uncertainty is very close to a normal distribution (see Figure~\ref{fig:rv_prec_gau}).

\subsubsection{[Fe/H] abundances}
\label{sec:pipeline_feh}

\begin{figure}
    \centering
    \includegraphics{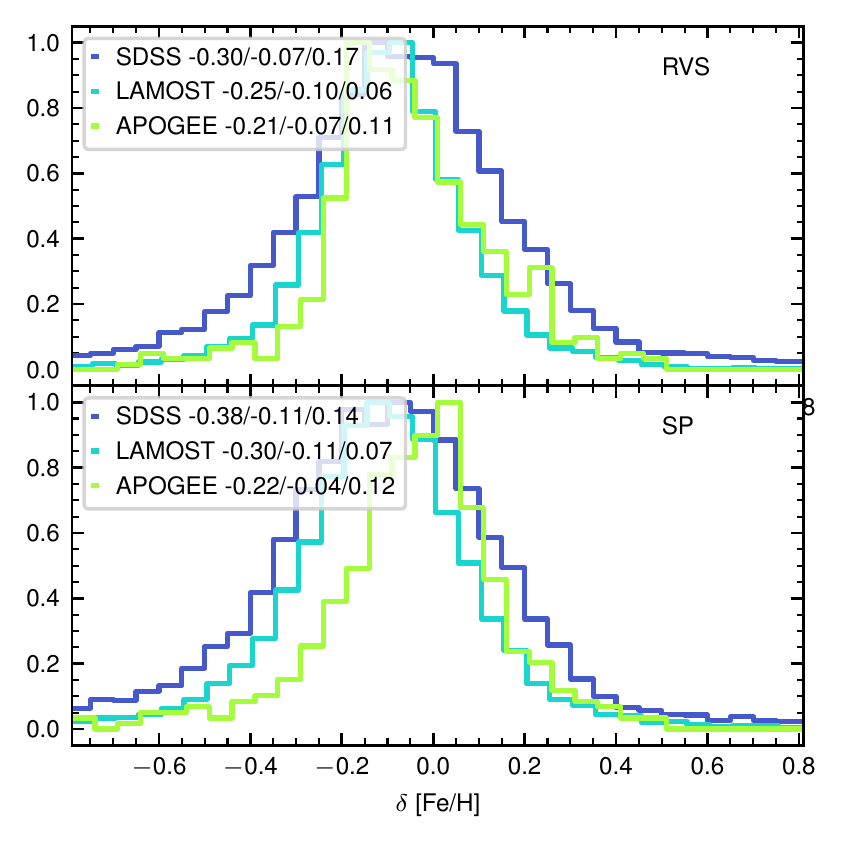}
    \caption{Comparison of abundances determined by the RVS (top panel) and \cwr{SP} (bottom panel) pipelines against data from other surveys. Dark blue curves show the residuals of DESI [Fe/H] measurements to the SDSS DR14 measurements, light blue curves show the comparison to LAMOST DR7 LRS measurements, and green curves show the comparison to  APOGEE DR17 [Fe/H] measurements. In the legend we provide the 16\textsuperscript{th}/50\textsuperscript{th}/84\textsuperscript{th} percentiles of the [Fe/H] residuals for each survey.}
    \label{fig:feh_prec}
\end{figure}

To assess the accuracy of iron abundances, we compare the measurements from the SV data with measurements of the same targets in APOGEE DR17, LAMOST DR7 LRS and SDSS DR14 SSPP. A histogram of the differences is shown in Figure~\ref{fig:feh_prec}. The overlap of DESI with APOGEE, LAMOST, and SDSS comprises $\sim$ 400, 16,000 and 10,000 stars, respectively. The top panel of the figure shows the comparison of $\mathrm{[Fe/H]}$ measurements from the RVS pipeline, while the bottom panel shows the comparison to the FERRE abundances. \cwr{This comparison shows that the [Fe/H] calibration of both the RVS and SP pipelines is within $0.1$---$0.2$~dex of that of other surveys. Some scatter in the metallicity residuals can be attributed to 
the SDSS, LAMOST, APOGEE, and DESI measurement errors.}

\begin{figure}
    \centering
    \includegraphics{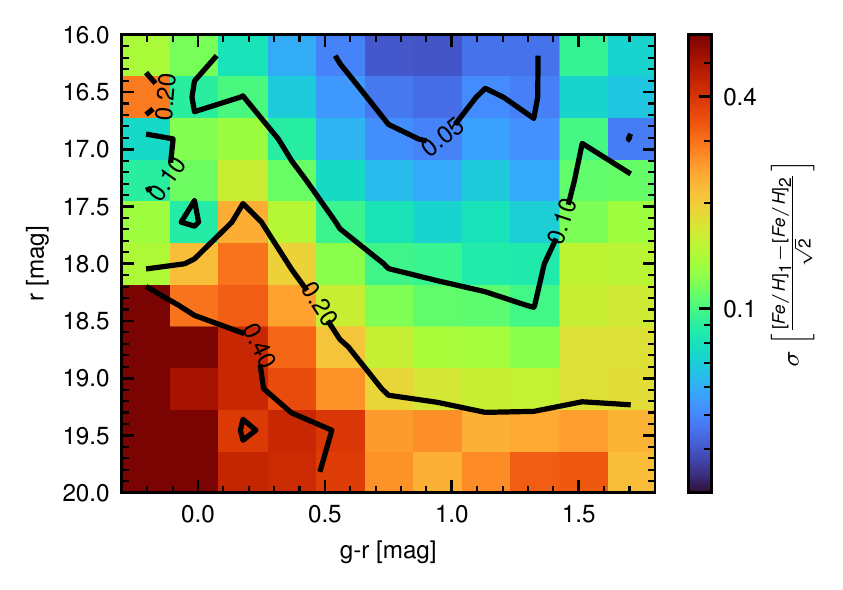}
    \caption{\cwr{The estimates of [Fe/H] random errors from the RVS pipeline as a function of color and magnitude. In each color--magnitude bin we show the typical random error estimated from the 16\textsuperscript{th}--84\textsuperscript{th} percentile range of the pairwise [Fe/H] differences between repeated measurements of the same stars.}}
    \label{fig:cmd_feh_accuracy}
\end{figure}

\cwr{To separate the effects of random errors in the DESI [Fe/H] measurements from systematic errors and errors in external catalogs, we also characterize [Fe/H] precision in bins of color and magnitude using repeated observations of the SV fields. 
Figure~\ref{fig:cmd_feh_accuracy} shows the distribution of pairwise [Fe/H] residuals for $\sim$ 3 million pairs of measurements, using observations with exposure times between 150 and 1000s. The [Fe/H] accuracy is estimated from the 16\textsuperscript{th} to 84\textsuperscript{th} percentile range of the distribution. The figure shows that, as expected, the [Fe/H] precision is a strong function of magnitude, with values around 0.05 at $r=16$, worsening to $0.2$--$0.4$ at $r=19$. For blue stars, $0.2<r<0.4$, the precision is noticeably worse than that for redder stars at the same magnitude. This is likely due to the impact of bright observing conditions, which degrade the signal most strongly in the blue arm of the instrument, and to the decreasing number of detectable absorption lines for hotter stars.}

\subsubsection{Stellar atmospheric parameters}
\reply{
We do not provide here a detailed comparison between our measurements of $\log g$ and $T_\mathrm{eff}$ and those from other surveys, but they show good agreement. The typical deviation (corresponding to the 16th/84th percentile range) of $T_\mathrm{eff}$ with respect to APOGEE and LAMOST is $-100 \lesssim \delta T_{\rm eff} \lesssim 250$~K, with a median offset of approximately 100~K. The RVS pipeline gives a broader range of $T_\mathrm{eff}$ offsets than the SP pipeline. The $\log g$ differences between DESI and other surveys are  $ -0.1 \lesssim \delta \log g \lesssim 0.5$  with a median offset of 0.2~dex and are similar between the RVS and SP pipelines.
}

\subsection{Abundances of star clusters and dwarf galaxies}

\begin{figure*}
    \centering
    \includegraphics{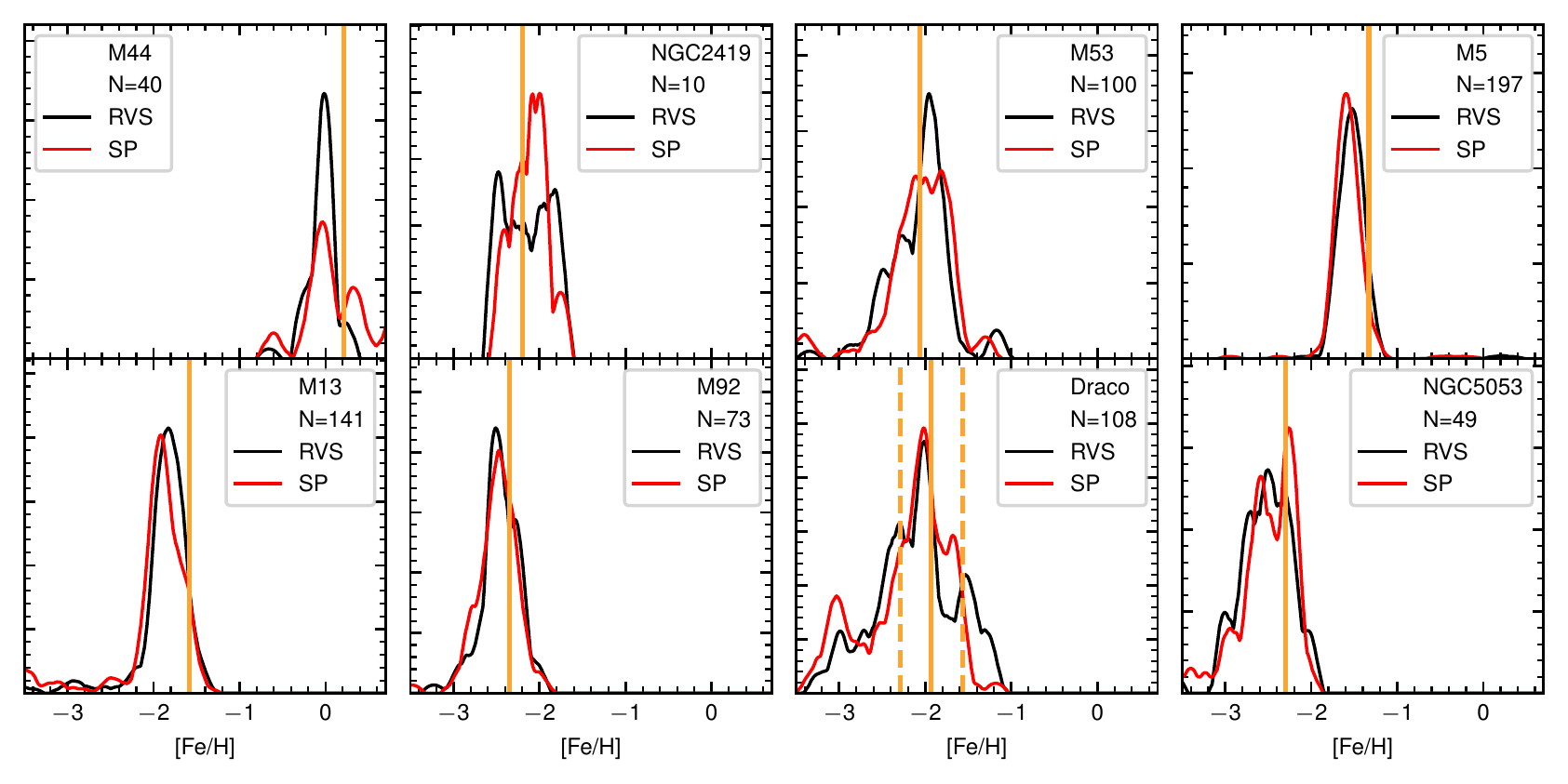}
    \caption{Comparison of DESI [Fe/H] measurements by the MWS RVS and SP pipelines of stars in star clusters and a dwarf galaxy to literature measurements from  \citet{Carretta2009}, \citet{kirby2011}, and \citet{dorazi2020}. Each panel shows a sample of stars observed as part of the DESI SV1 program (Table~\ref{tab:mws_sv_areas}). We only use stars with proper-motions and radial velocities matching the expected proper-motion (within $1\,\mathrm{mas\,yr^{-1}}$) and radial velocity  (within $10$-$15\mathrm{\,km\,s^{-1}}$ of the corresponding object). We only include measurements from spectra with \cwr{$S/N>6$} in at least one of the arms. The curves show \cwr{kernel density} estimates with \cwr{an} Epanechnikov kernel, \cwr{assuming a} bandwidth of 0.3~dex\cwr{, a conservative estimate of the precision of the} DESI metallicity measurements. Orange lines show literature values of the metallicities. For Draco we also show the 1$\sigma$ metallicity scatter in the system. The number of stars used for the analysis is provided in the legend of each panel.}
    \label{fig:clusters_feh_prec}
\end{figure*}

To validate the abundance measurements from the survey, 
\cwr{specifically their accuracy}, we also look at DESI observations of several globular clusters and OCs as well as dwarf galaxies for which there is a good external measurement of [Fe/H]. Fig.~\ref{fig:clusters_feh_prec} shows the distribution of measured abundances for several objects from Table~\ref{tab:mws_sv_areas} that were observed during SV. For this figure we do not attempt to identify carefully likely members of each object, only selecting stars that  have proper-motion within $1\,\mathrm{mas\,yr^{-1}}$ and radial velocity within 10-15$\mathrm{\,km\,s^{-1}}$ of the parent object. The plot provides a \cwr{broad} overview of the [Fe/H] accuracy of DESI. The expectation is that \cwr{cluster members should show} narrow distributions centered on their true [Fe/H] values, \cwr{while dwarf galaxy members are expected to show some [Fe/H] spread.  The observed [Fe/H] spreads on the figure are also expected to be affected by random [Fe/H] errors that are magnitude- and colour-dependent, as discussed in Section~\ref{sec:pipeline_feh}. We focus here on the [Fe/H] offsets. We clearly see that for several systems, like M13 and M5, DESI measurements (from both the RVS and SP pipelines) are systematically shifted from the literature values.  If we compare the median DESI metallicity for each system to other measurements from the literature, we see that the average offset $\mathrm{[Fe/H]_\mathrm{DESI}}- \mathrm{[Fe/H]_\mathrm{external}}$ is $-0.13$~dex for RVS and $-0.14$ dex for SP. The range spanned by the [Fe/H] offsets is $-0.27$ dex (M5) to 0.06 dex (M53) for the RVS pipeline, and  $-0.32$ (M13) to 0.1 dex (NGC 2419) for the SP pipeline.}


\subsubsection{Quality of the fits}
\label{sec:fit_quality}

\begin{figure*}
    \centering
    \includegraphics{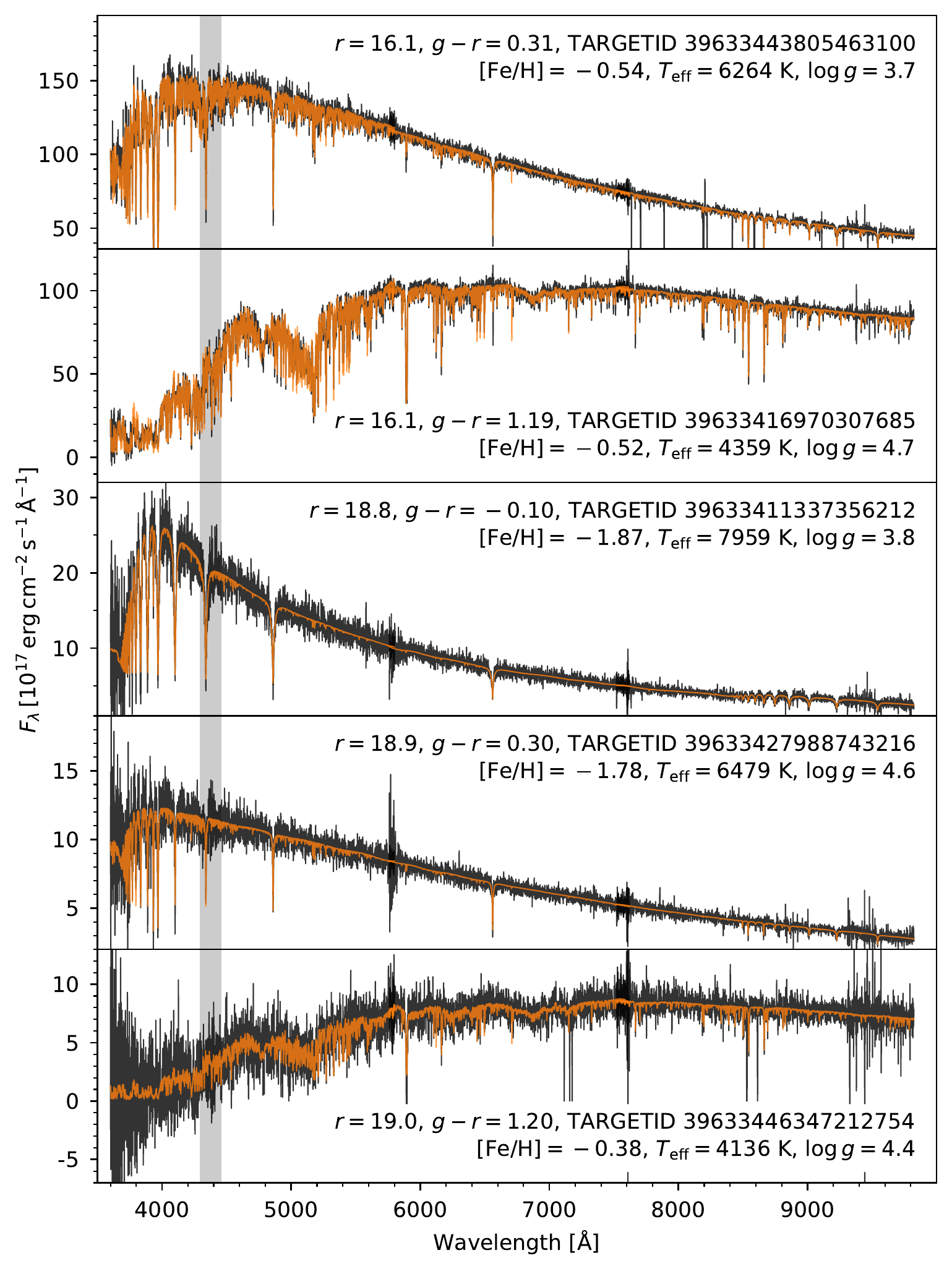}
    \caption{DESI spectra of several stars spanning the typical range of color and magnitude observed in the MWS main sample. The data are shown in black. The spikes in variance at $\sim5800$\AA{} and $\sim7500$\AA{} correspond to regions of overlap between spectrograph arms. In orange we overplot the best-fit models from the RVS pipeline. The gray band, $4300-4450$\AA, indicates wavelengths affected by a dichroic defect, which we mask in our fits \updatecite{\citep[see][]{guy22a}}.}
    \label{fig:spec_plot}
\end{figure*}

\begin{figure}
    \centering
    \includegraphics{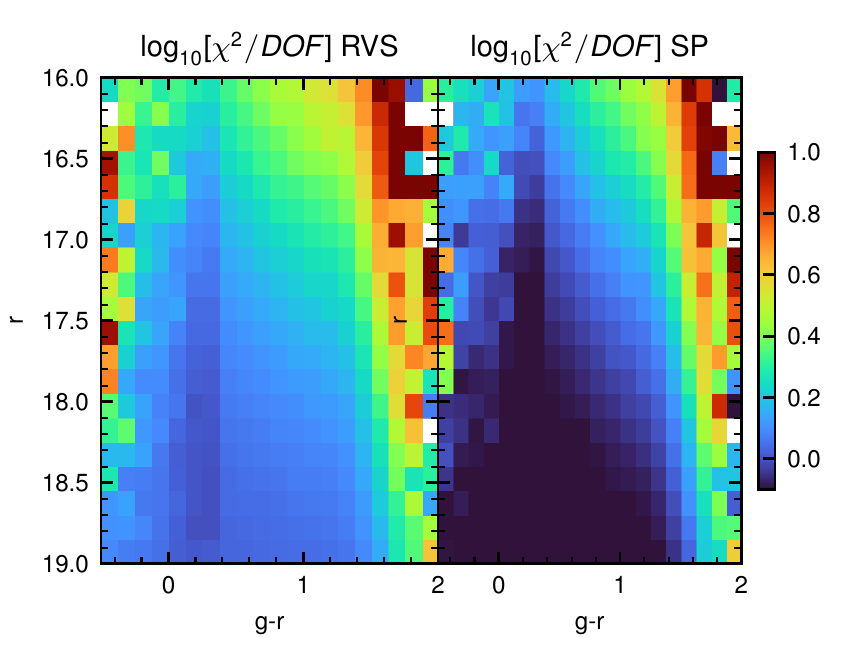}
    \caption{The median logarithm of $\chi^2$ per DOF for spectral fits with the RVS and SP pipelines, as a function of target color and magnitude.}
    \label{fig:chisq}
\end{figure}

\cwr{Fig.~\ref{fig:spec_plot} shows five typical spectra for MWS main sample targets and the corresponding fits from the RVS pipeline. The S/N is representative of what is expected for the survey at the corresponding magnitudes. The agreement between the data and the model is very good even at high S/N. We also note that, since the RVS model includes a polynomial multiplicative correction to the spectrum (see Eq.~\ref{eqn:rvs_model}), the agreement between the model and the data directly probes the accuracy of the flux calibration.} Fig.~\ref{fig:chisq} summarizes the quality of the spectral fits by showing the typical $\chi^{2}$ per degree of freedom (DOF) for the RVS pipeline (left) and the SP pipeline (right), as a function of the color and magnitude of the star. In both cases we see that $\chi^{2}$ has a strong dependence on magnitude and color. For the RVS fits, the chi-squares are close to unity for faint stars, as expected if the errors are treated correctly. For the SP fits, the chi-squares seem to be slightly below unity for faint stars. However, both pipelines show strong increases in $\chi^2/\mathrm{DOF}$ to $\sim 3$--$10$ at bright magnitudes and for stars with $g-r\sim 2$. This is primarily due to the limitations of the stellar atmospheric grids used, leading to template mismatches that are apparent at high S/N and for cool stars with many absorption lines and molecular bands.

\begin{figure}
    \centering
    \includegraphics[width=0.5\textwidth]{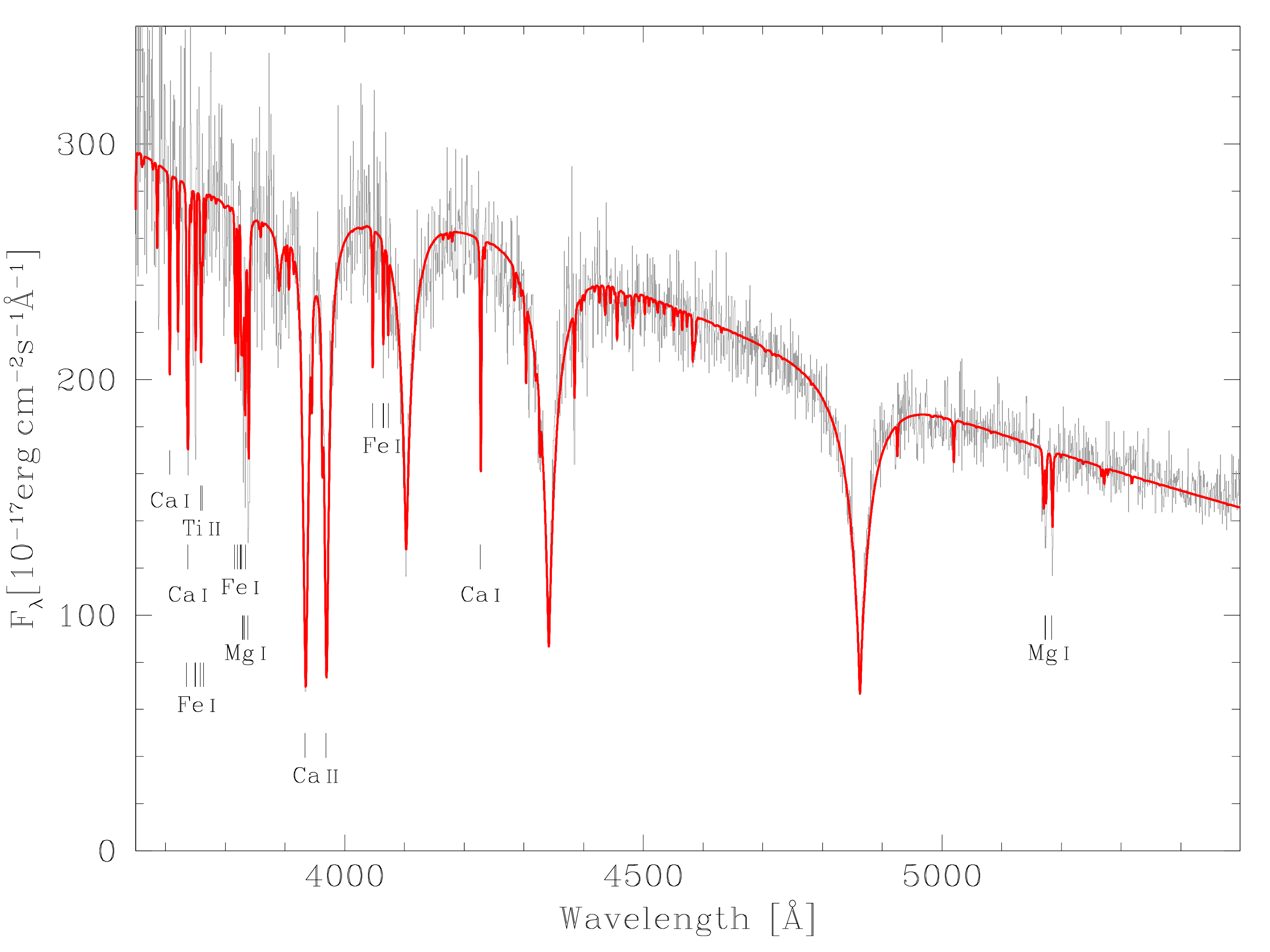}
    \caption{DESI spectrum (gray) of the white dwarf GD 362, revealing the metal pollution of the otherwise pristine helium and hydrogen atmosphere. A model fit is shown in red with prominent metal features labeled.}
    \label{fig:GD362}
\end{figure}

\subsection{White dwarf SV results: remnant planetary systems}

DESI has already identified several hundred planetary systems around white dwarfs via the metal pollution of their atmospheres. A dedicated study of these planetary systems is beyond the scope of this paper. Here we showcase the potential of this statistical sample using one of the most polluted white dwarfs so far observed by DESI, GD 362 \citep{gianninasetal04-1, zuckermanetal07-1, xuetal13-1}. GD 362 is extreme in both its pollution from planetary material (a total of 17 elements have been previously identified in the white dwarf's atmosphere; \citealt{zuckermanetal07-1,xuetal13-1}), as well as its fairly sizable hydrogen content in an otherwise helium-dominated atmosphere \citep[which can be indicative of water accretion][]{jura+xu10-1, kleinetal10-1, gentilefusilloetal17-1}.

The DESI spectrum of GD 362 obtained on 2021 June 21 is shown in Fig.\,\ref{fig:GD362}, revealing strong absorption lines from many metals, including Ca, Mg, and Fe. To analyze the DESI spectrum of GD 362, we compute a grid of white dwarf model spectra using the code of \cite{koester10-1} and by sampling the parameter space in \Teff, $\log g$ and number abundances $\log[\mathrm{H/He}]$ and $\log[\mathrm{Z/He}]$ around the values determined by \citet{xuetal13-1}. We fix the metal-to-metal ratios $\log\mathrm{[Z/Ca]}$ to those of \citet{xuetal13-1}. Using that grid, we find $\Teff=10,500$\,K, $\log g = 8.00$, $\log$[H/He] = $-1.75$, and $\log\mathrm{[Ca/He]}=-6.84$ as the best-fit parameters. The parameters differ somewhat from those of \citet{zuckermanetal07-1} and \citet{xuetal13-1} \cwr{($\Teff=10,540$\,K, $\log g = 8.24$, $\log$[H/He] = $-1.14$, and $\log\mathrm{[Ca/He]}=-6.24$)}, which is unsurprising as these authors based their analysis on deep Keck High Resolution Echelle Spectrometer data of this star. However, incorporating the Gaia photometry and astrometry in our analysis (Section~\ref{sec:wd_pipeline}) will offset the lower S/N of the DESI spectrum. DESI will identify $\simeq1000$ debris-polluted white dwarfs, allowing the first large-scale statistical analysis of the characteristics of these planetary systems as a function of host star mass and age. The \mwswd{} sample will also include many heavily polluted white dwarfs and probe the extreme parameter spaces of remnant planetary systems.

\subsection{MWS Targets in SV3 Fields}
\label{sec:sv3areas}

\begin{figure}
    \centering
    \includegraphics[width=\linewidth]{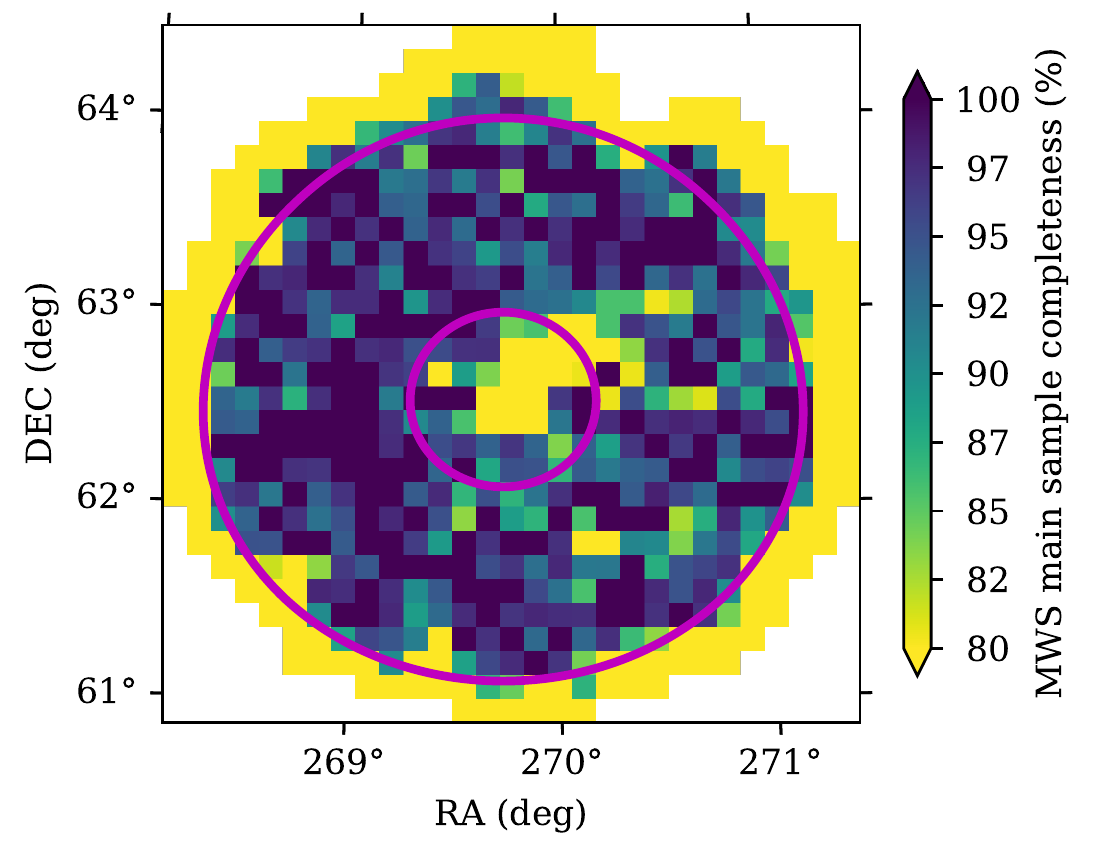}
    \caption{Fraction of all MWS main sample targets (\mainblue{}, \mainred{} and \mainbroad{}) observed in the SV3 region close to the north ecliptic pole, $(l,b)\approx (92\,\deg,30\,\deg)$, in $15' \times 15'$ pixels. Circles at $24'$ and $87'$ indicate the region of $\gtrsim 90\%$ completeness discussed in the text (the outer circle is slightly smaller than the nominal size of a DESI tile).}
    \label{fig:sv3_rose_example}
\end{figure}

The 20 SV3 regions (green circles in Fig.~\ref{fig:svfootprint}) each comprise 26-28 slightly dithered DESI pointings with an area of overlap equivalent to a single DESI tile. Compared to the four-pass MWS main survey, this dense coverage pattern provides much higher completeness on all target classes. Several of these regions overlap the Sagittarius stream, but the majority do not fall on known Milky Way halo structures.

For a typical SV3 region, nine observations were taken under standard bright-sky conditions comparable to those expected for MWS and BGS, 13 under dark conditions (in which the DESI cosmological surveys will operate), and five under extremely bright conditions (in which the main DESI survey will switch to the backup stellar program). Fibers on the bright SV3 tiles were allocated to targets from the BGS and MWS SV selections, with BGS having higher priority (this is roughly the same way in which fibers will be allocated in the main survey, although the BGS target density was higher in SV3). Fibers were also assigned to flux standard stars and secondary program stellar targets on the dark SV3 tiles.

Fig.~\ref{fig:sv3_rose_example} shows the completeness of MWS main survey targets (\mainblue{}, \mainred{} and \mainbroad{}) from bright-time SV3 observations in a typical SV3 region (this example is near the north ecliptic pole, $b\sim30\,\deg$). The completeness is $\gtrsim 90\%$ in the annulus $24' < r < 87'$. The dither pattern results in slightly lower ($\lesssim 80\%$) completeness near the center (slightly extended to the northeast) and at the edge of each region. 

This high completeness allows us to characterize the parent sample from which the MWS targets are drawn in these regions, using the measurements \cwr{made by the RVS and SP pipelines (Section~\ref{sec:pipeline})}. In the following comparisons, we use all stars from SV3 that would be selected in one of the three main survey categories (\mainblue{}, \mainred{}, \mainbroad{}). We consider only stars with a successful RVS pipeline velocity measurement ({\tt RVS\_WARN} = 0) and $S/N \gtrsim 2$ in all three spectrograph arms. These data quality cuts remove $\sim1\%$ of all MWS main survey targets in the SV3 dataset. We refer to this subset of SV3 targets as the MWS SV3 sample.

\begin{figure}
    \centering
    \includegraphics[width=8.5cm]{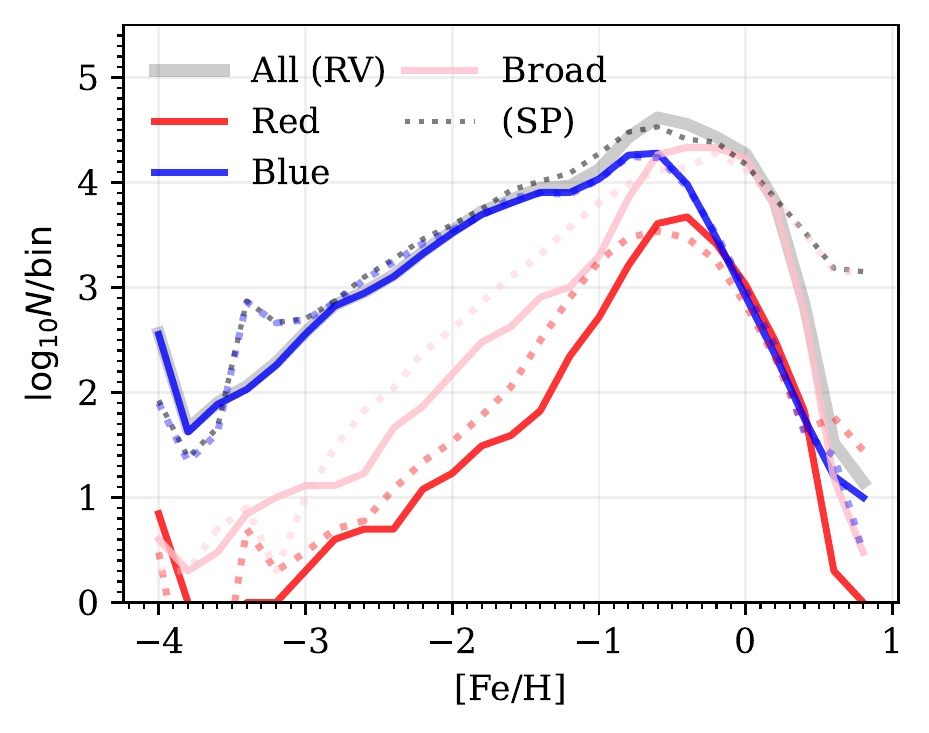}
    \caption{$\mathrm{[Fe/H]}$ distributions for all the stars in the SV3 dataset that belong to any of the three MWS main survey target categories. The gray lines show the sum of the three distributions. The bin width is 0.2 dex. Solid and dotted lines show results from our RVS and SP pipelines, respectively.}
    \label{fig:sv3_feh_split}
\end{figure}

Fig.~\ref{fig:sv3_feh_split} shows the MDF of the MWS SV3 sample in each of the three main survey target categories. As expected, the distribution peaks at $\feh \approx -0.5$, with the \mainblue{} selection (which samples metal-poor turnoff stars in the thick disk and halo) dominating at lower metallicity and the \mainbroad{} selection (mostly nearby metal-rich disk stars) dominating at higher metallicity. The \mainred{} selection has a similar distribution to \mainbroad{} below $\feh \approx -0.5$ and notably fewer stars around solar metallicity. An offset of $\lesssim 0.2$~dex is apparent between the RVS and SP pipelines for \mainred{} and \mainbroad{} targets, but not for \mainblue{} targets (see Sec.~\ref{sec:pipeline_feh}).

\begin{figure}
    \centering
    \includegraphics[width=8.5cm]{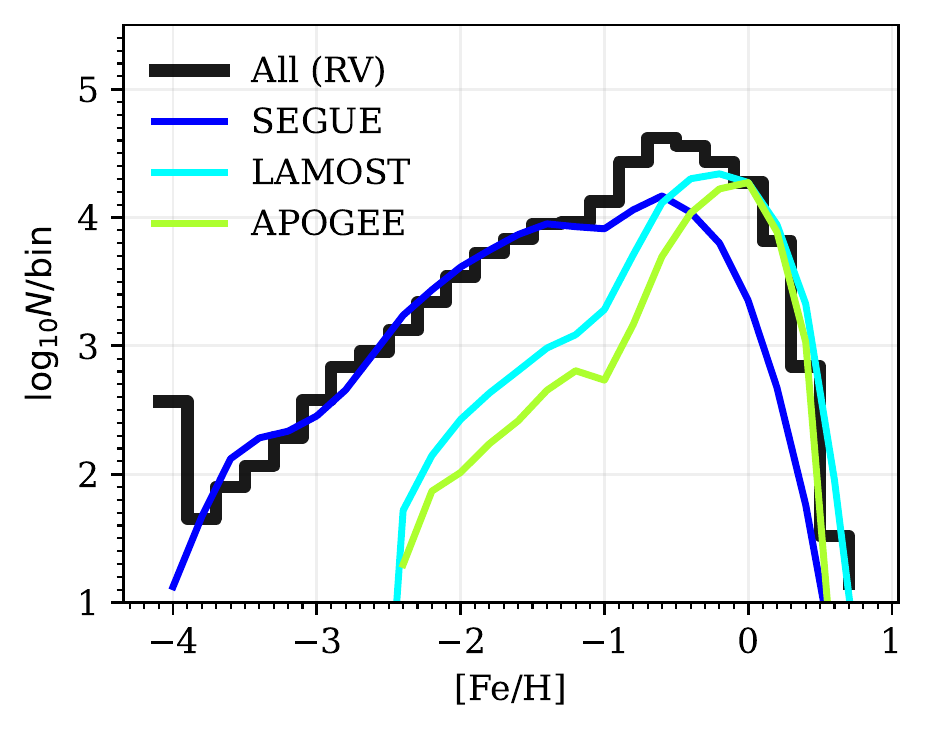}
    \caption{Comparison of the metallicity distribution of MWS main survey targets in the SV3 dataset, \cwr{measured by the RV pipeline} (black; as in Fig.~\ref{fig:sv3_feh_split}), to data from APOGEE, SEGUE, and LAMOST.}
    \label{fig:sv3_feh_others}
\end{figure}

Fig.~\ref{fig:sv3_feh_others} compares the combined MDF of all MWS main survey targets in the \svthree{} sample (as shown in Fig.~\ref{fig:sv3_feh_split}) to data from APOGEE \citep{APOGEE}, SEGUE \citep{yannyetal09-1,rockosi22} and LAMOST \citep{cuietal12-1}, restricted to the same range of high Galactic latitudes and imposing basic quality cuts for each survey\footnote{APOGEE: \cwr{SDSS DR17}, plotting $\mathrm{FE\_H}$ with $\mathtt{STARFLAG}=0$ and $0<\mathtt{M\_H\_ERR}<0.2$. SEGUE: \cwr{SDSS DR17}, plotting $\mathtt{FEH\_ADOP}$ with $\mathtt{ZWARNING} = 0 | 16$, $\mathtt{SNR}>10$, $\mathtt{ELODIERVFINALERR} > 0$, and $\mathtt{FEH\_ADOP}>-5$. LAMOST: DR7 v2, plotting $\mathtt{fe\_h}$ with $\mathtt{snrr}>10$, $0<\mathtt{feh\_err}<0.5$, and $\mathtt{rv\_err} > 0$.}. This comparison illustrates the much broader scope of the MWS target selection compared to these previous surveys \citep[similar considerations motivated the design of the medium-resolution H3 survey;][]{H3}. The \mainblue{} targets in our \svthree{} data have a similar distribution to the sample from SEGUE, which selected for the metal-poor halo. The \mainred{} and \mainbroad{} samples have distributions similar to the data from LAMOST and APOGEE, both of which comprise intrinsically brighter stars and include subsets with selection functions tuned to recovering halo giants. 

\begin{figure}
    \centering
    \includegraphics[width=8.5cm]{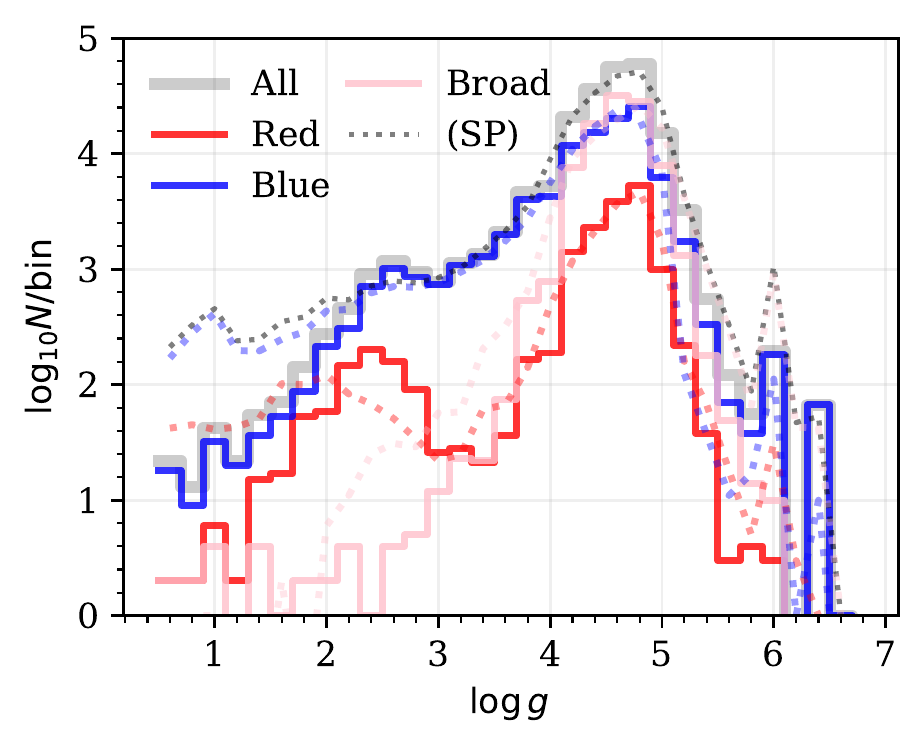}
    \caption{Surface gravity distributions for all stars in the SV3 dataset that would be selected in any of the three MWS main survey target categories. Solid lines show results from the RVS pipeline and dotted lines results from the SP pipeline.}
    \label{fig:sv3_logg}
\end{figure}

\begin{figure}
    \centering
    \includegraphics[width=\linewidth,trim=0 0 0 0, clip=True]{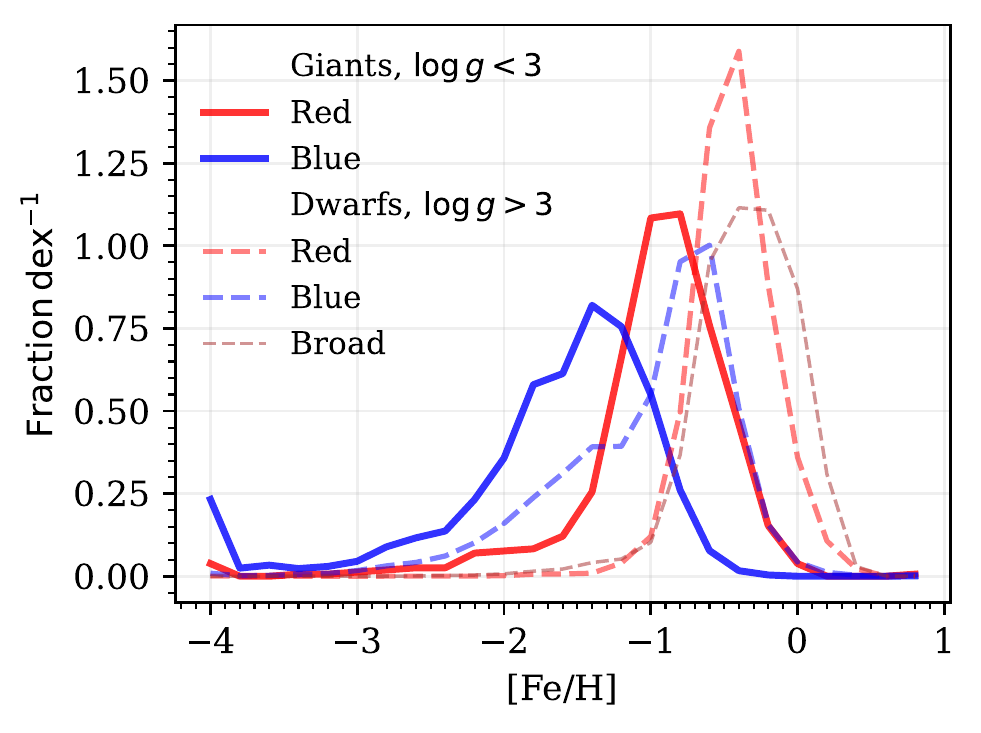}
    \caption{$\mathrm{[Fe/H]}$ distributions for \mainblue{}, \mainred{} and \mainbroad{} targets in the SV3 dataset, divided into "giant" ($\log g \le 3$; solid) and "dwarf" ($\log g > 3$; dashed) subsets (as shown in Fig.~\ref{fig:sv3_logg}, there are very few \mainbroad{} giants in the dataset).}
    \label{fig:sv3_feh_logg_split}
\end{figure}

Fig.~\ref{fig:sv3_logg} shows the $\log g$ distributions of main survey targets in the \svthree{} data. \cwr{As expected, the \mainbroad{} sample is dominated by dwarf stars. The \mainred{} sample was selected to reduce the contamination of foreground dwarfs and recover a higher proportion of giants. Fig.~\ref{fig:sv3_logg} shows that, although \mainred{} contains a large number of dwarf contaminants (a necessary consequence of the mild astrometric selection required to avoid strong kinematic bias) there is also a clear second peak at low $\log g$ values, which is absent in \mainbroad{}}. The \mainblue{} sample contains a large number of giants and subgiants; as shown in Fig.~\ref{fig:distances_cmd_galaxia_halo}, in the MWS magnitude range \mainblue{} probes the metal-poor main-sequence turnoff, subgiant branch, and horizontal branch at distances $20<r<60$~kpc. The $\log g$ distributions from the RVS and SP pipelines broadly agree, the most notable exception being a slightly higher number of \mainblue{} and \mainbroad{} stars with $\log g < 4$ in the SP results. 

\begin{figure*}
    \centering
    \includegraphics[width=\linewidth]{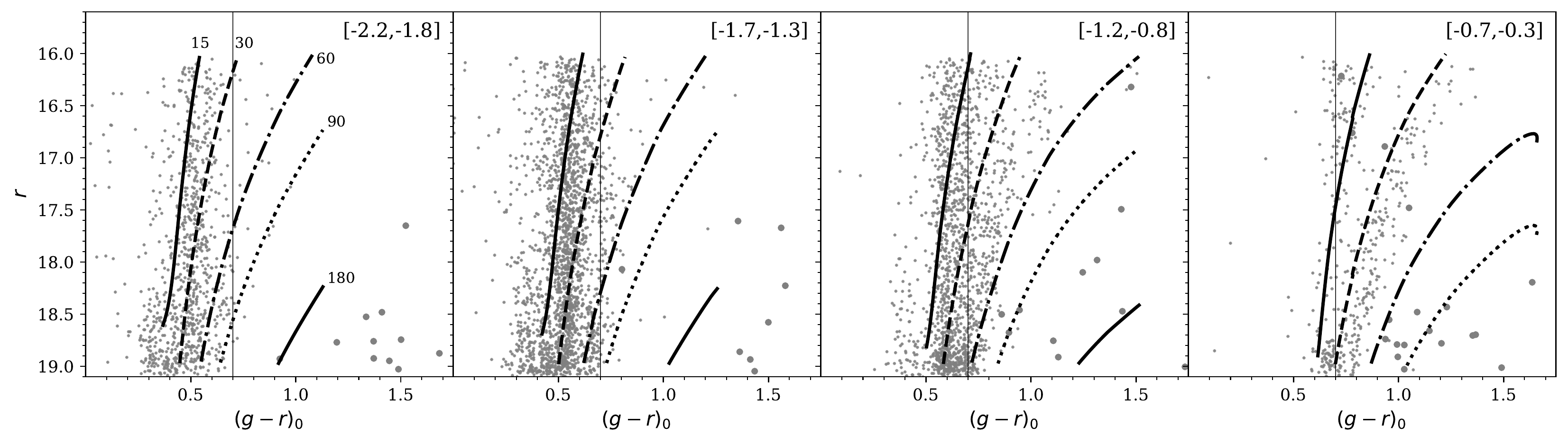}
    \caption{Color--magnitude diagrams for giants ($\log g < 3.5$) in the \svthree{} fields, separated into bins of $\feh$ (panels; range shown in each panel). We overplot MIST isochrones (corresponding to the central $\feh$ of each bin) at 15, 30, 60, 90, and 180~kpc (line styles, as shown in the leftmost panel). Larger points indicate the small number of giants in \mainbroad{}. The vertical line shows the separation in color between the \mainblue{} and \mainred{} selections.}
    \label{fig:cmd_giants_split_feh}
\end{figure*}

\begin{figure*}
    \centering
    \includegraphics[width=\linewidth]{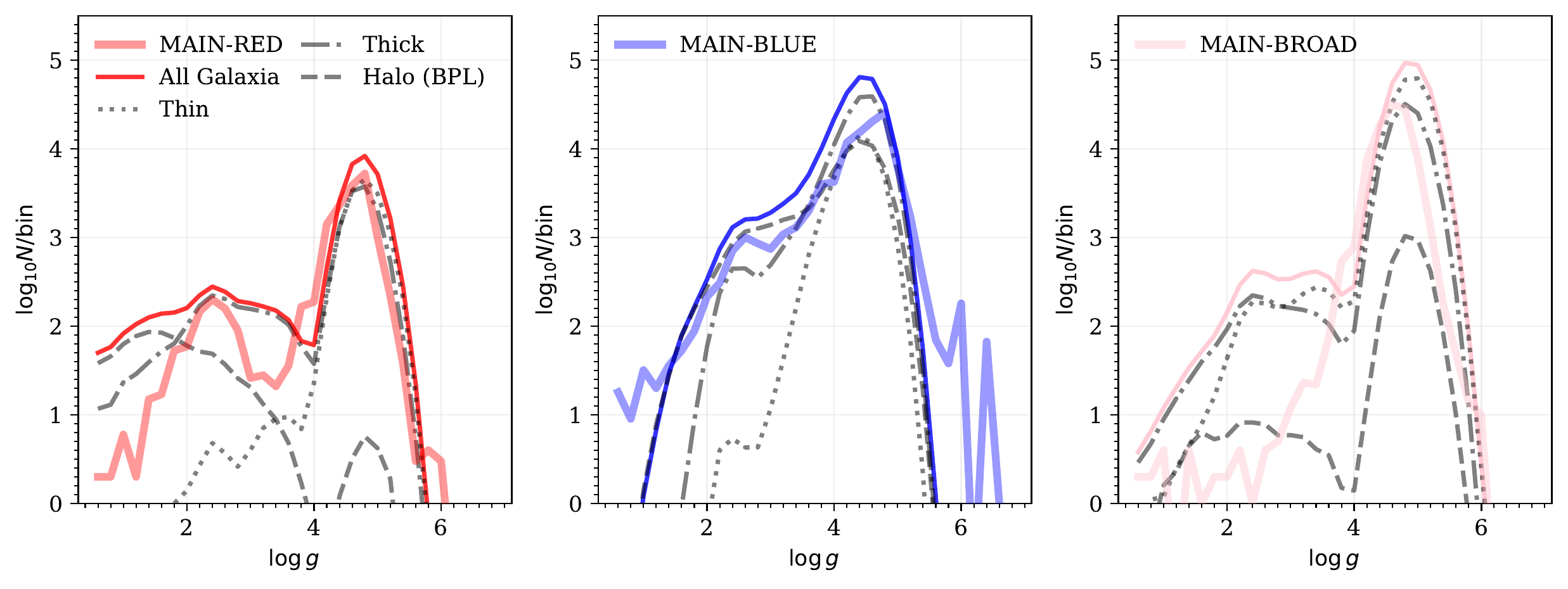}
    \caption{Comparison of our RVS pipeline measurements \cwr{of surface gravity from} the SV3 dataset (thick lines) to the predictions of our \cwr{modified (broken-power-law halo)} Galaxia model (thin colored lines) selected in the same regions. \cwr{From left to right, the panels show the} \mainblue{}, \mainred{} and \mainbroad{} selections. Individual structural components in Galaxia are shown by dotted (disk), dotted-dashed (thick disk), and dashed gray lines (halo, with our fiducial broken-power-law density profile; see text).}
    \label{fig:sv3_galaxia}
\end{figure*}

Fig.~\ref{fig:sv3_feh_logg_split} shows separate metallicity distributions for dwarfs and giants. \mainblue{} giants (median $\mathrm{[Fe/H]}\approx -1.5$) are clearly sampling the metal-poor halo. \mainred{} giants are notably more metal-rich \cwr{($\mathrm{[Fe/H]}\approx -0.9$)}, with a distribution similar to that of \mainblue{} dwarfs. The \mainred{} selection is weighted toward the lower RGB out to $\sim30$~kpc (at larger distances the density of giants falls rapidly), and is therefore likely to be dominated by the relatively metal-rich Gaia Enceladus feature \citep{Belokurov_2018, Helmi_2018_GSE, Naidu_etal_2020}. The \mainred{} and \mainbroad{} dwarf samples have median $\mathrm{[Fe/H]}\approx -0.4$ and $-0.3$ respectively, consistent with the expectation that they are dominated by the outer thin and thick disks. The \mainblue{} dwarf sample is biased toward the metal-poor tail of those populations.

Fig.~\ref{fig:cmd_giants_split_feh} provides another perspective on the same data, showing color--magnitude diagrams for giants in different ranges of metallicity. The isochrones in this figure (from the MIST library; \citealt{Paxton:2011tk, Paxton:2013vg, Paxton:2015tv, Choi:2016wb, Dotter:2016wg}) illustrate the types of stars entering the selection function at different distances. A metal-rich structure is clearly visible in the \mainred{} sample at 30-50~kpc. 

The figures above demonstrate that the MWS main target selection function will yield a representative survey of distant Galactic structures at high latitude. Comparison to predicted distributions for the same fields by the Galaxia model \citep{sharma2011} supports this conclusion. The metallicity and velocity distributions for the \svthree{} data have shapes and amplitudes in broad agreement with the default Galaxia model. Our fiducial broken-power-law halo variant (see section \ref{sec:main_targets}) improves this agreement by reducing the counts of distant red giants in the \mainred{} sample. Some discrepancy in detail between Galaxia and the \svthree{} data is expected: the MWS selection is sensitive to stars from the Gaia Enceladus progenitor and the Sagittarius stream (which are not included in the Galaxia model) and to the slope of the smooth stellar halo density profile.

Finally, Fig.~\ref{fig:sv3_galaxia} compares the $\log g$ distributions of the \svthree{} data and our Galaxia mock catalog with a broken-power-law halo (convolved with a fiducial Gaussian error $\sigma_{\log g}=0.2$). Galaxia predicts substantially more high-proper-motion giants in the \mainbroad{} selection than we see in the data. These are associated with the thin and thick disks. The counts of giants in the \mainred{} sample are in reasonable agreement at $\log g \approx 2$ but are overpredicted by Galaxia at both higher $\log\,g$ (an excess of thick disk giants) and lower $\log\,g$ (an excess of halo giants). The latter difference is sensitive to the outer slope of the density profile. 

\section{Summary}
\label{sec:conclusion}

We have presented the scientific motivation and final target selection criteria for DESI MWS, which will obtain radial velocities, stellar parameters, and chemical abundances for approximately seven million stars over contiguous footprints in the north and south Galactic caps totaling $\sim14,000\mathrm{deg^2}$. The overarching goal of \MWS{} is to provide new constraints on the assembly history of the Milky Way and the distribution of dark matter through measurements of chemical composition and radial velocity. The main \MWS{} sample focuses on stars in the outer thick disk and stellar halo. It is essentially magnitude-limited for colors $g-r < 0.7$ (the halo and thick disk turnoff; \mainblue{}) and has only a weak astrometric selection applied to redder stars to favor the targeting of distant giants (\mainred{}). These two samples will have $\simeq30\%$ spectroscopic completeness. Red stars failing the astrometric selection (\mainbroad{}) will still be observed, with spectroscopic completeness $\simeq19\%$. This straightforward and inclusive selection function will greatly simplify the forward modeling of the survey. It will also favor serendipitous discoveries; essentially every $16<r<19$ star in the DESI footprint has a finite probability of being observed by \MWS{}. Combined with Gaia, the main \MWS{} sample will be an excellent resource for mapping structures and substructures in the thick disk and accreted halo, probing distances from 1 to $\gtrsim100$~kpc with multiple overlapping tracer populations. \MWS{} will also provide highly complete samples of white dwarfs, faint M-dwarfs, BHBs, and RR Lyrae variables. These samples will address the star formation history of the Milky Way disk, provide additional probes of the Galactic structure and serve as a resource for a wide range of stellar and planetary physics. MWS will be supplemented by a backup \cwr{observing program} targeting brighter stars and \cwr{several} more specific secondary science programs, which will be described \reply{in a separate publication}. 

The first DESI data release \citep{dr} will include approximately 500,000 stellar spectra from the \MWS{} SV program, introduced in Sec.~\ref{sec:sv}. \cwr{We have used these data to show that, with our current analysis pipelines}, DESI can measure stellar radial velocities to an accuracy $\simeq1\,\mathrm{km\,s^{-1}}$ and [Fe/H] to $\sim0.2\,\mathrm{dex}$, for representative main-sequence turnoff and giant branch halo tracers in our main sample, given observing conditions comparable to those used for \MWS{} targets in the full bright-time \MWS{}. \cwr{This velocity precision meets the requirements of the scientific program described in Section~\ref{sec:goals}. It is sufficient to distinguish kinematic substructures (for example, tidal streams and shells) from the bulk stellar halo, to explore large-scale velocity perturbations associated with the LMC, and to characterize the internal velocity dispersion of features as dynamically cold as the tidal streams of globular clusters.}

We have quantified systematic offsets in our measurements with respect to previous spectroscopic surveys and between our two main analysis pipelines. The discrepancies are most significant \cwr{in the parameters and abundances measured for} very cool stars. \cwr{The results in Section~\ref{sec:pipeline_feh} give us confidence that [Fe/H] uncertainties from DESI are, at least, comparable to those from other surveys at similar resolution, such as SDSS SEGUE. The random [Fe/H] uncertainties span a range of 0.05 dex for our brightest targets to 0.4 dex for the faintest. When comparing to external high-resolution measurement of clusters and dwarf galaxies, we see a typical systematic offset of $\sim 0.15$~dex in the DESI measurements. However, this varies from system to system, likely indicating systematic errors that are a function of temperature, $\mathrm{[Fe/H]}$ or $\mathrm{[\alpha/Fe]}$. We plan to investigate and where possible correct these deficiencies in preparation for the first DESI data release.}

The radial velocity, metallicity, and surface gravity distributions we obtain from the highly complete and higher-S/N \svthree{} minisurvey over $\approx100\,\mathrm{deg}^{2}$ are broadly consistent with predictions from the Galaxia Milky Way model \citep{sharma2011}. The agreement in giant counts is greatly improved if we assume a steeper decline in the density of halo stars at Galactocentric distances $\gtrsim25$~kpc, relative to Galaxia's original "unbroken"-power-law halo density profile. A mock realization of the full \MWS{} survey based on this modified Galaxia model predicts $\simeq1$~million \MWS{} targets beyond 15~kpc, $\sim100,000$ beyond 30 kpc and $\sim1000$ beyond 100~kpc, of which we will be able to observe $\approx30\%$. We therefore expect \cwr{spectrophotometric} distance estimates based on \MWS{} data to provide strong constraints on the outer density profile of the stellar halo.

\MWS{} is one of several active or near-future large-scale Galactic spectroscopy projects, including H3 \citep[a smaller medium-resolution stellar survey with similar goals to MWS;][]{H3}, \reply{4MOST \citep[in the southern hemisphere;][]{4most} and WEAVE \citep[in the northern hemisphere;][]{Dalton2012}. The 4MOST and WEAVE spectrographs have a high-resolution mode ($R\sim20,000$) in addition to an $R\sim 5000$ mode capable of surveys comparable in scope to MWS}. These surveys complement one another. 
\MWS{} is particularly well suited to a large-scale mapping of the outer thick disk, the Gaia Enceladus debris, the Sagittarius stream, and other northern-sky overdensities. The MWS data should greatly improve our understanding of how these features contribute to the overall stellar halo density, metallicity, and velocity dispersion profiles, and provide a robust foundation for detailed comparisons with theoretical \cwr{models}.


\vspace{1.5\baselineskip}

\reply{We thank the reviewer, Matthias Steinmetz, for his thorough reading of our manuscript and constructive suggestions}. A.P.C. and N.K. are supported by a Taiwan Ministry of Education (MoE) Yushan Fellowship awarded to A.P.C., and by Taiwan National Science and Technology Council (NSTC) grant 109-2112-M-007-011-MY3. This work used high-performance computing facilities operated by the Center for Informatics and Computation in Astronomy (CICA) at National
Tsing Hua University. This equipment was funded by MoE, NSTC, and National Tsing Hua University. C.A.P. acknowledges financial support from the Spanish Ministry of Science and Innovation (MICINN) projects AYA2017-86389-P and PID2020-117493GB-I00. C.J.M. acknowledges financial support from Imperial College London through an Imperial College Research Fellowship grant. M.V. and L.B.e.S. acknowledge support from NASA-ATP award 80NSSC20K0509. The work of A.D. and J.N. is supported by NOIRLab, which is managed by the Association of Universities for Research in Astronomy (AURA) under a cooperative agreement with the National Science Foundation. T.S.L. acknowledges financial support from the Natural Sciences and Engineering Research Council of Canada (NSERC) through grant RGPIN-2022-04794.
B.T.G. was supported by grant ST/T000406/1 from the Science and Technology Facilities Council (STFC). This project has received funding from the European Research Council under the European Union’s Horizon 2020 research and innovation program (grant agreement No. 101020057). \proof{This research made use of computing time available on the high-performance computing systems at the Instituto de Astrofisica de Canarias. The authors thankfully acknowledge the technical expertise and assistance provided by the Spanish Supercomputing Network (Red Espanola de Supercomputacion), as well as the computer
resources used: the LaPalma Supercomputer and the Diva cluster, both located at the Instituto de Astrofisica de Canarias.}


This research is supported by the Director, Office of Science, Office of High Energy Physics of the U.S. Department of Energy (DOE) under contract No. DE–AC02–05CH11231, and by the National Energy Research Scientific Computing Center, a DOE Office of Science User Facility under the same contract. Additional support for DESI is provided by the U.S. National Science Foundation, Division of Astronomical Sciences, under contract No. AST-0950945 to the NSF’s National Optical-infrared Astronomy Research Laboratory; the Science and Technology Facilities Council of the United Kingdom; the Gordon and Betty Moore Foundation; the Heising-Simons Foundation; the French Alternative Energies and Atomic Energy Commission (CEA); the National Council of Science and Technology of Mexico (CONACYT); MICINN; and the DESI Member Institutions: \url{https://www.desi.lbl.gov/collaborating-institutions}.

The DESI Legacy Imaging Surveys consist of three individual and complementary projects: the Dark Energy Camera Legacy Survey (DECaLS), the Beijing--Arizona Sky Survey (BASS), and the Mayall $z$-band Legacy Survey (MzLS). DECaLS, BASS, and MzLS include data obtained, respectively, at the Blanco telescope, Cerro Tololo Inter-American Observatory, NSF’s NOIRLab; at the Bok telescope, Steward Observatory, University of Arizona; and at the Mayall telescope, Kitt Peak National Observatory, NOIRLab. NOIRLab is operated by the Association of Universities for Research in Astronomy (AURA) under a cooperative agreement with the National Science Foundation. Pipeline processing and analyses of the data were supported by NOIRLab and the Lawrence Berkeley National Laboratory (LBNL). The Legacy Surveys also use data products from the Near-Earth Object Wide-field Infrared Survey Explorer (NEOWISE), a project of the Jet Propulsion Laboratory/California Institute of Technology, funded by the National Aeronautics and Space Administration. The Legacy Surveys were supported by the Director, Office of Science, Office of High Energy Physics of the U.S. DOE; the National Energy Research Scientific Computing Center, a DOE Office of Science User Facility; the U.S. National Science Foundation, Division of Astronomical Sciences; the National Astronomical Observatories of China, Chinese Academy of Sciences, and the Chinese National Natural Science Foundation. LBNL is managed by the regents of the University of California under contract to the U.S. DOE. The complete acknowledgments can be found at \url{https://www.legacysurvey.org/}.


This work has made use of data from the European Space Agency (ESA) mission {\it Gaia} (\url{https://www.cosmos.esa.int/gaia}), processed by the {\it Gaia} Data Processing and Analysis Consortium (DPAC, \url{https://www.cosmos.esa.int/web/gaia/dpac/consortium}). Funding for the DPAC has been provided by national institutions, in particular the institutions participating in the {\it Gaia} Multilateral Agreement.

Funding for SDSS-IV has been provided by the Alfred P. Sloan Foundation, the U.S.\ DOE of Science, and the participating institutions. SDSS-IV acknowledges support and resources from the Center for High Performance Computing  at the University of Utah. The SDSS website is \url{www.sdss.org}.

SDSS-IV is managed by the Astrophysical Research Consortium for the Participating Institutions of the SDSS Collaboration including the Brazilian Participation Group, the Carnegie Institution for Science, Carnegie Mellon University, the Center for Astrophysics $|$ Harvard \& Smithsonian, the Chilean Participation Group, the French Participation Group, Instituto de Astrof\'isica de Canarias, the Johns Hopkins University, the Kavli Institute for the Physics and Mathematics of the Universe (IPMU) / University of Tokyo, the Korean Participation Group, LBNL, Leibniz Institut f\"ur Astrophysik Potsdam (AIP),  Max-Planck-Institut f\"ur Astronomie (MPIA Heidelberg), Max-Planck-Institut f\"ur Astrophysik (MPA Garching), Max-Planck-Institut f\"ur Extraterrestrische Physik (MPE), the National Astronomical Observatories of China, New Mexico State University, New York University, University of Notre Dame, Observat\'ario Nacional / MCTI, the Ohio State University, Pennsylvania State University, Shanghai Astronomical Observatory, the United Kingdom Participation Group, Universidad Nacional Aut\'onoma de M\'exico, University of Arizona, University of Colorado Boulder, University of Oxford, University of Portsmouth, University of Utah, University of Virginia, University of Washington, University of Wisconsin, Vanderbilt University, and Yale University.

The Guoshoujing Telescope (the Large Sky Area Multi-object Fiber Spectroscopic Telescope LAMOST) is a National Major Scientific Project built by the Chinese Academy of Sciences. Funding for the project has been provided by the National Development and Reform Commission. LAMOST is operated and managed by the National Astronomical Observatories, Chinese Academy of Sciences.

For the purpose of open access, the authors have applied a Creative Commons attribution (CC BY) license to any author-accepted manuscript version arising from this submission. 

%
The authors are honored to be permitted to conduct scientific research on Iolkam Du’ag (Kitt Peak), a mountain with particular significance to the Tohono O’odham Nation.

For more information, visit \url{desi.lbl.gov}
%

\software{NumPy \citep{numpy}, SciPy \citep{scipy}, Astropy \citep{astropy:2013,astropy:2018}, Matplotlib \citep{matplotlib}, PyGaia (A. Brown; Gaia Project Scientist Support Team and the Gaia DPAC; \url{https://github.com/agabrown/PyGaia}),  galstreams \citep{galstreams_software}.}

\facilities{Mayall (Mosaic3), Bok (90Prime), Blanco (DECam), WISE, Gaia.}

\bibliographystyle{aasjournal}
\bibliography{bibliography}

\end{document}